\pdfoutput=1
\documentclass[sigconf,edbt]{acmart-edbt2025}

\def\BibTeX{{\rm B\kern-.05em{\sc i\kern-.025em b}\kern-.08em
    T\kern-.1667em\lower.7ex\hbox{E}\kern-.125emX}}

\usepackage{booktabs} 

\setcopyright{rightsretained}

\acmDOI{}

\acmISBN{978-3-89318-099-8}

\acmConference[EDBT 2025]{28th International Conference on Extending Database Technology (EDBT)}{25th March-28th March, 2025}{Barcelona, Spain}
\acmYear{2025}

\usepackage{graphicx}
\usepackage{subfigure}
\usepackage{xspace}
\usepackage{url}
\usepackage{algpseudocode}
\usepackage{algorithm}
\usepackage{multirow}
\usepackage{soul}
\usepackage{titlesec}
\usepackage{tikz}
\usepackage{amsfonts}
\usepackage{amsmath}
\usetikzlibrary{calc,decorations.pathreplacing}
\usepackage{enumitem}
\usepackage{mdframed}
\usepackage{multicol}
\usepackage{array}
\usepackage{epstopdf}

\newcommand{\tbf}{\textbf{\textcolor[rgb]{1,0,0}{TBF}}\xspace}

\makeatletter
\DeclareRobustCommand*\cal{\@fontswitch\relax\mathcal}
\makeatother

\newcommand{\eat}[1]{}
\newcommand{\stab}{\rule{0pt}{8pt}\\[-1.6ex]}

\newcommand{\sstab}{\rule{0pt}{8pt}\\[-2.4ex]}

\newcommand{\bi}{\begin{itemize}}
\newcommand{\ei}{\end{itemize}}
\newenvironment{tbi}{\begin{itemize}
        \setlength{\topsep}{1.2ex}\setlength{\itemsep}{0ex}}
        {\end{itemize}\vspace{-1.5ex}}

\newcommand{\be}{\begin{enumerate}}
\newcommand{\ee}{\end{enumerate}}
\newcommand{\beqn}{\begin{eqnarray*}}
\newcommand{\eeqn}{\end{eqnarray*}}

\newcommand{\stitle}[1]{\vspace{.5ex}\noindent{\bf #1}}
\newcommand{\eetitle}[1]{\vspace{0.8ex}\noindent{\underline{\em #1}}}
\newcommand{\etitle}[1]{\vspace{0.8ex}\noindent{\em #1}}

\newcommand{\ie}{\emph{i.e.,}\xspace}
\newcommand{\eg}{\emph{e.g.,}\xspace}
\newcommand{\wrt}{\emph{w.r.t.}\xspace}

\DeclareMathOperator*{\argmax}{arg\,max}
\newcommand{\kw}[1]{{\ensuremath{\mathsf{#1}}}\xspace}

\newcommand{\eop}{\hspace*{\fill}\mbox{$\Box$}}     
\newcounter{example}
\renewenvironment{example}{
        \vspace{1.2ex}
        \refstepcounter{example}
        {\noindent\bf Example \theexample:}}{
        \eop
	\vspace{1.2ex}}

 \newcommand{\warn}[1]{\textcolor{red}{#1}}

\newcommand{\nthesection}{\arabic{section}}

\newcounter{prop}
\renewcommand{\theprop}{\arabic{theorem}}
\newcounter{lemma}

\newcounter{cor}
\renewcommand{\thecor}{\arabic{theorem}}
\newcounter{definition}[section]

\newcounter{alg}[section]
\renewcommand{\thealg}{\nthesection.\arabic{alg}}

\newcounter{arule}
\renewcommand{\thearule}{\arabic{arule}}

\newenvironment{proofS}{
        \vspace{1ex}
        {\noindent\bf Proof sketch:\ }}{\eop\vspace{1ex}}

\newcommand{\comwu}[1]{{\color{red}[com-Wu:~{#1}]}}

\newcommand{\revise}[1]{\textcolor{blue}{#1}}
\newcommand{\mengying}[1]{\textcolor{cyan}{MY: {#1}}}

\newcommand{\hanchao}[1]{\textcolor{orange}
{#1}}

\newcommand{\D}{{\mathcal D}}
\newcommand{\A}{{\mathcal A}}
\newcommand{\F}{{\mathcal F}}
\newcommand{\E}{{\mathcal E}}

\newcommand{\C}{{\mathcal C}}
\newcommand{\V}{{\mathcal V}}
\renewcommand{\O}{{\mathcal O}}
\renewcommand{\S}{{\mathcal S}}
\newcommand{\T}{{\mathcal T}}
\renewcommand{\dag}{\kw{DAG}}
\renewcommand{\P}{{\mathcal P}}
\newcommand{\ad}{\kw{adom}}
\newcommand{\mos}{\kw{MOSP}}
\newcommand{\modata}{\kw{MODis}}

\newcommand{\op}{\kw{op}}

\newcommand{\modis}{\kw{MODis}}
\newcommand{\apxmodis}{\kw{ApxMODis}}
\newcommand{\bimodis}{\kw{BiMODis}}

\newcommand{\nomodis}{\kw{NOBiMODis}}
\newcommand{\divmodis}{\kw{DivMODis}}

\newcommand{\swb}{\kw{BIB}}
\newcommand{\swp}{\kw{BIP}}
\newcommand{\upi}{\kw{UPareto}}
\newcommand{\opg}{\kw{OpGen}}

\newcommand{\kaggle}{\kw{Kaggle}}
\newcommand{\open}{\kw{OpenData}}

\newcommand{\hf}{\kw{HF}}
\newcommand{\kizoo}{\kw{KIZoo}}

\newcommand{\gbm}{\kw{GBmovie}} 
\newcommand{\rfh}{\kw{RFhouse}} 

\newcommand{\lgc}{\kw{LGCmental}}
\newcommand{\lgr}{\kw{LGRmodel}}

\newcommand{\modelone}{\kw{M1}} 
 
\newcommand{\modelthree}{\kw{M3}}

\newcommand{\metam}{\kw{METAM}}
\newcommand{\metammo}{\kw{METAM}-\kw{MO}}
\newcommand{\starmie}{\kw{Starmie}}
\newcommand{\sklearn}{\kw{SkSFM}}
\newcommand{\ho}{\kw{H2O}}
\newcommand{\relp}{\kw{rImp}}
\newcommand{\maxl}{\kw{maxl}}
\newcommand{\margin}{\kw{mg}}
\newcommand{\customsize}{\fontsize{7.75}{10.8}\selectfont}

\newcommand{\ltodo}[2]{\item \makebox[0pt][r]{$\square$\hspace{2em}}#1%
   \ifx\relax#2\relax\else\textbf{ (#2)}\fi}
\newcommand{\ldone}[2]{\item \makebox[0pt][r]{$\boxtimes$\hspace{2em}}\sout{#1}%
   \ifx\relax#2\relax\else\textbf{ (#2)}\fi}

\begin{document}
\title{Generating Skyline Datasets for Data Science Models}


\author{Mengying Wang}
\affiliation{%
  \institution{Case Western Reserve University}
  \city{Cleveland}
  \state{Ohio}
  \country{USA}}
\email{mxw767@case.edu}

\author{Hanchao Ma}
\affiliation{%
  \institution{Case Western Reserve University}
  \city{Cleveland}
  \state{Ohio}
  \country{USA}}
\email{hxm382@case.edu}


\author{Yiyang Bian}
\affiliation{%
  \institution{Case Western Reserve University}
  \city{Cleveland}
  \state{Ohio}
  \country{USA}}
\email{yxb227@case.edu}

\author{Yangxin Fan}
\affiliation{%
  \institution{Case Western Reserve University}
  \city{Cleveland}
  \state{Ohio}
  \country{USA}}
\email{yxf451@case.edu}


\author{Yinghui Wu}
\affiliation{%
  \institution{Case Western Reserve University}
  \city{Cleveland}
  \state{Ohio}
  \country{USA}}
\email{yxw1650@case.edu}



\renewcommand{\shortauthors}{}

\begin{abstract}
Preparing high-quality datasets required 
by various data-driven AI and machine learning 
models has become a cornerstone task in 
data-driven analysis. 
Conventional data discovery methods typically 
integrate datasets towards a single 
pre-defined quality measure 
that may lead to bias for 
downstream tasks.   
This paper introduces \textbf{MODis}, a 
framework that 
discovers 
datasets 
by optimizing {\em multiple} user-defined, 
model-performance measures. 
Given a set of data sources and a model, 
 \textbf{MODis} 
selects and integrates data sources  
into a skyline dataset, 
over which 
the model is expected to have the
desired performance in all the 
performance measures. 
We formulate \textbf{MODis} as 
a multi-goal finite 
state transducer,   
\eat{
in terms of 
finite state transducer with 
finer grained operators enhanced by 
selection conditions. }
and derive three feasible algorithms 
to generate skyline datasets. Our first 
algorithm adopts a ``reduce-from-universal'' strategy, 
that starts with a universal schema and 
iteratively prunes unpromising data. 
Our second algorithm further reduces 
the cost with a bi-directional strategy 
that interleaves 
data augmentation and 
reduction. We also introduce 
a diversification algorithm to mitigate 
the bias in skyline datasets. 
We experimentally verify the 
efficiency and effectiveness of 
our skyline data discovery algorithms, and showcase 
their applications in optimizing 
data science pipelines.   
\end{abstract}

\maketitle

\section{Introduction}
\label{sec:intro}

High-quality machine learning (ML) models have become criticale assets for various domain sciences research. 
A routine task in data-driven domain sciences is to prepare datasets that can be used to improve such data science models. 
Data augmentation~\cite{roh2019survey} and feature selection~\cite{li2017feature} have been 
studied to suggest data for ML models~\cite{doan2012principles}. 
Nevertheless, they 
typically generate data by favoring a pre-defined,  
single performance goal, such as 
data completeness or feature importance. 
Such data may be biased and not very useful to 
actually improve the model performance, 
and moreover, fall short at 
addressing multiple user-defined ML performance  
measures (\eg expected accuracy, training cost). 
Such need 
is evident in multi-variable experiment optimization~\cite{konakovic2020diversity, low2023evolution, paleyes2022challenges}, 
feature selection~\cite{li2017feature}, and
AI benchmarking~\cite{donyavi2020diverse}, among others. 

Discovering datasets that can improve 
a model 
over {\em multiple} user-defined 
performance measures remains 
to be desirable yet less studied 
issue. Consider the following real-world example.

\begin{example}
\label{exa-motivation}
To assess the impact and causes of harmful algal blooms (HABs) in a lake, a research team aims to forecast the chlorophyll-a index (CI-index), a key measure of algal blooms. 
The team has gathered over $50$ factors
(\eg fertilizer, water quality, weather)
of upstream rivers and watershed 
systems, 
and trained 
a random forest (RF) 
with a small, regional dataset.
The team wishes to find new 
data with important spatiotemporal and chemical 
attributes, to generalize the RF model. In particular,  
the model is expected to perform well 
over such dataset  
in terms of three performance measures: 
root mean square error ($RMSE$),  $R^2$ test, for ``Level 2 bloom'' CI-index, 
and training time cost. 
Desirably, 
the data generation process 
can inform {\em what} are crucial features to inspect,  
track {\em where} 
the feature values are from, and {\em how}  
they are integrated from the data sources.

\vspace{.5ex}
The research team may issue
a {\em skyline query}~\cite{chomicki2013skyline} that requests: 

\begin{figure}[tb!]
\vspace{-1ex}
\centerline{\includegraphics[width =0.45\textwidth]{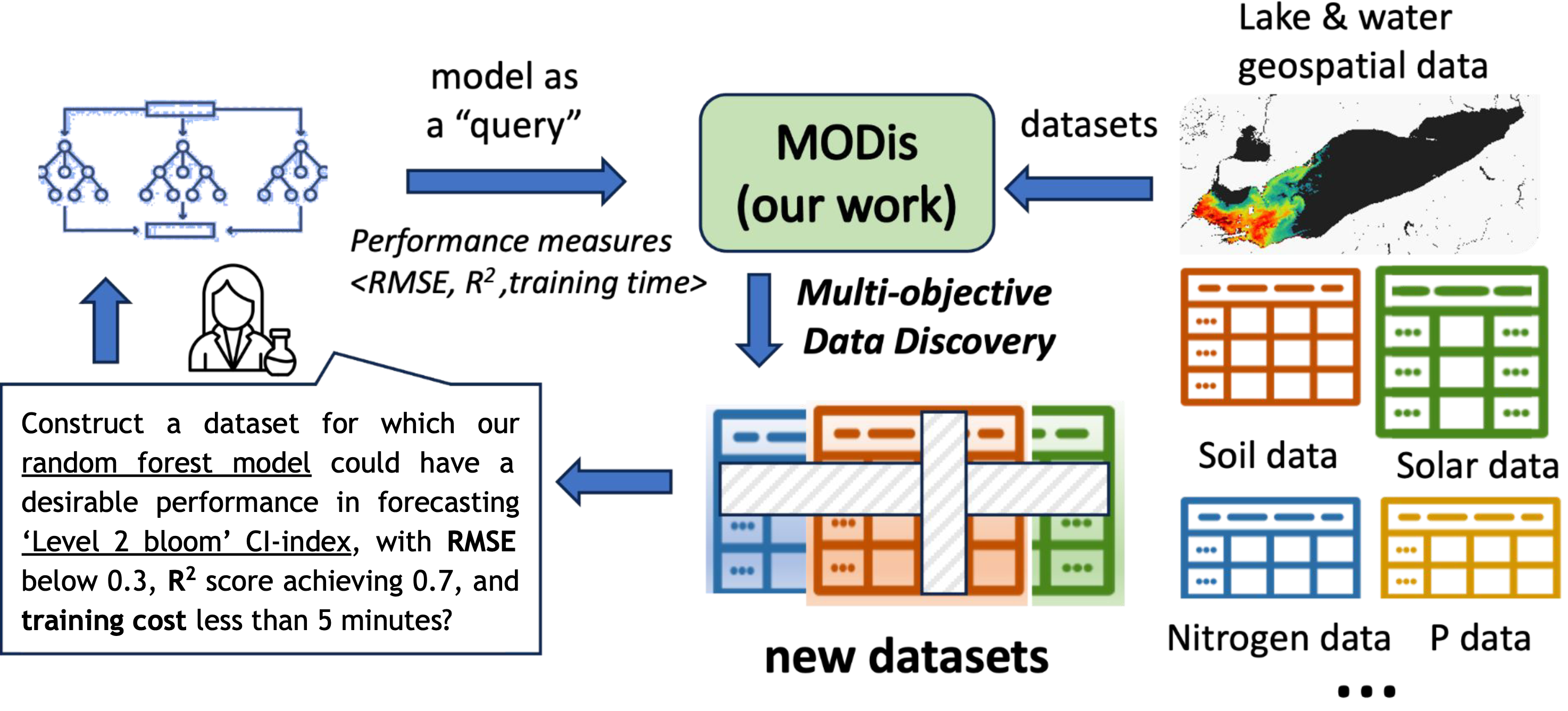}}
\centering
\vspace{-1ex}
\caption{
Data generation for CI index prediction addressing  
multiple user-defined ML performance criteria, in order to improve an input ML model.  
}
\label{fig:motivation}
\vspace{-2ex}
\end{figure}

\vspace{-1ex}
\begin{center}
\fbox{
\begin{minipage}{.46\textwidth}
``{\em Generate a \underline{dataset} for which our \underline{random forest model} for predicting 
`Level 2 bloom' CI-index
is expected to have a 
\underline{RMSE} below $0.3$, \underline{$R^2$} score at least $0.7$, and incur a \underline{training cost} in 5 minutes?}'' 
\end{minipage}
}
\end{center}

Here, the thresholds  
``$0.3$'', ``$0.7$'', and ``$5$ minutes'' 
are set based on historical  
performance of the  
RF model over 
a data sample. 

One may apply data integration, or 
perform feature engineering to 
refine existing datasets with important 
features. Nevertheless, these 
methods often fall short at consistently 
generate data towards optimizing 
user-defined ML performance, leaving 
alone the needs for addressing multiple measures 
\eg accuracy and 
training cost. 

Another approach is to 
introduce a utility function 
as a linear weighted sum 
of multiple measures. This 
turns the need into a single objective.  
However, achieving both high accuracy and 
low training cost 
can be ``conflicting''; moreover, a best dataset 
that optimizes such utility function may not 
necessarily satisfy the expected bounds as posed  
for each measure in the query. 

Ideally, a data generation process should provide 
a dataset that 
ensures the model achieves best expected performance on at least one measure, with compromisingly good performance on the rest, and all satisfying the user-defined bounds if any. 
\eat{
With an excessive 
number of trials that repeats  
a pipeline of (manual) hyper-parameter tuning,  
feature selection, data integration, 
cross-validation, and inference tests, 
the team has 
eventually verified that only two datasets (a basin area and 
an upstream river) suffices to 
ensure the desired performance of the model, among 
which only ``\texttt{Nitrogen}'' and ``\texttt{Phosphorus}'' 
 jointly determines CI-index,  
and
not all records are needed.  
For short-term prediction in 2023, 
it suffices to use the seasonal ``\textit{Spring}'' and the years ``\textit{2013}'' and ``\textit{2015}'' due to similar high nutrition loads~\cite{ai2023short}. Nevertheless, 
it took a substantial time to construct such a dataset.
}
\end{example}

\eat{
A typical process involves an excessive 
number of trials that repeats  
a pipeline of (manual) hyper-parameter tuning,  
feature selection, data integration, 
cross-validation, and inference tests. It turns out that 
 only $8$ key features 
with certain ranges of values are 
needed to achieve a regression 
with desirable accuracy at $89.6\%$
(quantified by RMSE and $R^2$ test). 
This process took the team 8 weeks to get the desired model and the following 
observations. 

\sstab
(1) Not all the water data tables 
from every upstream river system are needed, but only 
two -  a basin area and 
an upstream river. 

\sstab
(2) A single physicochemical attribute 
``\texttt{Phosphorus}'' alone does not determine 
C1-index well for recent years, but ``\texttt{Nitrogen}'' and ``\texttt{Phosphorus}'' 
 jointly determines CI-index, as an 
emerging pattern. 

\sstab
(3) The original data contains, for both columns ``\texttt{Nitrogen}'' and ``\texttt{Phosphorus}'', 
the values over the past $11$ years. However, for an accurate short-term prediction in 2023, 
the fraction of the seasonal ``\textit{Spring}'' data and in particular years ``\textit{2013}'' and ``\textit{2015}'' 
are important, due to high nutrition loads.
}

The above example calls for data generation approaches that can respond to 
the question by providing ``skyline'' 
datasets that can address multiple ML performance measures. 
More formally, given a query that specifies an input data science model $M$, 
a set of source tables $\D$ = $\{D_1, \ldots D_n\}$, and a set of user-defined performance measures $\P$ (\eg accuracy, training time), our task is to generate a new table from $\D$, over which the expected performances of $M$ simultaneously reaches desirable goals for all measures in $\P$. 
As remarked earlier, traditional data 
integration and feature engineering 
with a pre-defined, single optimization 
objective falls short of generating data 
for such needs. 


\eat{
As remarked earlier, manipulating datasets from 
scratch is expensive. One may apply 
data integration to generate tables 
from ``joinable'' tables, and verify 
them by testing the performance 
of the model.  Nevertheless, this 
incurs excessive join operations, leaving alone the cost from model inferences, 
fine-tuning, and 
additional in-lab validation 
(\eg experiments, simulations)~\cite{paleyes2022challenges}. 
Moreover, conventional data 
integration optimizes data completeness 
rather than model performance, 
leaving alone the need 
for data discovery that 
simultaneously satisfy 
multiple performance goals. 
}

Moreover, 
a desirable data generation process 
should 
(1) declaratively produce  
such a data by simple, primitive operators 
that are well supported by established  
query engines and 
data systems, 
(2) perform data discovery 
without expensive 
model inference and validation; and 
(3) ensure quality guarantees on 
the resulting skyline dataset, 
for multiple 
performance measures. 
In addition, the generation should be 
efficient. This remains a challenging 
issue, considering 
large-scale data sources and 
the space of new datasets that 
can be generated from them.

\vspace{.5ex}
\stitle{Contribution}. 
We introduce \modis, a 
multi-objective data discovery 
framework. \modis 
interacts data integration 
and ML model performance estimation 
to pursue a {\em multi-objective} 
data discovery paradigm. We summarize 
our contributions as follows. 

\stab
(1) 
We provide a formal computation model for 
the skyline data generation process in 
terms of a 
finite state transducer (FST). 
An FST extends finite automata by associating 
an output artifact that undergoes modifications 
via sequences of state transitions. 
The formal model is 
equipped with (1) simple and primitive 
operators, 
and (2) a model performance oracle 
(Section~\ref{sec-system}). 
We use FST as an abstract 
tool to describe 
data generation algorithms and 
perform formal analysis to 
verify costs and 
provable quality guarantees. 
\eat{
We study the expressiveness  
and property of the 
system. 
We show that the 
system resembles 
of data integration 
and feature selection. 
Better still, its 
computation demonstrates 
confluence property and 
a Church Russel property. }

\sstab
(2) Based on the 
formal model, 
we
introduce the skyline data generation 
problem, 
in terms of Pareto optimality 
(Section~\ref{sec-problem}) . 
The goal is to generate  
a skyline set of 
datasets, ensuring each has at least one 
performance measure where the model's expected performance is no worse than any other dataset.
While the problem is intractable, 
we present a fixed-parameter 
tractable result, for a polynomially bounded dataset exploration space from the running graph of an FST process, 
and a fixed measures set $\P$. 

\vspace{.5ex}
Based on the above formulation, we provide 
three feasible algorithms to generate 
skyline datasets. 

\sstab 
(3) Our first algorithm provides an approximation 
on Pareto optimal datasets by 
exploring and verifying a bounded number of 
datasets that can be generated from 
data sources 
(Section~\ref{sec-apxmodis}). 
The algorithm adopts a 
``reduce-from-universal'' strategy 
to dynamically drop  
values from a universal dataset towards 
a Pareto optimal set of tables. 
We show that this algorithm 
approximates Skyline set within 
a factor of $(1+\epsilon)$ 
for all performance metric, 
and ensures exact dominance 
for at least one measure.
In addition, we present a 
special case with a 
fully polynomial time 
approximation. 

\sstab 
(4) Our second 
algorithm further 
reduces unnecessary 
computation. It follows a bi-directional scheme to 
prune unpromising
data, and leverages   
a correlation analysis 
of the performance metrics 
to early terminate the 
search 
(Section~\ref{sec-bimodis}). 

\eat{
The first exploits 
known correlations of the 
performance metrics 
to perform early termination 
and pruning, via a bi-directional 
search scheme. }

\sstab 
(5)
Moreover, we  
introduce a diversification 
algorithm to mitigate the  
impact of data bias (Section~\ref{sec-divmodis}).  
We show that the algorithm achieves a  
$\frac{1}{4}$-approximation to an optimal 
diversified skyline dataset among 
all verified $(1+\epsilon)$ counterparts. 

\eat{
\sstab 
(4) We present  
an approximation algorithm that 
supports {\em configurable} trade-off between 
the utility of the dataset  
and their sizes. The algorithm 
fast computes a set of size-bounded
datasets that approximate a pareto optimal 
solution with a quality guarantee 
in terms of $\epsilon$-Pareto optimality. 
We also introduce optimization 
techniques that utilize 
the correlation among 
utilities for early 
pruning and termination. 
}

\eat{
\sstab
(5) We finally introduce an 
efficient maintenance algorithm to 
maintain the datasets 
upon the updates of the 
underlying source dataset. 
}

\vspace{.5ex}
Using real benchmark datasets and tasks, we experimentally 
verify the effectiveness of our data discovery 
scheme. We found that \modis is practical 
in use. For example, our algorithms take 
$30$ seconds to generate new data, 
that can improve input 
models by 1.5-2 times in accuracy 
and simultaneously reduces their training cost by 1.7 times. 
It outperforms baseline approaches  
that separately performs data integration  
or feature selection; 
and remains feasible for larger 
datasets. Our case study 
also verified its practical application 
in domain science tasks.

\stitle{Related works}. We categorize related works as follows. 

\eetitle{Feature Selection.} Feature selection removes irrelevant and redundant attributes and identifies important ones for model training~\cite{li2017feature}. Filtering methods rank features in terms of correlation or mutual information~\cite{peng2005feature, nguyen2014effective} and choose the top ones. They typically assume  
linear correlation among features, omitting collective 
effects from feature sets and hence are often 
limited to support directly optimizing 
model performance. 
\eat{Wrappers~\cite{kohavi1997wrappers} choose features based on model performance estimated by a predictive model. The search cost is often extensive. Moreover, 
they risk overfitting due to optimizing a single  
criteria \eg model accuracy. Embedded methods directly learn feature selection models such as Lasso Regression, by penalizing (layer-wise) feature weights~\cite{zhang2019feature}. }
Our method differs from 
feature selection in the following. 
(1) It generates skyline dataset with primitive data augmentation and reduction operators, 
beyond simply dropping the entire columns. 
(2) We generate data that improves the model over multiple ML performance measures, beyond retaining critical features;  
and (3) our method does not require internal knowledge of the models or incur learning overhead. 

\eetitle{Data Augmentation.} Data augmentation aims to create data from multiple data sources 
towards a unified view~\cite{doan2012principles, ziegler2007data,roh2019survey,esmailoghli2023blend}. It is often specified 
to improve data completeness and richness~\cite{roh2019survey}
and may be sensitive to the quality of schema.  
Our method aimes to generate data to improve the expected performance of  data-driven models. This is different from the conventional data integration which mostly focuses on improving the data completeness.
Generative data augmentation~\cite{desmet2024hydragan} synthesize new rows for multi-objective optimization with a predefined schema. In contrast, \modis generates data with both rows and columns manipulation.  
Also, HydraGAN requires a target column for each metric, while \modis supports user-defined metrics with configurable generation. 

\eetitle{Data Discovery}. 
\eat{
Closer to our work is goal-driven data discovery~\cite{galhotra2023metam}. 
The methods perform data augmentation  
(joins) to enrich input data to improve 
the performance of given tasks.}
Data discovery 
aims to prepare datasets for ML models~\cite{laure2018machine,roh2019survey,esmailoghli2023blend,huang2023kitana,galhotra2023metam}. 
For example, Kitana~\cite{huang2023kitana} computes data profiles (\eg MinHash) and factorized sketches for each dataset to build a join plan, and then evaluates the plan using a proxy model. 
\metam~\cite{galhotra2023metam} involves the downstream task with a utility score for joinable tables.  
Comparing with prior work,  we formalize data generation with cell-level operators, beyond joins. 
We target multi-objective datasets and provide 
formalization in terms of Pareto optimality. 
We also provide algorithms with quality guarantees and 
optimization techniques.

\eetitle{Model Estimation}. Model 
estimation aims to 
provide accurate estimation of 
a model's performance without 
incurring expensive re-training 
and inference cost. 
For example, AutoML~\cite{ hwang2018fast, nguyen2020avatar, yang2020automl} train 
-surrogate models to estimate model performance~\cite{ hwang2018fast, nguyen2020avatar, yang2020automl}, or predict the model performance by learning from past attempts~\cite{feurer2018practical} or Reinforcement Learning~\cite{drori2019automatic}. 
Model selection~\cite{wang2023selecting} 
leverages metadata and historical observations to build graph neural network-based estimator for estimating model performance. 
Our work leverage Multi-output Gradient Boosting as the surrogate model for fast and 
reliable estimation, and benefits 
from established ML performance  
estimation approaches or other surrogate models.

\vspace{-1ex}
\section{Models and Performance Evaluation}
\label{sec:problem}

We start with several notations used in \modis framework. 

\stitle{Datasets}.
A dataset $D(A_1, \ldots A_m)$ is a structured 
table instance that conforms to a local 
 schema $R_D(A_1,\ldots A_m)$.  
Each tuple $t\in D$ is a m-ary vector, 
where $t.A_i$ = $a$ ($i\in [1,m]$) means 
the $i$th attribute $A_i$ of $t$ is assigned a value 
$a$. A dataset may have missing values 
at some attribute $A$ (\ie $t.A$ = $\emptyset$). 

Given a set of datasets $\D$ = $\{D_1, \ldots D_n\}$, each dataset $D_i$ confirms to a local schema $R_i$. 
The {\em universal schema} 
$R_U$ is the union of the local schemas of 
datasets in $\D$, \ie a set of
all the attributes involved in $\D$. 
\eat{
\revise{Here, $U$ refers to the unified set of attributes derived from $R_U$, encompassing all attributes available across $\D$. This forms the foundation for constructing the universal table $D_U$, which includes all joinable rows and attributes from $\D$ and $R_U$, providing a comprehensive dataset for specific tasks and serving as the start point of the discovery process.}
}
The {\em active domain} of an attribute $A$ 
from $D_U$, denoted as $\ad(A)$, 
refers to the finite 
set of its distinct values occurring in $\D$. 
The size of $\ad(A)$, 
denoted as $|\ad(A)|$, is  
the number of distinct values of 
$A$ in $\D$.

\stitle{Models}. A data science model (or simply ``model'') is a function 
in the form of 
$M: D \rightarrow \mathbb{R}^d$, which takes as input a 
dataset $D$,  and outputs  
a result embedding in $\mathbb{R}^d$ 
for some $d\in \mathbb{N}$. Here 
$\mathbb{R}$ and $\mathbb{N}$ 
are real and integer sets. 
In practice, 
$M$ can be a pre-trained machine learning model, 
a statistical model, or a simulator.
The input $D$ may represent a feature matrix 
(a set of numerical feature vectors), or a tensor (from real-world physical systems), to be used 
for a data science model $M$ 
as training or testing data. 
The output embedding 
can be conveniently 
converted to 
task-dependent output (\eg labels for classification, 
discrete cluster numbers for
clustering, or Boolean values
for outlier 
detection) with post-processing.

\eat{
  For example, given an input 
image feature matrix $D$,  a classifier 
$M$ assigns labels converted from a label probability matrix $M(D)$. 
}

\eetitle{Fixed Deterministic models}. 
We say a model $M$ is {\em fixed}, if 
its computation process does not change  
for fixed input. 
For example, a regression model 
$M$ is fixed if any factors 
that determines its inference   
(\eg  number of layers, learned model 
weights) remain fixed. 
The model $M$ is {\em deterministic} 
if it always outputs the same result 
for the same input. 
We consider fixed, deterministic models for the needs of 
consistent performance, which is a 
desired property in ML-driven data analytical tasks. 

\vspace{.5ex}
\stitle{Model Evaluation}. 
A {\em performance measure} $p$ (or simply ``measure'')
is a performance indicator of a model $M$, such as accuracy 
\eg precision, recall, F1 score (for classification); 
or mean average error (for regression analysis). 
It may also be a 
cost measure such as 
training time, 
inference time, or memory consumption. 

\eat{For example, a test $t$ = $(M,D, \{`F_1', `\text{inference cost}'\})$ 
of an image classifier $M$ specifies 
its performance over image 
dataset $D$ 
in terms of $F_1$ score 
and inference time cost. }


\vspace{.5ex}
We use the following settings.

\sstab
(1) We unify $\P$ as a set of normalized measures to {\em minimize}, with a range $(0,1]$. 
Measures to be maximized (\eg accuracy)  
can be easily converted to an inversed counterpart (\eg relative error).

\sstab
(2) Each measure $p\in \P$ has 
an optional range $[p_l, p_u]\in (0, 1]$. 
It specifies desired 
lower bound $p_l$ or an upper bound $p_u$ for model 
performance, such as 
maximum training or inference time, memory capacity, 
or error ranges. 

\stitle{Remarks}. 
As we unify $\P$ as a set of measures to be minimized, it is intuitive that an upper bound $p_u$ specifies a ``tollerence'' for the estimated performance. 
We necessarily introduce a ``lower bound'' $p_l > 0$ 
for the convinience of (1) ensuring well-defined theoretical 
quality guarantees (as will be discussed 
in Section~\ref{sec-apxmodis}), and (2) leaving the option 
open to users for the configuration needs of downstream tasks such as testing, comparison or benchmarking. 

\begin{table}[tb!]
\begin{small}
\begin{tabular}{|c|c|}
\hline
\textbf{Symbol} & \textbf{Notation} \\ 
\hline
$\D$, $D$, $D_U$ & a set of datasets, a single dataset, universal table\\
\hline
$R_D$, $R_U$ & local schema of $D$, and universal schema\\
\hline
$\A$, $A$, $\ad(A)$ & attribute set, attribute, and active domain\\
\hline
$M$ & a data science model $D\rightarrow \mathbb{R}^d$\\
\hline
$\P$, $p$, $(p_l,p_u)$ & perform. measures, a measure, its range \\
\hline
$T$, $t$ = $(M,D,\P)$, $t.\P$ 
& test set; single test, its performance vector \\
\hline
$\T$ = $(s_M, \S, \O, \S_F, \delta)$ & a data discovery system\\
\hline 
$\E$ & a performance estimation model \\
\hline 
$C$ = $(s_M, \O, M, T, \E)$ & a configuration of data discovery system \\
\hline
$G_\T$ = $(\V,\delta)$ & running graph \\
\hline
$s\prec s'$, $D\prec D'$ & state dominance, dataset dominance \\ \hline
\hline
\end{tabular}
\caption{Table of notations}
\vspace{-8ex}
\label{tab:notation}
\end{small}
\end{table}

\eetitle{Estimators}. 
A performance measure $p\in\P$
can often be efficiently estimated by 
an estimation model $\E$ 
(or simply ``estimator''), in PTIME 
in terms of $|D|$ 
(the number of tuples in $D$). 
An estimator $\E$ makes use of a set of 
historically observed performance  
of $M$ (denoted as 
$T$) 
 to infer its performance over a new dataset. 
It can be a regression model that learns from historical 
tuning records $T$ to predict 
the performance of $M$ given a new 
dataset $D$. 

By default, we use 
a multi-output Gradient Boosting Model~\cite{scikit-learn} 
that allows us to obtain  
the performance vector 
by a single call with 
high accuracy (see Section~\ref{sec:exp}). 

\eetitle{Tests}. 
Given a model $M$ and a  
dataset $D$, a {\em test} $t$ 
is a triple $(M,D,\P)$, which  
specifies a test dataset $D$, 
an input model $M$, and 
 a set of user-defined measures 
$\P$ = $\{p_1, \ldots p_l\}$. 
 A test tuple $t$ = $(M,D,\P)$ 
 is {\em valuated} by 
 an estimator $\E$ if 
 each of its measure $p\in \P$ is 
 assigned a (estimated) value by $\E$.


\eat{
Below we provide a set of commonly 
seen utilities, all computable in 
low PTIME in terms of 
input size. 
\mengying{Add a table for optional measurements for each utility measure. }
}

\begin{example}
\label{exa-models}
Consider
Example~\ref{exa-motivation}. A pre-trained 
random forest (\kw{RF}) model 
$M$ that predicts CI-index is evaluated by three measures 
$\P$ = $\{\kw{RMSE}$, $\kw{R^2}$, $\kw{T_{train}}\}$, which specifies 
the root mean square error, the $R^2$ score, 
and the training cost. A user specifies 
a desired normalized range of 
\kw{RMSE} to be within 
$(0,0.6]$, \kw{R^2} in $[0, 0.35]$ \wrt a ``inversed'' lower bound $1-0.65$,
and $\kw{T_{train}}$ 
in $(0,0.5]$  \wrt an upper bound of 
``3600 seconds'' (\ie 
no more than 1800 seconds).
\end{example}

We summarize the main notations 
in Table~\ref{tab:notation}.

\eat{
Feature-related objectives~\cite{li2017feature} will be computed exactly, and a surrogate model will estimate model-related objectives.

\eetitle{Benefits.}
1. Model performance

2. Feature separability (similarity-based): Fisher Score~\cite{fisher1936use}

3. Feature correlation (feature and target, information theoretical based): mutual information~\cite{lewis1992feature}

\eetitle{Costs.}

1. Feature redundancy: Variance Inflation Factor (VIF)

2. Model training time

3. Model complexity: Bayesian information criterion (BIC)~\cite{schwarz1978estimating}
}

\section{Skyline Dataset Generation: A Formalization}
\label{sec-system}

Given datasets $\D$, an input 
model $M$ and a set of measures $\P$, 
we formalize the generation process of a 
skyline dataset 
with a ``multi-goals'' {\em finite state transducer} (FST). An FST extends 
extends finite automata by associating outputs with transitions. 
We use FST to abstract and characterize the generation  
of Skyline datasets as a data transformation process. 
We introduce this formalization, with a counterpart 
for data integration~\cite{lenzerini2002data, doan2012principles}, 
to help us characterize the computation 
of skyline 
dataset generation.

\stitle{Data Generator}. 
A skyline dataset generator 
is a finite-state transducer, 
denoted as $\T$ = $(s_M, \S, \O, \S_F$, $\delta)$,  where (1) $\S$ is a set of states, 
(2) $s_M\in \S$ is a designated start state, 
(3) $\O$ is a set of operators of types $\{\oplus, \ominus\}$; 
(4) $\S_F$ is a set of output states; and 
(5) $\delta$ refers to a set of transitions. 
We next specify its components. 

\eetitle{States}. 
A {\em state} $s$ 
specifies a table $D_s$ that conforms to 
schema $R_s$ and active domains $\ad_s$. 
For each attribute $A\in R_s$, 
$\ad_s(A) \subseteq \ad(A)$ 
refers to a fraction of values $A$ can take at state $s$. $\ad_s(A)$ 
can be set as empty set $\emptyset$, 
which indicates that 
the attribute $A$ 
is not involved for training or testing $M$; or a wildcard `\_' (`don't care'), 
which indicates that $A$ 
can take any value in $\ad(A)$. 

\eetitle{Operators}.
A skyline data generator adopts two 
primitive 
polynomial-time computable operators, 
{\em Augment} and {\em Reduct}. 
These operators can be 
expressed by SPJ (select, project,  
join) queries, or implemented as user-defined 
functions (UDFs). 

\sstab
(1) {\em Augment}  
has a general form of $\oplus_c(D_M, D)$, 
which augments dataset $D_M$ with another  
$D\in \D$ subject to a literal $c$. Here $c$ is a literal in form of $A = a$
(an equality condition). An augmentation $\oplus_{c}(D_M, D)$ executes the following queries: 
(a) augment schema $R_M$ of $D_M$ 
with attribute $A$ from 
schema $R_D$ of $D$, {\em if $A\not\in R_M$};  
(b) augment $D_M$ with 
tuples from $D$ 
satisfying constraint $c$; and 
(c) fill the rest 
cells with ``null'' for unknown values.

\sstab
(2) {\em Reduct} $\ominus_c(D_M)$: 
this operator
(a) selects from $D_M$ the tuples 
that satisfy the selection condition posed by 
the literal $C$ posed on attribute $R_M.A$; and 
(b) 
removes all such tuples from $D_M$. 
\eat{
mask cells of the tuples in $D_M$ that satisfy the literal $c$ on attribute $R_M.A$ with ``null".
} 
Here $c$ is a single literal defined on $R_M.A$ as in (1). 


\eetitle{Transitions}. A {\em transition} $r$ = $(s,op, s')$ is a triple that 
specifies a state $s$ = $(D, R, \ad_s)$, an operator $\op\in \O$ over $s$, and a {\em result} state $s'= (D', R', \ad'_s)$, where 
$D'$ and $R'$ are obtained 
by applying \op over 
$D$ and $R$ conforming to 
domain constraint  $\ad'_s$, 
respectively. 
\eat{
Given a state $s$ 
and a set of operators $\O$ 
of the two classes ($\{\oplus, \ominus\}$),  applying an 
operator $\op\in \O$ over $s$ creates a new 
{\em result} state $s'= (D', R', \ad'_s)$, 
$D'$ and $R'$ are obtained 
by applying \op over 
$D$ and $R$ conforming to 
domain constraint  $\ad'_s$, 
respectively. 
We represent this as a {\em transition} $r$ = $(s,op, s')$, 
where $s'$ is the result of applying 
$op$ to $s$. 
}

In practice, the operators 
can be enriched by task-specific UDFs 
that perform additional data imputation, 
or pruning operations, to further improve 
the quality of datasets. 

\eat{
Given a set of datasets $\D$ = $\{D_M, D_1, \ldots, D_n\}$, a model $M$, and a set of observed tests $T$, 

where $s_M$ is an initialized 
start state $(D_M, R_M, \ad_M)$, 
$\O$ is a set of queries 
from query classes $\oplus$ 
or $\ominus$, $\D_F$ is a 
set of output states, and 
$\delta\subseteq \D \times\O \times \D$ 
is a set of transitions. 
}

\stitle{Running}. A {\em configuration} of $\T$, 
denoted as $C$ = $(s_M, \O, M, T, \E)$, initializes a start state $s_M$ with 
a dataset $D_M$, a finite set of operators $\O$,  
a fixed deterministic model $M$, an 
estimator $\E$, 
and a test set $T$, where each 
test $t\in T$ has a valuated performance vector $\F(t)$. 
Both $D_M$ and $T$ can be empty set $\emptyset$. 
A {\em running} 
of $\T$ \wrt a configuration $C$ = $(s_M, M, T, \E)$ follows a general, deterministic process below.

\sstab 
(1) Starting from $s_M$, and at each state $s$, $\T$ iteratively applies operators from $\O$ to update a table with new attributes and tuples or mask its tuple values. This spawns a set of child states. 

\sstab 
(2) For each transition $r$ = $(s, \kw{op}, s')$ 
spawned from $s$ with a result $s'$ by applying  $\kw{op}$, $\T$ 
(a) initializes a test tuple $t(M, D_{s'}, \P)$ 
if $t\not\in \T$ and invokes 
estimator $\E$ {\em at runtime} to 
valuate the performance vector of
$t$; or (b) if $t$ is already in $T$, 
it directly loads $t.\P$. 
\eat{
otherwise, it invokes 
estimator $\E$ {\em at runtime} to 
valuate  
the performance vector of
$t$, and adds $t$ to $T$. 
}

Consistently, we say a state node $s\in \V$ is {\em valuated}, 
if a corresponding test $t(M, D_s, \P)$ is valuated by 
$\E$. We denote  
its evaluated performance vector as $s.\P$.  

\vspace{.5ex}
The above process {\em terminates} at a set of output states $\S_F$,  
under 
an external termination condition, 
or no transition can be spawned 
(no new datasets can be generated with $\O$).  


\eat{
\begin{figure}[tb!]
\centerline{\includegraphics[width =0.46\textwidth]{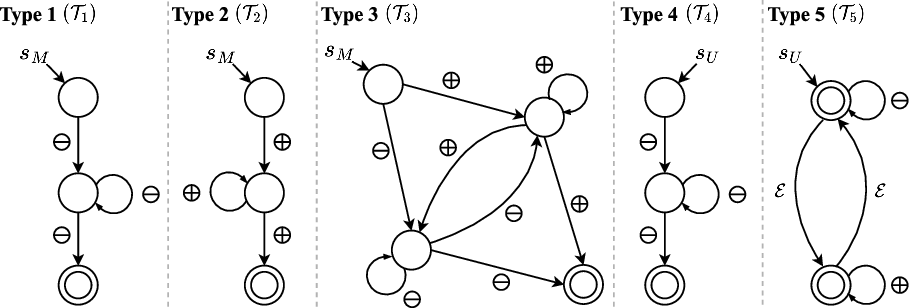}}
\centering
\caption{Data Discovery Systems: Specifications} 
 \vspace{-4ex}
\label{fig:dds}
\end{figure}
}

\begin{figure}[tb!]
\centerline{\includegraphics[width =0.45\textwidth]{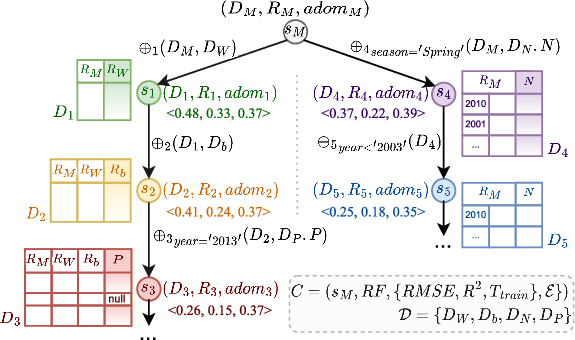}}
\centering
\vspace{-1ex}
\caption{A skyline data generation process, with a part of running graphs, and result datasets.}
\label{fig:transition}
\vspace{-2ex}
\end{figure}

The {\em result} of a running $\T$ refers to the set of corresponding datasets $\D_F$ 
induced from the output states $\S_F$. 
As each output state $s\in \S$ uniquely determines 
a corresponding output dataset $D_s$, 
for simplicity, 
we shall use a single general term 
``output'', denoted as $\D_F$, to 
refer to output states or datasets.

\eetitle{Running graph}. 
A running of $\T$ can be 
naturally represented 
as the dynamic generation of a {\em running graph} $G_\T$ = $(\V, \delta)$, which is a 
directed acyclic graph 
(\dag) with a set of state nodes 
$\V$, and a set of transition edges 
$r$ = $(s, op, s')$. 
A {\em path} of length $k$ is a sequence of 
$k$ transitions $\rho$ = 
$\{r_1, \ldots r_k\}$ 
such that for 
any $r_i$ = $(s_i,op,s_{i+1})$, 
$r_{i+1}$ = $(s_{i+1},op,s_{i+2})$; \ie 
it depicts a sequence of transitions 
that converts an initial state 
$s_1$ with dataset $D_1$ to 
a result $s_k$ with $D_k$.

\begin{example}
\label{exa-tg}
Following Example~\ref{exa-motivation}, 
Fig~\ref{fig:transition} shows a fraction of a running graph with input set $\D$=
$\{D_w, D_b, D_N, D_P\}$ (water, basin, nitrogen, and phosphorus tables, respectively). 
The augmentation 
$\oplus$ uses
spatial joins~\cite{vsidlauskas2014spatial},
a common query 
that join tables with tuple-level spatial similarity. 
With a configuration 
$C$ = $(s_M, \kw{RF}, \{RMSE, R^2, T_{train}\}, \E\})$ (where $\E$ is an MO-GBM estimator), a running starts by 
joining 
$D_w$ and $D_b$
to get $D_2$. 
$D_2$ is then augmented with the attribute 
``\texttt{Phosphorus}'' 
under 
a literal 
``\texttt{year} = {\em 2013}'', resulting in $D_3$ via a path $\{\oplus_1, \ldots, \oplus_3\}$. 
In each step, a test $t$ is initialized; 
and the estimator $\E$ 
is consulted to valuate the 
performance vector of $t$, and enrich $T$. 

Consider another path 
from $D_M$ that results a dataset $D_5$ in Fig.~\ref{exa-motivation}.  
It first
augmentes $D_M$ to $D_4$ 
with data in ``Spring''. 
A reduction with a condition 
``\texttt{year}<'2003'' selects 
and removes all the tuples 
in $D_4$ with historical data before $2003$, 
which leads to 
dataset $D_5$ that retains only 
the data since $2003$. 
\end{example}
\vspace{-1ex}
\section{Skyline Data Generation Problem} 
\label{sec-problem}


Given $\T$ and a configuration $C$, \modis  
aims find a running 
of $\T$ that ideally leads to a ``global'' optimal dataset, where $M$ is expected to deliver the highest performance over all metrics. Nevertheless, a single optimal solution may not always exist. 
First, 
two measures in $\P$ may 
in nature conflict 
due to trade-offs (\eg training cost versus accuracy, precision versus recall). 
Moreover, the ``no free lunch'' 
theorem~\cite{sterkenburg2021no} 
indicates that there may not exist a single test that demonstrate best performance 
over all measures. 
We thus pursue {\em Pareto optimality} for $\D_F$. We start with
a dominance relation below. 

\stitle{Dominance}. Given a data discovery system $\T$ and performance measures $\P$, 
a state $s$ 
= $(D_s, R_s, \ad_s)$ is {\em dominated} by 
$s'$ = $(D_{s'}, R_{s'}, \ad_{s'})$, 
denoted as $s\prec s'$, 
if there are valuated tests $t$ = $(M, D_s)$ 
and $t'$ = $(M, D_{s'})$ in $T$, such that 
\bi
\item for each $p\in \P$,  
$t'.p \leq t.p$; and 
\item there exists a measure $p^*\in \P$, such that 
$t'.p^*< t.p^*$.
\ei 
A dataset $D_s$ is dominated 
by $D_{s'}$, denoted as $D_s\prec D_{s'}$, if 
$s \prec s'$. 

\eetitle{Skyline set}. 
Given $\T$ 
and a configuration $C$, let $\D_F$ be the set of all the possible output datasets from 
a running of $\T$,  a set of datasets $\D_F^*\subseteq\D_F$ is 
a {\em skyline set} \wrt $\T$ and $C$, if 
\begin{itemize}
\item for any dataset $D\in \D_F^*$, 
and any performance measure $p\in \P$, 
there exists a test $t\in T$, such that 
$t.p\in [p_l,p_u]$; 
\item  there is no pair 
$\{D_1,D_2\} \subseteq \D_F^*$   
such that $D_1 \prec D_2$ or $D_2 \prec D_1$;  
and 
\item for any other 
$D\in \D_F\setminus\D_F^*$, 
and any $D'\in \D_F^*$, 
$D\prec D'$.
\end{itemize} 
We next formulate the skyline data generation problem. 

\eat{
\stitle{Encoding of the Search Space}

Configuration state graph.

Partial configuration/ Full configuration.

Augment edge/ Modify edge.

Start state/ Terminating state.
}

\eat{
\stitle{Path ($\rho$) Dominance}

$$(<P_1, P_2, ..., P_n>|<f_1, f_2, ..., f_n>)$$
$$\forall P_i \in \vec{P}, P_i = \text{{compute}} \vee P_i = \text{{estimate}}$$

\begin{equation*}
\begin{split}
\rho_1 \leq_{\epsilon} \rho_2 \iff
&
\left(\forall p^b \in P_{benifit}, \rho_1(p^b) \leq (1+\epsilon)\rho_2(p^b)\right) 
\land \\
&
\left(\forall p^c \in P_{cost}, (1+\epsilon)\rho_1(p^c) \geq \rho_2(p^c)\right)
\end{split}
\end{equation*}
}

\stitle{Skyline Data Generation}. 
Given 
a skyline data generator $\T$ 
and its configuration $C$ = $(s_M, \O, M, T, \E)$, 
the {\em skyline data generation} problem 
is to 
compute a skyline set $\D_F$ 
in terms of $\T$ and $C$. 

\begin{example}
\label{exa-modis}
Revisiting prior example 
and consider 
the temporal results $\D_F$= 
$\{D_1, \ldots, D_5\}$ with the following performance 
vectors valuated by 
the estimator $\E$ so far: 

\begin{center}
\begin{small}
    \begin{tabular}{|c|c|c|c|}  
        \hline
       T: (D, M, $\P$, $\E$)  & RMSE & $\hat{R^2}$ & $T_{train}$ \\  \hline
            $t_1:(D_1,\kw{RF},\P,  \kw{MO-GBM})$ & 0.48 & 0.33 & 0.37 \\
       \hline
            $t_2:(D_2,\kw{RF},\P, \kw{MO-GBM})$ & 0.41 & 0.24 & 0.37 \\
       \hline
            $t_3:(D_3,\kw{RF},\P, \kw{MO-GBM})$ & 0.26 & \underline{0.15} & 0.37 \\
       \hline
            $t_4:(D_4,\kw{RF},\P, \kw{MO-GBM})$ & 0.37 & 0.22 & 0.39\\
       \hline
            $t_5:(D_5,\kw{RF},\P, \kw{MO-GBM})$ & \underline{0.25} & 0.18 & \underline{0.35}\\
       \hline
    \end{tabular}
    \end{small}
\end{center}
Here $\hat{R^2}$ is inversed as 1-$R^2$:  
the smaller, the better. 
All the measures are 
normalized in $(0,1]$ \wrt 
user-specified upper and lower bounds, 
and the optimal values are underlined. 
One can verify the following dominance relation among the 
datasets: 
(1) $D_1 \prec D_2 \prec D_3$, 
and $D_4 \prec D_5$;
(2) $D_3 \not\prec D_5$ and vice versa.  
Hence a Skyline set 
$\D_F$ currently
contains $\{D_3, D_5\}$. 
\end{example}

We present the following hardness result.  

\begin{theorem}
\label{them-np}
Skyline data generation is (1) \kw{NP}-hard; 
and (2) fixed-parameter tractable, if 
(a) $\P$ is fixed, and (b) $|\D_F|$ is polynomially bounded 
by the input size $|\D|$. 
\end{theorem}

\begin{proofS}
The \kw{NP}-hardness can be verified by a reduction 
from the Multiobjective Shortest Path problem (\mos). 
Given an edge-weighted graph $G_w$, where each edge $e_w$ has a $d$-dimensional cost vector $e_w.c$, the cost of a path $\rho_w$ in $G_w$ is defined as $\rho_w.c$ = $\sum_{e_w\in\rho_w}$ $e_w.c$. 
The dominance relation between paths 
is determined by comparing their costs.  
\eat{
Specifically, $\rho_w$ dominates $\rho'_w$ if $\rho_w$ has equal or lower costs than $\rho'_w$ in all dimensions and is strictly better in at least one dimension. }
\mos is to compute a Skyline set of paths from start node $s$ to target node $t$. 

Given an instance of \mos, we construct an instance 
of our problem as follows. 
(1) We assign an arbitrally ordered index to the edges 
of $G_w$, say $e_1, \ldots e_n$.  
(2) We initialize a configuration $\T$ as follows. 
(a) $s_M$ has a single dataset $D_0$, 
where for each edge $e_i$ = $(v,v')$, 
there is a distinct tuple $t_i\in D_0$. 
(b) $\O$ contains a set of 
reduction operators, where each operator $o_i$  
removes tuple $t_i$ from $D_0$, 
and incurs a pre-defined 
performance measure $e_i.c$. 
(c) $M$ 
 maps each tuple $t_i$ in $D_0$ to a 
fixed embedding in $\mathbb{R}^d$. 
(d) The test set $T$ is $\emptyset$. 
We enforce the running graph of $\T$ to be 
 the input $G_w$, by setting 
the initial state as $s$ with associated 
dataset $D_0$,  a unique termination 
state as the node $t$, and the 
applicable transitions as the edges 
in $G_w$. 
One can verify that a solution of 
\mos is a Pareto set of paths 
from $s$ to $t$,  
each results in a dataset by 
sequentially applying 
the reduction operators 
following the edges of the path. 
This yields a set of datasets 
that constitutes 
a corresponding skyline set $\D_F$
as a solution for 
our problem. 
As \mos is shown to be 
NP-hard~\cite{hansen1980bicriterion, serafini1987some}, 
the hardness of skyline data generation follows. 
%

To see the fixed-parameter tractability, 
we outline an exact algorithm. 
(1) The algorithm exhausts the runnings of a skyline 
generator $\T$, and invokes a PTIME inference 
process of the model $M$ and valuate at most 
$N\leq |\D_F|$ possible states (datasets). 
(2) It invokes a multi-objective 
optimizer such as Kung's algorithm~\cite{kung1975finding}. 
This incurs $O(N\log N)^{|\P|-2}$ 
valuations when $|\P|\geq 4$, 
or $O(N(\log N))$ if $|\P|\textless 4$. 
As $N\leq |\D_F|$, and $|\D_F|$ is in $O(|D|)$, 
and $P$ is a fixed constant, 
the overall cost is in PTIME (see~\cite{full} for details). 
\end{proofS}

While the above exact algorithm is able to 
compute a skyline dataset, it remains infeasible 
even when enlisting $N$ 
states as a ``once-for-all'' 
cost is affordable. Moreover, 
a solution may contain 
an excessive number of datasets to be 
inspected. 
We next present 
three feasible algorithms, 
that 
generate datasets that approximate skyline sets with  
bounded size and quality guarantees.



\eat{
\comwu{We need to introduce fixed parameter results 
to clarify the polynomial boundedness of 
the size of the Skyline set. Then we need to 
clarify a ``FPTAS'' under this hypothesis.}
}


\section{Computing Skyline Sets}
\label{sec-methods}


\subsection{Approximating Skyline Sets}
\label{sec-apxmodis}


We next present our first algorithm that generates a size-bounded set, which 
approximates a Skyline set in $\D_F$.  
To characterize the approximation quality, 
we introduce a notion of 
$\epsilon$-skyline set. 

\stitle{$\epsilon$-Skyline set}. 
Given a data discovery system $\T$ with 
a configuration $C$, 
Let $\D_\S$ be a set of $N$ valuated datasets 
in the running of $\T$. 
Given a pair of datasets $(D, D')$ from $\D_\S$, and a constant $\epsilon\textgreater 0$,  
we say $D'$ {\em $\epsilon$-dominates} 
$D$, denoted as 
$D'\succeq_\epsilon D$, 
if for the corresponding 
tests $t$ = $(M, D)$ and $t'$ = $(M, D')$, 
\tbi
\item $t'.p \leq (1+\epsilon) t.p$ 
for each $p\in \P$, and  
\item there exists a measure $p^*\in \P$, 
such that $t'.p^* \leq t.p^*$. 
\ei

In particular, we call $p^*$ a 
{\em decisive measure}. Note that 
$p^*$ can be any $p\in \P$ and 
may not be fixed. 

\vspace{.5ex}
A set of datasets $\D_\epsilon\subseteq \D_\S$ 
is an {\em $\epsilon$-Skyline set} of 
$\D_\S$, if 
\tbi 
\item for any dataset $D\in \D_\epsilon$, 
and any performance measure $p\in\P$, 
there exists a corresponding 
test $t\in T$, such that 
$t.p\in [p_l, p_u]$; and 
\item 
for every dataset $D'\in \D_\S$, 
there exists a dataset 
$D\in \D_\epsilon$ such that 
$D \succeq_\epsilon D'$. 
\ei 


\stitle{$(N,\epsilon)$-approximation}. We say an algorithm 
is an {\em $(N,\epsilon)$-approximation} for \modis, 
if it satisfies the following: 
\tbi 
\item it explores and valuates at most $N$ states; 
\item for any constant $\epsilon\textgreater 0$, the system correctly outputs an $\epsilon$-Skyline set, as an approximation of a Skyline set defined over $N$ {\em valuated} states; and 
\item the time cost is polynomial determined by $|\D|$, $N$, and$\frac{1}{\epsilon}$.
\ei

Below we present our 
main result. 

\begin{theorem}
\label{thm-fptas}
Given 
datasets $\D$, 
 configuration $C$, 
 and a number $N$, 
there exists an $(N,\epsilon)$-approximation for \modis in 
$O\left(\min(N_u^{|R_u|}, N)\cdot \left(\left(\frac{\log(p_m)}{\epsilon}\right)^{|\P|-1}+I\right)\right)$ time, where $|R_u|$ is the size of the universal 
schema, 
$N_u$ = $|R_u|+|\ad_m|$ 
($\ad_m$ the largest active domain), $p_m$ = 
$\max\frac{p_u}{p_l}$
as the measure $p$ ranges over $\P$; and $I$ is the 
unit valuation cost per test. 
\end{theorem}

\stitle{Remarks}. 
The above result captures a {\em relative}
guarantee \wrt $\epsilon$ 
and $N$. When 
$N = \D_F$, an $(N,\epsilon)$-approximation ensures to output a $\epsilon$-Skyline set. 
The paradigm is feasible as one can explicitly trade the `closeness' of the output 
to a Skyline set with affordable time cost, by explicitly tuning 
$\epsilon$ and $N$. 
Moreover, 
the worst-case factor $|\ad_m|$ 
can also be ``tightened'' by 
a bound 
determined by
the value constraints posed by 
the literals. For example, an attribute 
$A$ 
may contribute up to two necessary values 
in the search if 
the literals involving $A$ only enforce 
two equality conditions ``A=2'' and ``A=5'', 
regardless of how large $|\ad(A)|$ is 
(see 
Sections~\ref{sec-system} and~\ref{sec:exp}).

\eat{
\revise{Note that $\vert\ad(A_i)\vert$ is not necessarily the total distinct literals in $A_i$, but can be kept finite by grouping values, such as using comparison constraints. For example, $\ad(A_i) = \{2, 5\}$  if values $A_i$ are grouped by comparisons with $2$ and $5$, ensuring efficient representation and processing.}
}

\begin{figure}
\centering
\begin{algorithm}[H]
\caption{:\apxmodis}
\label{alg:forward}
\begin{algorithmic}[1]
\algtext*{EndFor}
\algtext*{EndIf}
\algtext*{EndWhile}
\algtext*{EndFunction}
\algtext*{EndProcedure}

\State \textbf{Input:} 
    Configuration $C$ = $(s_U, \O, M, T, \E)$, 
    a number $N$, \\ a constant $\epsilon>0$, 
    a decisive measure $p_d$;
    user-specified upper bound $p_u$ for $p \in \P$;
\State \textbf{Output:} 
     $\epsilon$-Skyline set $\mathcal{D}_F$.

     \vspace{1ex}
    \State \textbf{Queue} $Q := \varnothing$, integer $i$ := 0, $Q$.enqueue(($s_U$, 0));\label{apx:initialize}  
    \While{$Q \neq \varnothing$ 
    \textbf{and}
    number of valuated states $< N$} \label{apx:main:s} 
        \State ($s$, i) := $Q$.dequeue();
        ${\D_F}^{i+1} := {\D_F}^i$; 
        \For{each ($s'$, $D_{s'}$) $\in$ \opg($s$)}
            \State $Q$.enqueue(($s'$, i + 1));
            \State ${\D_F}^{i + 1}$ := \upi($s'$, 
            ${\D_F}^{i + 1}$, $\epsilon$);
        \EndFor
    \EndWhile \label{apx:main:e}
 \State \Return ${\D_F}$ \label{apx:return}

\vspace{1ex}
\Procedure{\opg}{s}
    \State \textbf{set} $Q':=\varnothing$; \label{apx:opg:s}
    \For{each entry $l \in s.L$} 
        \If{$l = 1$}
            \State $l := 0$;  
            \State create a new state $s'$; 
            $s'.L := s.L$; 
            \State generate dataset $D_{s'}$ accordingly; 
            \State $Q'$.append(($s'$, $D_{s'}$)); 
        \EndIf
    \EndFor
    \State \Return $Q'$
\EndProcedure \label{apx:opg:e}

\vspace{1ex}
\Procedure{\upi}{$s'$, 
${\D_F}^{i + 1}$, $\epsilon$} \label{apx:upi:s}
\State {update 
$s'.\P$ with estimator $\E$;
\For{each $p \in \P$} 
    \If{$s'.\P(p) > p_u$} 
    \Return ${\D_F}^{i + 1}$;
    \EndIf
\EndFor
        \State update $\operatorname{pos}(s')$ with Equation~(\ref{eq:pos});
 \State 
        retrieve state 
        $s''$ where $\operatorname{pos}(s'')$ =  $\operatorname{pos}(s')$; 
        \If{no such 
        $s''$ exists} 
        \State ${\D_F}^{i + 1}$ := ${\D_F}^{i + 1}\cup \{D_s'\}$; 
        \EndIf
        \State \textbf{else if} $s'.\P(p_d) < s''.\P(p_d)$ \textbf{then}
        \State \hspace{3ex} ${\D_F}^{i + 1}$ := ${\D_F}^{i + 1}\setminus \{D_{s''}\} \cup \{D_{s'}\}$;
    \State \Return ${\D_F}^{i + 1}$
\EndProcedure \label{apx:upi:e}

}
\end{algorithmic}
\end{algorithm}
\vspace{-3ex}
\caption{\apxmodis: Approximating Skyline sets}
\vspace{-3ex}
\label{fig:approx}
\end{figure}

\subsection{Approximation Algorithm}
\label{sec-approx}
 
As a constructive 
proof of Theorem~\ref{thm-fptas}, 
we next present an $(N,\epsilon)$-approximation algorithm, denoted as \apxmodis.   

\stitle{``Reduce-from-Universal''}. 
Algorithm \apxmodis simulates the running of $\T$ from a start state $s_U$. 
The start state is initialized with a ``universal'' dataset $D_U$, 
which carries the universal schema $R_U$, and is populated by 
joining all the tables (with outer join to preserve 
all the values besides common attributes, by default). 
This is to enforce the search 
from a set of rows that 
preserve all the attribute values 
as much as possible to maximize 
the chance of sequential 
applications of reduction only. 
It transforms state dominance into a ``path dominance'' counterpart. 
For a transition edge $(s,\ominus, s')$, 
a weight is assigned 
to quantify the gap between 
the estimated model performance 
over datasets $D_s$ and its ``reduced'' 
counterpart $D_s'$. 
The ``length'' of a path from $s_U$ to $s$ aggregates 
the edge weights towards  
the estimated performance of 
its result.

\eat{
\mengying{During each transition (edge), one attribute or active domain is reduced, and the edge is weighted based on the performance variance caused by the transition.
The ``depth'' refers to the number of transitions from the initial state to the current state, while the ``shortest path'' denotes the path from the initial state to a target state that yields the dataset with the best performance measures for the given model.
\apxmodis "transforms" state dominance into a "path dominance" counterpart by assigning the valuated measures of end nodes to the paths from the initial state at runtime.}
}
\eat{
\warn{How do we characterize a 
``shortest path'': give the 
definition/intuition of the 
path length, and define a 
version of shortest path.}
\warn{``depth'': undefined.}\mengying{[MY: path length is defined in Running Graph in section III]}
}

\eetitle{Advantage}. We justify the ``reduce-from-universal'' 
strategy in the following context. 
(1) As the measures are 
to be minimized, 
we can extend ``shortest'' paths 
by prioritizing the valuation of datasets 
towards user-defined upper bounds 
with early pruning, to avoid   
unnecessary reduction. 
(2) Starting from a universal dataset allows early exploration of ``dense'' datasets, over which the model always tends to have higher accuracy in practice.


We next present the details of our algorithm. 

\stitle{Auxiliary structure}. 
~\apxmodis follows a dynamic 
levelwise state generation 
and valuation process, 
which yields
a running graph $G_\T$ 
with up to $N$  
nodes. 
It 
maintains the following. 

\sstab 
(1) $Q$ is a queue that maintains 
the dynamically generated and 
valuated state nodes. 
Each entry of $Q$ is a pair 
$(s, i)$ that records 
a state and the level it resides.

\sstab
(2) $\D_F$ is a list of datasets.  
$\D_F^i$ specifies the datasets 
processed at level $i$. 
Each state node $s$ is associate with a bitmap $L$ 
to encode if its schema $s.R_s$ 
contains an attribute $A$ in $D_U$, and if $D_s$ 
contains a value from its 
active domain $\ad(A)$. 
The map is consulted to assert the applicability of reduct operators at runtime. 

\sstab 
(3) Each state $s$ is associated with 
a position $pos(s)$ in a {\em discretized} $|\P-1|$-ary 
space, which is defined as 
\begin{small}
\begin{equation}
\operatorname{pos}(s) = 
\left[\left\lfloor\log _{1+\epsilon} \frac{s.\P(p_1)}{p_{l_1}}\right\rfloor, \ldots, \left\lfloor\log _{1+\epsilon} \frac{s.\P(p_{|\P|-1})}{p_{l_{|\P|-1}}}\right\rfloor\right]
\label{eq:pos}
\end{equation}
\end{small}

By default, we set the last measure 
in $\P$ as {\em a} decisive measure. 
We remark that one can choose 
any measure as decisive measure, 
and our results carry over.

\stitle{Algorithm}. The algorithm \apxmodis is illustrated in Fig.~\ref{fig:approx}. 
It initializes a queue $Q$ with a start state $s_U$, and set a position to $s_U$ (line~\ref{apx:initialize}). In lines~\ref{apx:main:s} to \ref{apx:main:e}, it update the Skyline set $\D_F$ for each level iteratively. 
At level $d$, for each state $s \in Q$, 
procedure \opg (line~\ref{apx:opg:s} to \ref{apx:opg:e}) explores all one-flip transitions in $s.L$ and generates a set with applicable reduct operators. 
\apxmodis 
enqueue 
new states and update the Skyline set ${\D_F}^{d+1}$ at next level accordingly by invoking Procedure \upi. 
This process terminations until 
$N$ states are valuated, or 
no new state can be generated. 

\begin{figure}[tb!]
\centerline{\includegraphics[width =0.45\textwidth]{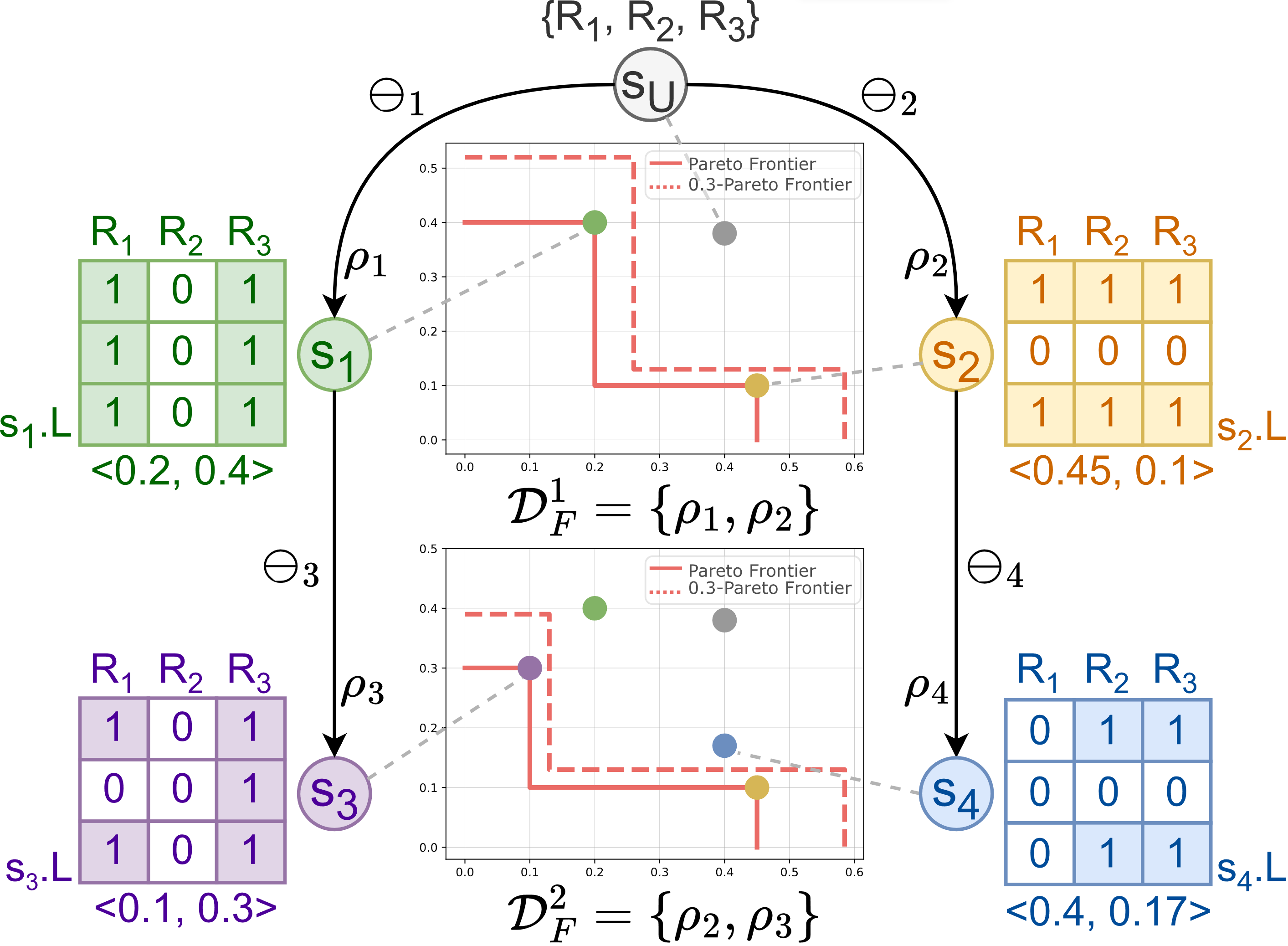}}
\centering
\vspace{-1ex}
\caption{``Reduct-from-Universal'': an illustration of two-level computation. It performs multiple level-wise spawns and updates the $\epsilon$-Skyline set.}
 \vspace{-3ex}
\label{fig:reduct}
\end{figure}

\stitle{Procedure \upi}. Given a 
new state $s'$, procedure \upi 
determines if $s'$ should be included 
in the current Skyline set. 
(1) It updates $s'.\P$ by consulting the estimator $\E$ 
(line 1), and decide an early skipping 
if its performance fails to satisfy 
the upperbound $p_u$ for some measure $p\in \P$. 
(2) Otherwise, \upi updates the 
position of state $s'$, and decides  
if $s'$ ``replaces'' 
a valuated state $s''$ at the same position 
due to $(1+\epsilon)$-dominance (lines 6-9). 
The Skyline set at current level 
is updated accordingly with 
dataset $D_{s'}$. 

\begin{example}
\label{exa-apxmodis}
Fig.~\ref{fig:reduct} illustrates data discovery 
of 
\apxmodis with $N$=$5$ and $\epsilon$=$0.3$, 
over a table set $\D$ = $\{D_1, \ldots, D_3\}$ and 
measures $\P$ = $\textless p_1, p_2\textgreater$. The operator set $O$ contains 
four reduct operators $\{\ominus_1, \ldots \ominus_4\}$. 
(1) It first constructs a 
universal dataset $D_U$ with universal schema $R_U$. $D_U$ can be 
obtained by 
optimized multi-way join~\cite{zhao2020efficient}, augmentation~\cite{li2021data}, or UDFs~\cite{dong2017arrayudf}. 
The bitmap $D_U.L$ is 
initialized accordingly. 
Procedure \kw{OpGen} 
then generates applicable reductions by ``flipping'' the entries in $D_U.L$.  
This spawns states $s_1$ and $s_2$ obtained by 
applying reduct $\ominus_1$ and $\ominus_2$, 
respectively
It then consults 
the estimator $\E$ to valuate model performances, and identified that 
$D_1\not\succeq_{0.3} D_2$ and vice versa. Thus 
\apxmodis sets the current $0.3$-Skyline set 
as $\{D_1, D_2\}$.  

\apxmodis next spawns states with 
applicable reductions $\ominus_3$ and $\ominus_4$, 
extending $\rho_1$ that leads to 
$s_1$, and $\rho_2$ that leads to $s_2$, 
This generates new extended paths $\rho_3$ and $\rho_4$ 
with results $D_3$ and $D_4$, respectively. 
It verfies that 
$\D_3 \succeq_{0.3} D_1$, but 
$D_2 \succeq_{0.3} D_4$; and $D_2 \not\succeq_{0.3} D_3$ 
and vice versa. This yields an updated  
$0.3$-Skyline set $\{D_2, D_3\}$, 
after valuating $5$ states. 
\end{example}

\stitle{Correctness \& Approximability}. ~\apxmodis 
terminates as it spawns $N$ nodes with at most  
$|R_U||\ad_m|$ distinct reduction, 
where $|\ad_m|$ refers to the size 
of the largest active domain. 

\vspace{2ex}
For approximability, we present the result below. 

\begin{lemma}
\label{lm-approximability}
For any constant $\epsilon$, \apxmodis correctly computes an $\epsilon$-Skyline set $\D_F$ as an approximated Skyline set 
defined on the $N$ states it valuated. 
\end{lemma}

\begin{proofS}
We verify the $\epsilon$-approximability, 
with a reduction to the multi-objective 
shortest path problem (\mos)~\cite{tsaggouris2009multiobjective}. 
Given an edge-weighted graph $G_w$, 
where each edge carries a $d$-dimensional 
attribute vector $e_w.c$, it computes a Skyline set of 
paths from a start node $u$. 
The cost of a path $\rho_w$ in $G_w$ is 
defined as $\rho_w.c$ = $\sum_{e_w\in\rho_w}$ $e_w.c$. 
The dominance relation between two paths 
is determined by 
the dominance relation of their cost. 
Our reduction (1) constructs $G_w$ 
as the running graph $G_\T$ with 
$N$ valuated state nodes and 
spawned transition edges; and 
(2) for each edge $(s, s')$, 
sets an edge weight as 
$e_w$ = $s.\P - s'.\P$. 
Given a solution $\Pi_w$ 
of the above instance of \mos, 
for each path $\rho_w\in\Pi$, 
we set a corresponding path $\rho$ 
in $G_\T$ with result dataset $D$, 
and adds it into $\D_F$. 
We can verify that 
$\Pi_w$ is an $\epsilon$-Skyline set 
of paths $\Pi_w$, if and only 
if $\D_F$ is an $\epsilon$-Skyline set 
of $\D_S$ that contains the
datasets from the set of $N$ valuated 
states in $G_\T$. 
We then show that~\apxmodis 
is an optimized process of 
the algorithm in~\cite{tsaggouris2009multiobjective}, 
which correctly computes 
$\Pi_w$ for $G_w$. 
\end{proofS}

\stitle{Time cost}. 
Let $|R_u|$ be the total number of 
attributes in the universal schema 
$R_u$ of $D_u$, and $|\ad_m|$ be the size of the largest active domain. \apxmodis performs 
$|R_u|$ levels of  
spawning, and at each node, 
it spawns at most $|R_u|$ + $|\ad_m|$ 
children, given that it 
``flips'' one attribute for each reduction, and for 
each attribute, at most one domain value 
to mask. Let $N_u$ be $|R_u|$+$|\ad_m|$. 
Thus, \apxmodis valuates 
at most $\min(N_u^{|R_u|}, N)$ 
nodes (datasets), taking $I\cdot \min(N_u^{|R_u|}, N)$ 
time, where $I$ 
refers to a polynomial time 
valuation cost of $\E$ per test. 
For each node, it then takes at most $\prod_{i=1}^{|\P|-1}\left(\left\lfloor\log _{1+\epsilon}\frac{p_{u_i}}{p_{l_i}}\right\rfloor+1\right))$ 
time to update the $\epsilon$-Skyline set. 
Given $\epsilon$ is small, $\log (1 + \epsilon) \approx \epsilon$, and the total cost is 
in $O\left(\min(N_u^{|R_u|}, N)\cdot \left(\left(\frac{\log(p_m)}{\epsilon}\right)^{|\P|-1}+I\right)\right)$ time. 
Given $|R_u|$ and $|\P|$ are small constants, 
the cost is polynomial in the 
input size $|D_u|$, $N$ and $\frac{1}{\epsilon}$. 
Theorem~\ref{thm-fptas} thus follows. 

\stitle{An FPTAS case}. 
We next present a case when \apxmodis 
ensures a stronger optimality guarantee. 
We say an $(N, \epsilon)$-approximation 
is a {\em fully polynomial time} approximation (FPTAS) 
for \modis, if 
(1) it computes 
an $\epsilon$-Skyline set 
for $\D_S$, where $\D_S$ 
refers to all possible datasets 
that can be generated from $D_U$, 
and (2) it runs in 
time polynomial in the size of 
$|D_U|$ and $\frac{1}{\epsilon}$. 


\begin{lemma}
\label{cor-fptas}
Given a skyline dataset generator $\T$ with configuration $C$, 
if $|\D_S|$ has a size that is polynomially bounded in $O(f(|D_U|))$, 
then \apxmodis is an FPTAS for \modis. 
\end{lemma}

\begin{proofS}
We show this by a reduction from 
\modis to \mos, similarly as 
in the approximability analysis. \mos 
is known to have a fully polynomial 
time approximable (FPTAS) in terms of 
$\epsilon$-dominance. We set \apxmodis to run as 
a $(|\D_S|, \epsilon)$-approximation, 
which is a simplified implementation of an  
FPTAS in~\cite{tsaggouris2009multiobjective} with multiple 
rounds of ``replacement'' 
strategy following path dominance. 
As 
$|\D_S|$ is bounded by a polynomial of  
the input size $|D_U|$, 
it approximates 
a Skyline set for all in PTME. 
\end{proofS}

The size bound of $\D_S$ is  
pragmatic and practical due to 
that the attributes often bear 
active domains that are 
much smaller than 
dataset size. 
Indeed, data science applications  
typically consider data with values 
under task-specific constraints.  
These suggest practical 
application of \apxmodis 
with affordable setting of 
$N$ and $\epsilon$. 
We present the detailed analysis in~\cite{full}. 

\eat{
it processes at most $N$ 
state nodes. At each node, it takes $O(|\ad_m|)$ 
time to spawn the children, and let each spawned edge 
take a valuation time of $I_\E$. 
Thus the total verification time takes 
$O(N|R_u||\ad_m|I_\E)$. 

in the main iteration (line~\ref{its:main} to \ref{ite:main}), for each depth from 1 to n, we evaluate all possible incoming transitions into the state at that depth. In the Extend\&Merge function, we may reach out to all paths in the Skyline set. So the Time Complexity is $O\left(n|D_U|^n\prod_{i=1}^{|\P|-1}\left(\left\lfloor\log _{1+\epsilon}\frac{p^{max}_i}{p^{min}_i}\right\rfloor+1\right)\right)$. Let's denote $P^{max}$ as the maximum value of $\frac{p_i^{max}}{p_i^{min}}$, given $\epsilon$ is small enough, so $\log (1 + \epsilon) \approx \epsilon$, then Time Complexity is $O\left(n|D_U|^n\left(\frac{\log(P^{max})}{\epsilon}\right)^{|\P|-1}\right)$.
}

\subsection{Bi-Directional Skyline Set Generation}
\label{sec-bimodis}
\eat{
Algorithm \apxmodis tends to exhibit enhanced efficiency when users specify higher expectations of the model performance, indicated by smaller thresholds $\boldsymbol{t}$, allowing for earlier termination. 
Yet, in cases involving extensive data discovery with a large $|\D|$, its ``reduction-only'' approach can lead to an 
enormous number of valuating. 
To mitigate this, we introduce \bimodis, which implements a bi-directional search strategy, 
aligning with Type-5 systems in Fig~\ref{fig:dds}, 
incorporating a {\em Correlation-Based Pruning} strategy with early detection of dominance.
It covers the space more rapidly and can offer a more balanced view of the solution, especially in high-dimensional datasets where many features may have a minimal impact.
}

Given our cost analysis, for skyline data generation 
with larger (more ``tolerate'') ranges $(p_l,p_u)$ 
and 
larger $|\D|$, \apxmodis 
may still need to valuate a large number of 
datasets. To further reduce valuation cost, 
we introduce \bimodis, its bi-directional variant. 
Our idea is to interact both augment and reduct 
operators, with a ``forward'' search from universal dataset, and a ``backward'' counterpart from a single dataset in $\D$. We also introduce a pruning strategy based on an early 
detection of dominance relation. 

\eat{
that covers the space more rapidly and can offer a more balanced view of the solution, especially 
for datasets with many 
less critical features. 
In addition, it exploits {\em Correlation-Based Pruning} to 
detect dominance early. 
}

\eat{
\stitle{Auxiliary structure}.
Algorithm \bimodis builds a dynamic spawning graph $G_\T$ akin to \apxmodis.
For forward search, it maintains a forward frontier $Q_f$, which initiates from a universal schema $s_U$ and progresses by reduction. 
Meanwhile, a backward frontier $Q_b$ is introduced for backward search, from a state $s_b$, which includes a minimal subset of $s_U$, \eg ensures no classes will be lost if $M$ is a classifier, and advancing step-by-step through augmentations. 
}

\begin{figure}
\centering
\begin{algorithm}[H]
\caption{:\bimodis
}
\begin{algorithmic}[1]
\algtext*{EndFor}
\algtext*{EndIf}
\algtext*{EndWhile}
\algtext*{EndFunction}
\algtext*{EndProcedure}

\State \textbf{Input:} 
    Configuration $C$ = $(s_U, \O, M, T, \E)$, 
    a constant $\epsilon>0$;
\State \textbf{Output:} 
     $\epsilon$-Skyline set $\mathcal{D}_F$.
     \vspace{1ex}
    
\State \textbf{set} $s_b = \text{BackSt}(s_U)$; 
    \textbf{queue} $Q_f := \{(s_U, 0)\}$, 
                   $Q_b := \{(s_b, 0)\}$; 
                   integer $i$ := $0$; 
    \label{bi:ini}

\While{$Q_f \neq \varnothing$, 
       $Q_b \neq \varnothing$ \textbf{and} 
       $Q_f \cap Q_b = \varnothing$}
\label{bi:start}
\State $(s', i)= Q_f$.dequeue();
\Comment{Forward Serach}
\State $(s'', i)= Q_b$.dequeue(); 
\Comment{Backward Serach} 
\State ${\D_F}^{i+1} = {\D_F}^i$;

\For{\textbf{all} $s^f \in$ \opg($s'$) \textbf{and} $s^b \in$ \opg($s''$)}
    \State ${\D_F}^{i + 1}$ = \upi($s^f$, ${\D_F}^{i + 1}$, ${\D_F}^i$, $\epsilon$);
    \State ${\D_F}^{i + 1}$ = \upi($s^b$, ${\D_F}^{i + 1}$, ${\D_F}^i$, $\epsilon$);
    
    \If{canPrune$(s^f,s^b)$} \label{bi:prunes}
        \State \kw{prune}($\C,s^f,s^b$); 
    \EndIf \label{a2:pos_prune} \label{bi:prunee}
    
    \State $Q_f$.enqueue(($s^f$, $i+1$)), 
           $Q_b$.enqueue(($s^b$, $i+1$));
\EndFor \label{a2:fse}
\EndWhile
\label{bi:end}
\State \Return $\D_F$
\end{algorithmic}
\end{algorithm}
\vspace{-3ex}
\caption{\bimodis: Bi-directional Search}
\vspace{-3ex}
\label{alg:bimodis}
\end{figure}

\eetitle{Algorithm}. 
Algorithm \bimodis, as shown in Fig.~\ref{alg:bimodis}, 
has the following steps. (1) {\em Initialization (lines~\ref{bi:ini})}. It first invokes a procedure~\kw{BackSt} to initialize a back-end start state node $s_b$. 
Two queues $Q_f$ and $Q_b$ are initialized, seeded with 
start state $s_U$ for forward search, and a back state $s_b$ 
for backward search, respectively. They serve as the forward and backward frontiers, respectively. 
(2) {\em Bi-directional Search (lines~\ref{bi:start}-\ref{bi:end}).}
\bimodis conducts an exploration from both directions, controlled by $Q_f$ for forward search, and $Q_b$ for backward search.  
Similar to \apxmodis, a Skyline set $\D_F$ is maintained 
in a levelwise manner. The difference is that 
it invokes a revised procedure \opg (with original counterpart in \apxmodis in Fig.~\ref{fig:approx}), which generates reduct operators for the forward search, and augment operators for the backward search.
\eat{
{\em Correlation-Based Pruning} is applied in line~\ref{bi:prunes} to \ref{bi:prunee}. For any two states from opposite directions, if a parameterized $\epsilon$-dominance relation is identified by procedure \kw{canPrune}, the intermediate states are pruned by the \kw{prune} procedure.
}
The search process terminates when both $Q_f$ and $Q_b$ are empty, or when a path is formed, the result $\D_F$ is returned.

\eetitle{Procedure~\kw{BackSt}}. This procedure 
initializes a backend dataset $D_b$  for augmentation. 
This procedure can be tailored to the specific task. 
For example, 
for a classifier $M$ with input features  
and a target attribute $A$ to be classified, 
we sample a small (minimal) set of tuples in $D_U$ to 
$D_b$ that covers all 
values of the active domain $\ad$ of $A$, to ensure that 
no classes will be ``missed'' in dataset $D_b$. 
Other task-specific strategies can also be applied here. 

\vspace{.5ex}
To reduce the valuation cost, \bimodis leverages correlation analysis 
over historical performance $T$, to 
assert ``non-$\epsilon$-dominance'' early, without a full valuation of their measures $\P$.  

\stitle{Correlation-Based Pruning}. At runtime, 
\bimodis dynamically maintains 
a correlation graph $G_\C$, where 
each node represents a measure in $\P$, and there is an edge $(p_i, p_j)$ in $G_\C$ if $p_i$ and $p_j$ are {\em strongly correlated}, with an associated weight $|\kw{corr}(p_i, p_j)|$~\cite{zheng2019towards}. Here 
we say two measures are strongly correlated, 
if their Spearman correlation coefficient $\kw{corr}(p_i, p_j)\geq \theta$, given their 
value distribution in the current 
set of tests $T$, for a user-defined threshold $\theta$. 
$G_\C$ is dynamically updated, 
as more valuated tests are added to $T$.

\eat{
For each pair $p_i$, $p_j \in \P \cup \{|D|\})$, \bimodis 
maintains the {\em Spearman correlation coefficient} $\kw{corr}(p_i, p_j)$ based on the up-to-date testset $T$. 
A perfect Spearman correlation of $\pm 1$ occurs when both variables are perfect monotone functions of each other~\cite{zheng2019towards}. The pair 
$(p_i, p_j)$ is considered strongly correlated if $|\kw{corr}(p_i, p_j)|$ exceeds a user-defined threshold $\theta$. 
We maintain a correlation graph $G_\C$, where each node represents a measure in $\P$, and there is an edge $(p_i, p_j)$ in $G_\C$ if $p_i$ and $p_j$ are strongly 
correlated, with an associated weight 
$|\kw{corr}(p_i, p_j)|$. 
$G_\C$ is continuously updated to capture the strength of correlations among measures as more valuated tests are added to $T$.
}

\eetitle{Parameterized Dominance}. 
\bimodis also ``parameterize'' any unvaluated measures in the performance vector $s.\P$ 
of a state $s$ with a potential range $[\hat{p_l}, \hat{p_u}]\subseteq [p_l, p_u]$. This range is derived from the 
valuated measures that are most strongly correlated, by consulting $G_\C$ and test sets $T$. 
The entire vector $s.\P$ is incrementally 
updated, for each $p\in \P$, by setting 
(1) $s.\P(p)$ as $t.p$ (valuated), if there is a corresponding test $t=(M, D_s)\in T$ with $t.p$ valuated; or (2) $s.\P(p)$ as a variable with an estimated range $[s.\hat{p_l}, s.\hat{p_u}]$, if no test over $p$ of $D_s$ is valuated.
 
A state $s$ is {\em parameterized $\epsilon$-dominated by} another state $s'$, denoted as $s' \succapprox_{\epsilon} s$,  if for each $p \in \P$, 
\tbi
\item $s'.\P(p)\leq (1+\epsilon) s.\P(p)$, if both 
are valuated; 
\item $s'.\hat{p_u} \leq (1+\epsilon) s.\hat{p_l}$, if neither 
is valuated; or
\item $s'.\P(p) \leq (1+\epsilon) s.\hat{p_l}$ 
(resp. $s'.\hat{p_u} \leq (1+\epsilon) s.\P(p)$), 
if $s'.\P(p)$ (resp. $s.\P(p)$) is valuated but 
$s.\P(p)$ (resp. $s'.\P(p)$) is not. 
\ei

Based on the above construction, \bimodis 
monitors a monotonicity condition as follows. 

\eetitle{Monotonicity Condition}. 
Given the current test set $T$, we say a state $s$ (resp. $s'$) with  
a performance measure $p$ at a path $\rho$ 
has a {\em monotonicity property}, if for 
any state $s''$ reachable from $s$ (resp. can reach $s'$) via $\rho$, 
$s.\hat{p_u}\textless \frac{s''.\hat{p_l}}{1+\epsilon}$ (resp. $s'.\hat{p_u}\textless \frac{s''.\hat{p_l}}{1+\epsilon}$). 

Given two states $s$ and $s'$,  
where $s' \succapprox_{\epsilon} s$, 
a state $s''$ on a path $\rho$ from $s$ or to $s'$ 
{\em can be pruned} by Correlation-based Pruning, if for every $p\in \P$, $s''$ has $p$ 
at $\rho$ with a monotonicity property 
\wrt $s$ (resp. $s'$). 
We present the following pruning rule. 

\begin{lemma}
\label{lm-prune}
Let $s \in Q_f$ and $s' \in Q_b$. 
If $s' \succapprox_{\epsilon} s$, 
then for any state node $s''$ 
on a path from $s$ or to $s'$  
that can be pruned by Correlation-Based Pruning, 
$D_{s''}$ is not in any $\epsilon$-Skyline set  
of the datasets that can be generated from valuated states.
\end{lemma} 

\begin{proofS}
We verify the pruning rule with a case study of 
$s$ and $s'$, subject to 
the monotonicity property.  
\textbf{Case 1: Both $s'.\P(p)$ and $s.\P(p)$ are valuated.}
If $s' \succapprox_{\epsilon} s$, then by definition, $s'.\P(p)\leq (1+\epsilon) s.\P(p)$ for all $p \in \P$. This 
readily leads to $\epsilon$-dominance, \ie $s'\succeq_\epsilon s$. As $s''$ has every performance measures 
$p\in \P$ with a monotonicity property \wrt $s$, 
$s\succeq_\epsilon s''$. Hence $s''$ can be safely pruned without valuation. 
\textbf{Case 2: Neither $s'.\P(p)$ nor $s.\P(p)$ is valuated.} By definition, as $s' \succapprox_{\epsilon} s$, 
then for every $p\in \P$, $s'.\hat{p_u}\leq (1+\epsilon) s.\hat{p_l}$. 
Given that $s''$ has every performance measures 
$p\in \P$ with a monotonicity property \wrt $s$, 
then by definition, for each $p\in \P$, we have 
$s'.p\leq s'.\hat{p_u}\leq (1+\epsilon) s.\hat{p_l}\leq (1+\epsilon)s.\hat{p_u}\textless (1+\epsilon)\frac{s''.\hat{p_l}}{1+\epsilon} \leq s''.p$, for 
every $p\in \P$. By definition of state dominance, $s' \succ s''$, for unevaluated $s''$.  
Following a similar proof, 
one can infer that $s \succ s''$ 
for a state $s$ in the forward front 
of \bimodis. 
Hence $s''$ can be safely pruned. 
\textbf{Case 3: One of $s'.\P(p)$ or $s.\P(p)$ is valuated.}
Given that $s' \succapprox_{\epsilon} s$, we have  
 (a) $s'.\P(p) \leq (1+\epsilon) s.\hat{p_l}$, if only $s'.\P(p)$ is valuated; or  (b) $s'.\hat{p_u} \leq (1+\epsilon) s.\P(p)$, if only $s.\P(p)$ is valuated.
Consider case 3(a). 
As $s$ can reach $s''$ via a path $\rho$, and 
$s''$ satisfiies the pruning condition, 
we can infer that 
$s'.\P(p) \leq (1+\epsilon) s.\hat{p_l} \leq 
(1+\epsilon) s.\hat{p_u} \textless (1+\epsilon) 
\frac{s''.\hat{p_l}}{1+\epsilon}\leq s''.p$, 
hence $s'\succ s''$. Similarly for case 3(b), we can infer that   
$s'.\hat{p_u} \leq (1+\epsilon) s.p \leq 
(1+\epsilon) s.\hat{p_u} \textless (1+\epsilon) 
\frac{s''.\hat{p_l}}{1+\epsilon}\leq s''.p$. 
hence $s'\succ s''$.  
For both cases, $s''$ can be pruned 
without evaluation. 
Lemma~\ref{lm-prune} hence follows. 
\end{proofS}

Procedures~\kw{canPrune} and~\kw{prune} (lines~11-12; omitted) 
asserts the Correlation-Based Pruning condition, 
and perform the maintenance of 
$G_\C$, $T$ and other auxiliary structures, 
respectively. Note that the above rule is checkable in PTIME 
\wrt input size $|\D_S|$. When $|\D_S|$ 
is large, one can generate a path with 
its states unevaluated, 
and check at runtime if the condition holds between 
evaluated states in the forward and backward 
frontier. 

We present the detailed analysis in~\cite{full}.

\eat{
if any state reachable from $S_U$ or 
can reach $s_b$ via the transition path can be 
skipped without valuation, with early detection of 
$\epsilon$-dominance. 
We maintain a bounds list that stores pairs of performance vectors with validated $\epsilon$-dominance relationships, where each vector in a pair is associated with states from opposite directions. 
For any feasible states $s$ and $s'$ from opposite directions, we first check if they fall within an existing bound. If they do, they are skipped; otherwise, the bounds list is updated or expanded accordingly. 
}

\eat{
\begin{proofS}
(1) We show that 
the parameterized $\epsilon$-dominance  
includes cases as necessary or sufficient 
conditions for $\epsilon$-dominance. 
This can be verified by contradiction, 
and the definition of 
parameterized dominance. 
(2) We then show that 
for any state $s''$ on a path from $s$ or to $s'$, if $s''$ is pruned according to Correlation-Based Pruning 
, there must exists a state $s_p \in G_\T$ that $\epsilon$-dominates 
$s''$. 
\end{proofS} 
}

\begin{example} 
We illustrate {\em Correlation-Based Pruning} in the figure below. From left to right, it depicts a set of test records $R$, the correlation graph $G_\C$, and part of the running graph $G_\T$.  
$G_\C$ is constructed from $T$ with measures as nodes and Spearman correlations as edge weights. 
For each $s_n \in G_\T$, the associated $p \in \P_{s_n}$ is obtained by test $t_{s_n} = (M, D_{s_n})$. 
\begin{center}
\begin{small}
    \hspace*{5pt}
    \begin{minipage}[c]{0.27\textwidth}
        \centering
        \begin{tabular}{|c|c|c|c|c|}
            \hline
             & Label & $p_1$ & $p_2$ & $p_3$ \\
             \hline
             \tikz[remember picture] \node[anchor=center, inner sep=0] (su){$s_U$}; & (1, 1, 1, 1) & \underline{0.42} & \underline{0.18} & 0.9 \\
             \hline
             $s_1$ & (1, 1, 1, 0) & 0.4 & 0.17 & 0.1 \\
             \hline
             \tikz[remember picture] \node[anchor=center, inner sep=0] (s2){$s_2$};  & (1, 0, 0, 1) & \underline{0.5} & \underline{0.22} & / \\
             \hline
             \tikz[remember picture] \node[anchor=center, inner sep=0] (s3){$s_3$}; & (0, 1, 0, 0) & \textbf{0.45} & \textbf{/} & / \\
             \hline
             $s_b$ & (0, 0, 0, 0) & 0.6 & 0.4 & 0.3 \\
             \hline
        \end{tabular}
    \end{minipage}%
    \begin{minipage}[c]{0.23\textwidth}
        \centering        \includegraphics[height=0.87in]{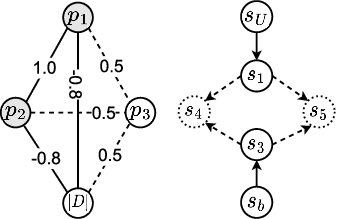}
    \end{minipage}
\end{small}
\end{center}
\begin{tikzpicture}[remember picture, overlay]
  \coordinate (suwest) at (su.west);
  \coordinate (s2west) at (s2.west);
  \coordinate (midpoint) at ($(suwest)!0.5!(s2west)$);

  \draw[line width=0.6pt] 
    ([xshift=-1.73pt]suwest) -- 
    ([xshift=-8.73pt]suwest) -- 
    ([xshift=-10pt]s2west) -- 
    ([xshift=-3pt]s2west);

  \draw[->, line width=0.6pt] 
  ([xshift=-9.7pt]midpoint) -- 
  ([xshift=-14pt]midpoint) -- 
  ([xshift=-14.5pt]s3.west) -- 
  ([xshift=-6.2pt]s3.west);
\end{tikzpicture}

At $\theta=0.8$, $p_1$ and $p_2$ are positively correlated with each other and negatively correlated with $|D|$, so $\P^k = \{p_1, p_2\}$.
From $s_U$ and $s_b$, the forward and backward frontiers derive states $s_1$ and $s_3$, respectively.
To estimate $s_3.\P(p_2)$, note that $s_3.\P(p_1) = 0.45$, which lies between $s_U.\P(p_1) = 0.42$ and $s_2.\P(p_1) = 0.5$.
Given the strong correlation between $p_1$ and $p_2$,
we infer $s_3.\P(p_2)$ to be within the interval 
$[0.18, 0.22]$, with $\hat{p_{l_2}} = s_U.\P(p_2)$ and $\hat{p_{u_2}} = s_2.\P(p_2)$. 
With $\epsilon=0.3$, we find $s_3 \succapprox_{\epsilon} s_1$ because $0.45 \leq (1+0.3)\cdot0.4$ and $0.22 \leq (1+0.3)\cdot0.17$. 
For intermediate states $s_4$ (bitmap entry $(1, 1, 0, 0)$) and $s_5$ (bitmap entry $(0, 1, 1, 0)$), which are not recorded in $T$ and have $|D_{s_4}| = |D_{s_5}| = 2$, a similar inference process shows they fall within the bounds set by $[s_1.\P, s_3.\P]$. As a
result, $s_4$ and $s_5$ can be pruned.
\end{example}

\stitle{Time Cost}. 
\bimodis 
takes the same time complexity as \apxmodis. 
The $(N, \epsilon)$-approximation holds 
for \bimodis given that it correctly updates 
the $\epsilon$-Skyline set by definition. 
Our experimental study verifies that it is much faster in practice and particularly suitable for larger $\epsilon$ or search spaces (represented by maximum path length). It also scales more efficiently for large datasets (see Exp-3 in Section~\ref{sec:exp}).




\eat{
\eetitle{Discussion on Estimators}. 
The lower and upper bound for the performance of a surrogate model~\cite{mousavi2020minimax}.

\begin{lemma} 
\label{theo:surrogate}
If an estimator the performance value, assume $\forall \text{estimator}, P_i(\text{accuracy}) \leq (1+\epsilon)\theta \cdot P$, then the actual approximation ratio would be $(1+f(\epsilon))$
\end{lemma}

\warn{Practically show that it is small and cite some papers.}

Exponential time, with a strong approximate guarantee. The size of the Skyline set is polynomially bounded~\cite{tsaggouris2009multiobjective}.
}

\eat{
Issues:

1. Enumerate the exponential number of cases. It will only apply to small cases where we can enumerate all the possible configurations in the worst case. 

2. Lack of diversity in values.
}

\vspace{-1ex}
\subsection{Diversified Skyline Dataset Generation} 
\label{sec-divmodis}


A Skyline dataset may still contain data that largely overlap or are similar, hence leading to bias and reducing the generality of the model if adopted. 
This may occur due to skewed value distribution  
in the active domains,  
common attributes, over
specific performance metrics in the skyline 
data generation process. 
It is often desirable to explore a diversified variant of skyline set generation to create varied datasets that mitigate such bias~\cite{konakovic2020diversity, low2023evolution}. 

Given $\T$ and a configuration $C$, a set of datasets $\D$, constants $N$, $\epsilon$ and $k$,  
the {\em diversified skyline data generation} 
is to compute a set $\D^*_F$ 
of at most $k$ tables, such that 
(1) $\D^*_F$ is an $\epsilon$-Skyline set 
of $N$ valuated states by an $(N, \epsilon)$-approximation of \modis, 
and (2) among all $\epsilon$-Skyline sets 
over $N$ states valuated in (1), it 
maximizes a diversification score 
defined as: 
\begin{equation}
\kw{div}(\D_F) = \sum_{i=1}^{k-1} \sum_{j=i+1}^{k} \kw{dis}(D_i, D_j)
\label{eq:div}
\end{equation}
where a distance function $\kw{dis}$ quantifies   
the difference of datasets 
in terms of both value distributions and estimated performance, and is defined as:
\[\kw{dis}(D_i, D_j) = 
\alpha\frac{1-\kw{cos}(s_i.L, s_j.L)}{2}  + 
(1-\alpha)\frac{\kw{euc}(t_i.\P, t_j.\P)}{\kw{euc_{m}}}\]
We adopt Cosine similarity $\kw{cos}$ 
and Euclid Distance ($\kw{euc}$). 
The latter 
is normalized by the maximum Euclid Distance 
$euc_m$ among the historical performances in $T$. 

\vspace{.5ex}
We next outline an algorithm, 
denoted as \divmodis, that extends 
an $(N, \epsilon)$-approximation to 
computes an 
a diversified $\epsilon$-Skyline set $\D_F$  
of at most $k$ datasets. 

\eat{
Given a configuration $C$, a set of datasets $\D$, an integer $k$, and constant $\epsilon$, 
the goal is to compute an $\epsilon$-Skyline set $\D_F \subset \D$
of at most $k$ datasets, maximizing a diversification measure as follows:
\begin{equation}
div(\D_F) = \sum_{i=1}^{k-1} \sum_{j=i+1}^{k} dis(D_i, D_j)
\label{eq:div}
\end{equation}
}

\eetitle{Algorithm}. 
\divmodis revises \modis by incrementally diversify an input $\epsilon$-Skyline set ${\D_F}^{i}$ at level $i$ (partially shown in Fig.\ref{alg:divmodis}). It derives 
$\D^P_F$ by a greedy selection and replace strategy 
as follows. 
(1) It initializes $\D^P_F$ as a random $k$-set from ${\D_F}^{i}$, and updates $\D^P_F$ by incrementally replacing tables with the highest marginal gain in diversification, hence an improved $div(\D^P_F)$. (2) $\D^P_F$ is passed to be processed at level $i+1$, upon the arrival of new states. \divmodis returns the diversified set $\mathcal{D}_F$, 
following the same termination condition 
as in \apxmodis. 

\begin{figure}
\vspace{-2ex}
\centering
\begin{algorithm}[H]
\caption{:Diversification step at level $i$}
\begin{algorithmic}[1]
\algtext*{EndFor}
\algtext*{EndIf}
\algtext*{EndWhile}
\algtext*{EndFunction}
\algtext*{EndProcedure}

\State \textbf{Input:} 
    $\epsilon$-Skyline set ${\D_F}^{i}$ (from \upi),  integer $k$;
\State \textbf{Output:} 
     a diversified $k$-subset of ${\D_F}^{i}$ (to be passed to level $i+1$).
     \vspace{1ex}
    
\If{$|{\D_F}^{i}| \leq k$} \Return ${\D_F}^i$
\EndIf 
\State initialize $\D_F^P$ with $k$ 
random dataset in ${\D_F}^i$; 
\State score := $div(\D_F^P)$; 
\For{\textbf{all} $D \in \D_F^P$}
\For{\textbf{all} $D' \in {\D_F}^{i}$}
    \If{$D' \in \D_F^P$} Continue;
    \EndIf
    \State $\D_F^{P'} := (\D_F^P \setminus \{D\}) \cup \{D'\}$;
    \State score' := $div(\D_F^{P'})$
    \If{score' $>$ score} 
    \State $\D_F^P := \D_F^{P'}$, score $:=$ score';
    \EndIf
\EndFor
\EndFor
\State \Return $\D_F^P$
\end{algorithmic}
\end{algorithm}
\vspace{-4ex}
\caption{Level-wise diversification of \divmodis}
\vspace{-3ex}
\label{alg:divmodis}
\end{figure}

We show that the diversified \modis 
can be approximated, 
for a submodular diversification function $\kw{div}$. 
Our result holds for the specification of 
$\kw{div}$ in Equation~\ref{eq:div}. 

\begin{lemma} 
\label{lemma:div}
Given $N$ and $\epsilon$, 
\divmodis achieves a $\frac{1}{4}$ approximation for 
diversified \modis, \ie 
(1) it correctly computes a $\epsilon$-Skyline set $D^P_F$
over $N$ valuated datasets, and 
(2) $\kw{div}(D^P_F)\geq \frac{1}{4}\kw{\kw{div}(\D^*_F)}$. 
\end{lemma}

\begin{proofS}
We show an induction on the levels. 
(1) We verify the guarantee at a single level, 
by constructing an approximation preserving 
reduction to the stream submodular maximization problem~\cite{chakrabarti2015submodular}. Given a  
 stream $E = \{e_0, \ldots e_m\}$, an integer $k$, and a submodular 
diversification function $f$, it
computes a $k$-set of elements $S$ 
that can maximize $f(S)$. 
Our reduction constructs a stream of 
datasets $\D_S$ following the level-wise 
generation. We show that 
the function $\kw{div}$ is 
a submodular function. 
(2) By integrating a 
greedy selection and replacement 
policy, 
\divmodis keeps a $k$-set with the most diverse and representative datasets to mitigate the biases in the Skyline set.
\divmodis achieves a $\frac{1}{4}$-approximation of an $\epsilon$-Skyline set with maximized diversity at each level $i$. 
\eat{
To show the approximation 
ratio, we construct a
reduction from  \modis diversification variant
to {\em stream submodular maximization problem}~\cite{chakrabarti2015submodular}. Given a streaming of elements $E = \{e_0,e_1, ..., e_m\}$, an integer $k$, and a submodular function $f$, stream submodular maximization problem
maintains a set of elements $S$ with size $k$ at any time such that 
$f(S) \leq \frac{1}{4}f(S^*)$ where 
$S^*$ is the globe optimal solution.
}
Please see the detailed proof in~\cite{full}. 
\end{proofS}

\stitle{Analysis}.
\divmodis incurs an overhead to update the diversified $k$-set. 
As \modis valuates up to $\min(N_u^{|R_u|}, N)$ nodes (datasets), 
the total additional overhead is in $O(\min(N_u^{|R_u|}, N) \cdot k \cdot T_{\mathcal{S}})$, 
where $T_{\mathcal{S}}$ refers to the unit valuation cost for a single table, which is in PTIME. 
As both $k$ and $T_{\mathcal{S}}$ are relatively small, 
the practical overhead for \divmodis remains small (see Sec.~\ref{sec:exp}). 

\eat{
\vspace{.5ex}
\stitle{Interpretability}. Our \modis algorithms 
can be readily extended to 
answer provenance questions such as 
``what'' data sources are used, 
``where'' the columns are 
from, and how a skyline set is 
generated. This can be done by 
retrieving, for each result 
dataset $D$ in the skyline set,
the paths and associated 
operators from the running graph. We provide more discussion in~\cite{full}. 
}

\stitle{Remarks}. 
Alternatives that solve multi-objective 
optimization may be applied, such as 
evolutionary algorithms such as NSGA-II~\cite{deb2002fast}, 
or reinforcement-learning based methods~\cite{liu2014multiobjective}. 
The former rely on costly stochastic processes (e.g., mutation and crossover) and may require extensive parameter tuning. 
The latter are effective for general state exploration but require 
high-quality training samples 
and may not converge over ``conflicting'' 
measures. In contrast, \modis is training 
and tuning free. Our experiments verified 
that \apxmodis provides early 
generation of high-quality datasets from a few large input datasets, 
due to ``reduce-from-Universal'' 
strategy, \bimodis enhances efficiency through bidirectional exploration and  pruning, hence 
benefits for larger number of small-scale datasets, and \divmodis 
benefits most for datasets with skewed distribution.

\section{Experiment Study}
\label{sec:exp}

We next experimentally verify 
the efficiency and effectiveness of 
our algorithms. We 
aim to answer three questions: 
\textbf{RQ1}: 
How well can our algorithms improve the performance of models in multiple measures? 
\textbf{RQ2}: 
What is the impact of generation settings, such as data size?
\textbf{RQ3}: 
How fast can they generate skyline sets, and how scalable are they? 
We also illustrate the applications of our approaches with case studies\footnote{
Our codes and datasets are available at 
github.com/wang-mengying/modis}.

\stitle{Datasets.}
We use three sets of tabular datasets: 
kaggle~\cite{KaggleYourHome}, \open~\cite{DataGov}, and \hf~\cite{HuggingFaceAI} (summarized in Table~\ref{tab-data}). 

\eat{
\stitle{Datasets}. 
We use the following datasets summarized below:

\sstab
(1) \kaggle~\cite{KaggleYourHome}, collected from 
a set of tables involving movie information;  

\sstab
(2) \open: a fraction of a large open public dataset~\cite{DataGov}. We sampled $2K$ tables, involing schools recording, school evaluations and school types, houses recording, housing price in New York and Chicago, geographical location, among others; 

\sstab
(3) \hf: a set of 
tables involving Avocado prices and relevant information,  
sampled from Hugging Face~\cite{HuggingFaceAI}.  
}


\begin{table}
    \centering
    \begin{small}
    \begin{tabular}{|c|c|c|c|}
    \hline
       Dataset Sets  & \# tables & \# Columns & \# Rows  \\ \hline
       \kaggle  & 1943 & 33573 & 7317K 
       \\ 
       \hline
       \open  & 2457 & 71416 & 33296K
       \\ 
       \hline
       \hf & 255  & 1395 & 10207K 
       \\ \hline
       
    \end{tabular}
    \end{small}
    \caption{Characteristics of Datasets}
    \vspace{-5ex}
   \label{tab-data}
      \vspace{-3ex}
\end{table}

\stitle{Tasks and Models.}
A set of tasks are assigned for evaluation. 
We trained: (1) a Gradient Boosting model (\ul{\gbm}) to predict movie grosses using \kaggle for Task $T_1$; 
(2) a Random Forest model (\ul{\rfh}) to classify house prices using \open with the same settings in~\cite{galhotra2023metam} for Task $T_2$; and 
(3) a Logistic Regression model (\ul{LRavocado}) to predict Avocado prices using \hf for Task $T_3$. 
(4) a LightGBM model (\lgc)~\cite{ke2017lightgbm} to classify mental health status using \kaggle for Task $T_4$. 
We also introduced 
task $T5$, a link regression task for
recommendation. This task takes as input a bipartite graph 
between users and products, and links indicate their interaction. 
A LightGCN~\cite{he2020lightgcn} (\lgr), 
a variant of graph neural networks (GNN) optimized for 
fast graph learning, 
is trained 
to predict top-$k$ missing edges in an input bipartite graph 
to suggest products to users. 
A set of $1873$ bipartite graphs is constructed from \kaggle for 
$T_5$.  
The ``augment'' (resp. ``reduct'') operators are  
defined as edge insertions (resp. edge deletions) 
to transform a bipartite graph to another.  

We use the same training scripts for each task and all methods 
for a fair comparison.
We assigned measures $\P_1$ through $\P_5$ for tasks $T_1$ to $T_5$, respectively
, as summarized in Table~\ref{tab-measures}.
We also report the size of the data ($p_{DSize}$) in terms of (total $\#$ of rows total $\#$ of columns), excluding attributes with all cells masked.

\eat{
\stitle{Tasks and Models}. 
We have trained the following 
models: 
(1) a random forest models \rfh 
for classifying house price 
(Task $T_1$), 
using 
\open and the settings consistently in~\cite{galhotra2023metam, fan2022semantics};   
(2) a Gradient Boosting Model 
(\gbm) for predicting the movies' worldwide gross sale, using \kaggle (Task $T_2$); and 
(3) a regression model 
for predicting Avocado price, 
using \hf (Task $T_3$).  

We trained all these models 
with scikit-learn~\cite{scikit-learn}.
For a fair 
comparison, we use the original 
training scripts provided by the baseline 
methods and validated that the 
reproduced models have 
consistent performance 
as reported. 
}

\eetitle{Estimator $\E$}. We 
adopt MO-GBM~\cite{scikit-learn}  
as a desired model performance estimator. It 
outperforms other candidate models even with a simple training set 
For example, for $T_1$, MO-GBM performs inference for all objectives on one state in at most $0.2$ seconds, with a small MSE of $0.0003$ when predicting ``Accuracy''. 
\eat{
We trained  
a multi-output Gradient Boosting Model (MO-GBM)~\cite{scikit-learn} 
as our estimator. 
}

\eat{
\mengying{To simplify estimation and improve accuracy, we used the bitmap encoding $s.L$ as input to predict the performance measures $\P_s$ for state $s$ over the given model $M$. 
}
}
\eat{which outputs predicted 
values for multiple variables, allowing us to valuate the performance vector for each test with one call.}

\stitle{Algorithms}. 
We implemented the following methods. 

\sstab
(1) \textbf{\underline{MODis}}: Our multi-objective data discovery algorithms, including \apxmodis, \bimodis, and \divmodis. We also implemented \nomodis, a counterpart of \bimodis without correlation-based pruning. 
(2) \underline{\metam}~\cite{galhotra2023metam}: 
a goal-oriented data discovery algorithm
that optimizes a single utility score with consecutive joins of tables. 
We also implemented an extension \metammo, 
by incorporating multiple measures into a single 
linear weighted utility function. 
(3) \underline{\starmie}~\cite{fan2023semantics}: a data discovery method that focuses on table-union search and uses contrastive learning to identify joinable tables.
For \metam and \starmie, we used the code from 
original papers. 
(4) \underline{\sklearn}~\cite{scikit-learn}: 
An automated feature selection method 
in scikit-learn's \kw{SelectFromModel}, which recommends important features with a built-in 
estimator.
(5) \underline{\ho}~\cite{h2o_platform}: an AutoML platform; we used its feature selection module, which fits features and predictors into a linear model.

\stitle{Construction of 
$D_U$ and Operators}.
To prepare 
universal datasets $D_U$ 
for \modis, we 
preprocess \kaggle, \open and \hf 
into joinable tables and construct 
$D_U$ with multi-way joins. 
This results in $D_U$ 
datasets with a size (in terms of 
\# of columns and \# of rows):
$(12, 3732)$, $(27, 1178)$, $(13, 18249)$ 
and $(20, 140700)$, for 
tasks $T_1$ to $T_4$, respectively. 
Specifically, we applied 
$k$-means clustering over 
the active domain of 
each 
attribute (with a maximum $k$ 
set as $30$), and 
derived equality literals, 
one for each cluster. 
We then compressed the 
input tables by replacing 
rows into tuple clusters, reducing 
the number of rows. 
This pragmatically help us 
avoid starting from large 
$D_U$ by only 
retaining 
the values of interests,   
and still yield desired 
skyline datasets. 
For $T_5$, a 
large bipartite graph 
is constructed 
with a size of 
$(7925, 34)$ (\# of edges, \# of nodes' features). 
The generation of graphs consistently aligns with its table data counterpart, 
by conveniently replacing augment and reduction to their graph counterpart 
that performs link insertions and deletions. 

\eat{
\revise{
We also include task $T_5$, where the base table is a graph. 
The universal dataset $D_U$ for $T_5$ is constructed by augmenting the base table with information from \kizoo, resulting in a size of $(7925, 34)$ (\# of edges, \# of nodes' features). Other settings for $T_5$ follow the default configurations used in other tasks, showcasing \modis's applicability across diverse data modalities.}
\eat{
We identified related attributes based on a loose standard with attributes' names and overlaps to maximize the information included in $s_U$, with the size of $(12, 3732)$, $(27, 1178)$ and $(13, 18249)$.
}
}

\eat{
\warn{add description.}
Distributed and scalable machine learning and predictive analytics platform that allows the building of machine learning models on big data and provides easy productionalization of those models in an enterprise environment.
}




\eat{
\item
\automl~\cite{}: \warn{add description.} 

\item 
\fselect~\cite{}: \warn{add description.}}


\begin{table}
    \centering
 \begin{center}
 \begin{small}
     \begin{tabular}{|c|c|c|} \hline
       Notation &  Measures    & Used In  \\ \hline
       $p_{Acc}$   & Model Accuracy &   $\P_1$, $\P_2$, $\P_4$ \\ \hline
       $p_{Tr}$   & Training Time Cost  &   $\P_1$-$\P_4$ \\ \hline
       $p_{F1}$   & $F_1$ score  &   $\P_2$, $\P_4$ \\ \hline
       $p_{AUC}$   & Area under the curve  &   $\P_4$ \\ \hline
       $p_{Nc(n)}$   & NDCG(@n)  &   $\P_5$ \\ \hline
       $p_{MAE}$, $p_{MSE}$   & Mean Absolute / Squared Error &   $\P_3$\\ \hline
       $p_{Pc(n)}$, $p_{Rc(n)}$   & Precision(@n), Recall(@n) &   $\P_5$\\ 
       \hline \hline
       $p_{Fsc}$   & Fisher Score~\cite{li2017feature} &   $\P_1$, $\P_2$ \\ \hline
       $p_{MI}$   & Mutual Information~\cite{li2017feature,galhotra2023metam} &   $\P_1$, $\P_2$ \\ \hline
     \end{tabular}
     \end{small}
    \caption{Performance Measures}
     \label{tab-measures}
     \end{center}
\vspace{-5ex}
\end{table}

\stitle{Evaluation metrics}. 
We adopt the following metrics 
to quantify the effectiveness of data discovery approaches. Denote $D_M$ as an initial dataset, and 
$\D_o$ a set of output datasets from 
a data discovery algorithm. 
(1) We define the {\em relative improvement} 
$\relp(p)$ for a given 
measure $p$ achieved by a
method as $\frac{M(D_M).p}{M(D_o).p}$.
As all metrics are normalized to be minimized,
the larger $\relp(p)$ is,
the better $D_p$ is in improving $M$ \wrt $p$. 
Here $M(D_M).p$ and $M(D_p).p$ are obtained 
by actual model inference test. 
This allows us to fairly compare all 
methods in terms of the quality of data 
suggestion. 
For efficiency, we compare the time cost of data discovery upon 
receiving a given model or task 
as a ``query''. 

\eat{
\eetitle{Task scenarios}. We created the following 
task scenarios with specified task, model, and performance 
measures (Perf.), summarized in Table~\ref{tab-task}. \warn{Give the table}. Here $\P_1$ - $\P_3$ are 
defined as follows. \warn{No need to use the same 
set of performance metrics. Anything larger than one is good. 
Accordingly for radar graph -- you can have one with Axis of three, four, or five.}
}

\eat{
\eetitle{TUS~\cite{nargesian2018table}}} 


\begin{table*}[tb!]
\customsize
\centering
\renewcommand{\arraystretch}{1.05}
\begin{small}
\begin{tabular}{|c|c|c|c|c|c|c||c|c|c|c|}
\hline
$T_2$: House & Original & \metam & \metammo & \starmie & \sklearn & \ho & \apxmodis & \nomodis & \bimodis & \divmodis \\ \hline
$p_{F1}$ & 0.8288 & 0.8510 & 0.8310 & 0.8351 & 0.7825 & 0.8333 & 0.9044 & \ul{\textbf{0.9125}} & \ul{\textbf{0.9125}} & 0.8732 \\ \hline
$p_{Acc}$ & 0.8305 & 0.8322 & 0.8333 & 0.8331 & 0.7826 & 0.8305 & 0.9050 & \ul{\textbf{0.9121}} & \ul{\textbf{0.9121}} & 0.8729 \\ \hline
$p_{Train}$ & 0.2000 & 0.21 & 0.19 & 0.2100 & 0.2000 & 0.2000 & 0.1533 & \ul{\textbf{0.1519}} & \ul{\textbf{0.1519}} & 0.2128 \\ \hline
$p_{F_{sc}}$ & 0.0928 & 0.0889 & 0.0894 & 0.0149 & 0.2472 & 0.0691 & 0.2268 & \ul{\textbf{0.2610}} & \ul{\textbf{0.2610}} & 0.2223 \\ \hline
$p_{MI}$ & 0.126 & 0.1109 & 0.1207 & 0.0243 & \ul{\textit{0.2970}} & 0.1054 & 0.2039 & 0.2018 & 0.2018 & \textbf{0.3164} \\ \hline
Output Size  & (1178, 27)  & (1178, 28)  & (1178, 28)  & (1178, 32)  & (1178, 4)  & (1178, 15)  & (835, 17)  & (797, 17)  & (797, 17)  & (1129, 5)\\ \hline

\hline
\hline



$T_4$: Mental & Original & \metam & \metammo & \starmie & \sklearn & \ho & \apxmodis & \nomodis & \bimodis & \divmodis \\ \hline
$p_{Acc}$      & 0.9222 & 0.9468 & 0.9462  & 0.9505 & 0.8839 & 0.9236 & \textbf{0.9532} & 0.9471 & \ul{0.9525} & 0.9471 \\
\hline
$p_{Pc}$     & 0.7940 & 0.7991 & 0.8070 & 0.8106 & 0.6577 & 0.7892 & \textbf{0.8577} & 0.8454 & \ul{0.8549} & 0.8454 \\
\hline
$p_{Rc}$        & 0.7722 & 0.7846 & 0.7959 & 0.8030 & 0.7523 & 0.7879 & \textbf{0.8097} & \ul{0.8092} & 0.8075 & \ul{0.8092} \\
\hline
$p_{F1}$           & 0.7829 & 0.7918 & 0.8014 & 0.8068 & 0.7018 & 0.7885 & \textbf{0.8330} & 0.8269 & \ul{0.8305} & 0.8269 \\
\hline
$p_{AUC}$           & 0.9618 & 0.9757 & 0.9774 & 0.9784 & 0.9326 & 0.9615 & \textbf{0.9792} & 0.9755 & \ul{0.9789} & 0.9755 \\
\hline
$p_{Train}$ & 0.4098 & 0.3198 & 0.4027 & 0.3333 & \textbf{0.2359} & \ul{0.2530} & 0.3327 & 0.2818 & 0.3201 & 0.2818 \\
\hline
Output Size  & ($10^5$, 14) & ($10^5$, 15) & ($10^5$, 15) & ($10^5$, 16) & ($10^5$ 8) & ($10^5$, 8) & (128332, 16) & (116048, 16) & (128332, 17) & (116048, 16) \\
\hline

\end{tabular}
\end{small}
\caption{Comparison of Data Discovery Algorithms in Multi-Objective Setting ($T_2$, $T_4$)}
\label{tab:comparison}
\vspace{-5ex}
\end{table*}

\eat{
\begin{table*}[tb!]
\customsize
\centering
\begin{tabular}
{|c|c|c|c|c|c|c||c|c|c|c|}
\hline
$T_1$: Movie Gross (C) & Original & \metam & \metammo & \starmie & \sklearn & \ho & \apxmodis & \nomodis & \bimodis & \divmodis \\ \hline
$p_{Acc}$ & 0.8560 & 0.8743 & 0.8676 & 0.8606 & 0.8285 & 0.8545 & 0.9291 & \ul{\textbf{0.9874}} & \ul{\textit{0.9755}} & 0.9427 \\ \hline
$p_{Train}$ & 1.4775 & 1.6276 & 1.1785 & \ul{\textbf{1.2643}} & 0.6028 & 0.9692 & 0.9947 & \ul{\textit{0.8766}} & 0.8027 &  \\ \hline
$p_{Fsc}$ & 0.0824 & 0.0497 & 0.0801 & 0.1286 & 0.7392 & 0.3110 & 0.6011 & \ul{\textit{0.7202}} & \ul{\textbf{0.9240}} & 0.8010\\ \hline
$p_{MI}$ & 0.0538 & 0.0344 & 0.0522 & 0.1072 & 0.3921 & 0.1759  &\ul{\textbf{0.4178}} & \ul{\textit{0.3377}} & 0.3839 & 0.4165\\ \hline
$p_{VIF}$ & 1.5831 & 1.9669 & 1.9782 & 1.2980 & 1.6742 & 1.5096 & \ul{\textbf{2.4092}} & \ul{\textit{2.1331}} & 1.7688 & 2.0188 \\ \hline
Output Data Size  & (3264, 10) & (3264, 11) & (3264, 11) & (3264, 23) & (3264, 3) & (3264, 8) & (2958, 9) & (1980, 12) & (1835, 11) & (2176, 10) \\ \hline
\end{tabular}
\vspace{0.5ex}
\caption{Comparison of Data Discovery Algorithms in Multi-Objective Setting}
\label{tab:comparison}
\end{table*}
}

\begin{table}[tb!]
\customsize
\centering
\renewcommand{\arraystretch}{1.05}
\begin{small}
\begin{tabular}{|>{\centering\arraybackslash}p{1.32cm}|>{\centering\arraybackslash}p{0.93cm}|>{\centering\arraybackslash}p{1.16cm}|>{\centering\arraybackslash}p{1.05cm}|>{\centering\arraybackslash}p{1.05cm}|>{\centering\arraybackslash}p{1.05cm}|}
\hline
$T_5$: Model & Original 
& ApxMODis & NOMODis & BiMODis & DivMODis \\
\hline
$p_{Pc_5}$    & 0.7200 
& \textbf{0.8200} & 0.8000 & \textbf{0.8200} & 0.8000 \\
\hline
$p_{Pc_{10}}$  & 0.6600 
& 0.8100 & 0.8000 & \textbf{0.8200} & 0.8000 \\
\hline
$p_{Rc_5}$       & 0.1863 
& \textbf{0.2072} & 0.2022 & \textbf{0.2072} & 0.2022 \\
\hline
$p_{Rc_{10}}$      & 0.3217 
& 0.3866 & 0.3816 & \textbf{0.3977} & 0.3816 \\
\hline
$p_{Nc_5}$        & 0.6923 
& \textbf{0.7935} & 0.7875 & 0.7924 & 0.7875 \\
\hline
$p_{Nc_{10}}$        & 0.6646 
& 0.7976 & 0.7891 & \textbf{0.8033} & 0.7891 \\
\hline
Output Size           & (7925, 0) 
& (5826, 30) & (1966, 6) & (2869, 4) & (1966, 6) \\
\hline
\end{tabular}
\end{small}
\caption{Comparison of \modis Methods on $T_5$}
\label{tab:modsnet}
\vspace{-6ex}
\end{table}

\begin{figure}[tb!]
\vspace{-2ex}
\centerline{\includegraphics[width =0.43\textwidth]{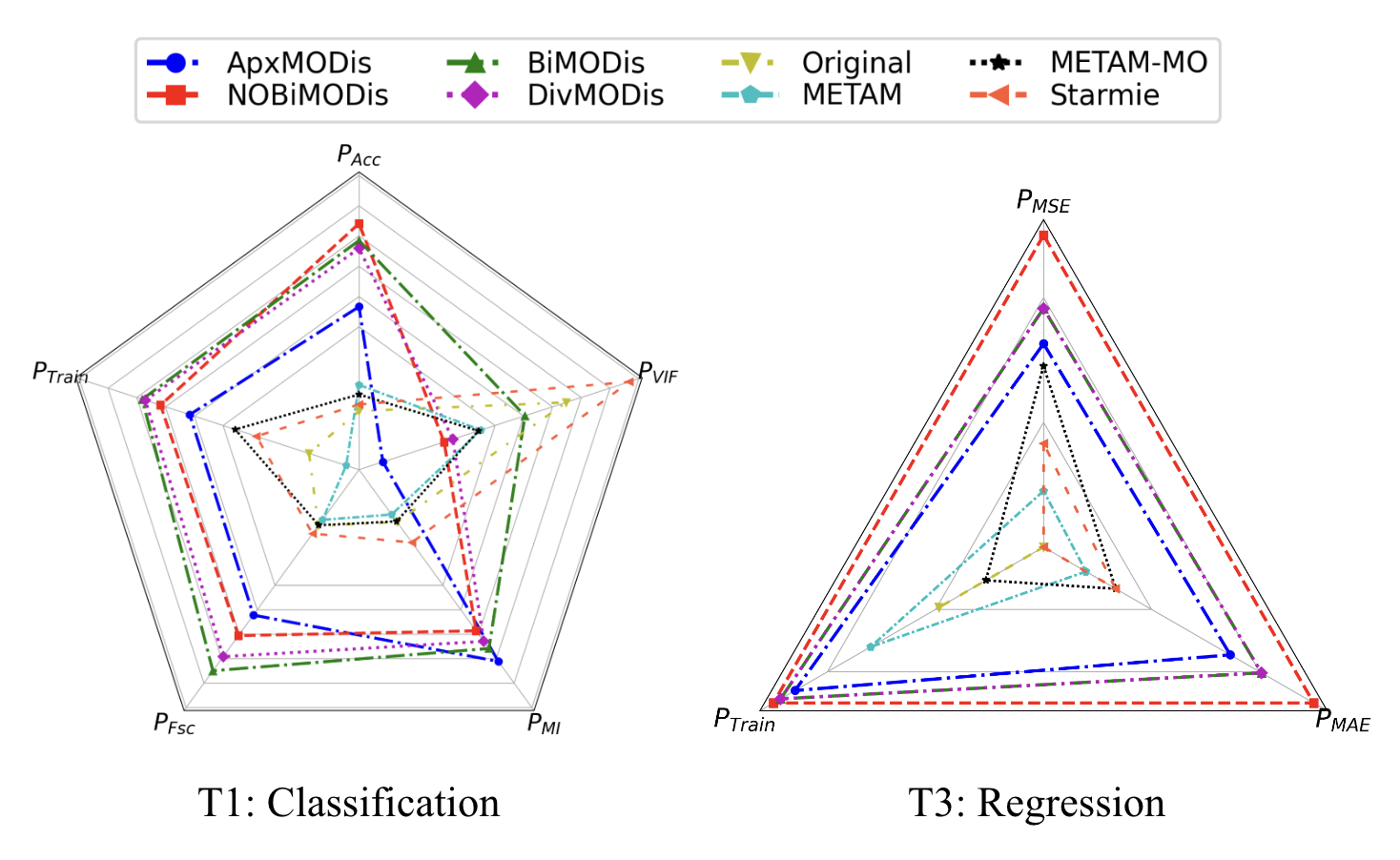}}
\centering
\vspace{-1ex}
\caption{Effectiveness: Multiple Measures}
\vspace{-3ex}
\label{fig:mo-eff}
\end{figure}

\stitle{Exp-1: Effectiveness}. 
We first evaluate \modis methods 
over five tasks. 
Results for $T_1$ and $T_3$ are shown
in Fig.~\ref{fig:mo-eff} (the outer, the better). 
While results for $T_2$ and $T_4$
are presented in Table~\ref{tab:comparison}.
Results for $T_5$ are in Table~\ref{tab:modsnet}.
We also report the model performance over 
 the input tables as a ``yardstick'' 
 (``Original'') for all methods. 
As all baselines output a single table, to compare \modis algorithms, we select the table in the Skyline set with the best estimated $p_{Acc}$, $P_{F1}$, $P_{MSE}$, $p_{Acc}$ and $p_{Pc_5}$ for $T_1$ to $T_5$, respectively. 
As \metam optimizes a single utility score, we choose the same measure for each task as the utility. 
We apply model inference to all the output tables to report actual performance values. 
We have the following observations. 

\sstab
(1) \modis algorithms outperform all the baselines in all tasks. 
As shown in Table~\ref{tab:comparison}, for example, for $T_4$, the datasets that bear best $p_{Acc}$ and the second best are returned by \apxmodis (0.9535) and \bimodis (0.9525), respectively, and all \modis methods generated datasets that achieve $0.87$ on $p_{F1}$ in $T_2$. 

\sstab
(2) Over the same dataset and for other measures, \modis algorithms outperform the baselines in most cases. 
For example, in $T_1$, the result datasets that most optimize $p_{Fsc}$ and $p_{MI}$ are obtained by \bimodis and \apxmodis, respectively; also in $T_2$  and $T_3$, \nomodis and \bimodis show absolute dominance in most measures. 
Table~\ref{tab:modsnet} also verifies 
that \modis easily generalizes to suggest 
graph data for GNN-based analytics, beyond tabular data.  

\sstab
(3) Methods with data  
augmentation (\eg \metam and \starmie) enriches 
data to improve model accuracy, at a cost of training time, while feature selection methods (\eg \sklearn and \ho) reduce data at the cost of accuracy with improved training efficiency. \modis 
methods are able to balance these trade-offs better 
by {\em explicitly} performing multi-objective optimization. 
For example, $p_{Acc}$ and $p_{Train}$ in $T_4$, The best result for training cost (0.2359s) 
is contributed from \sklearn, yet at a cost of 
lowest model accuracy (0.8839). 

We also compared $p_{Acc}$ on $T_4$ with HydraGAN, a generative data augmentation method, which achieves $0.9355$ with $330$ rows but fell short of data discovery methods. Increasing the number of rows further reduced performance, reflecting the limitations of generative approaches in this context, which cannot utilize verified external data sources, and synthetic data often lacks inherent reliability and contextual relevance of discovered data.

\begin{figure}[tb!]
\centerline{\includegraphics[width =0.5\textwidth]{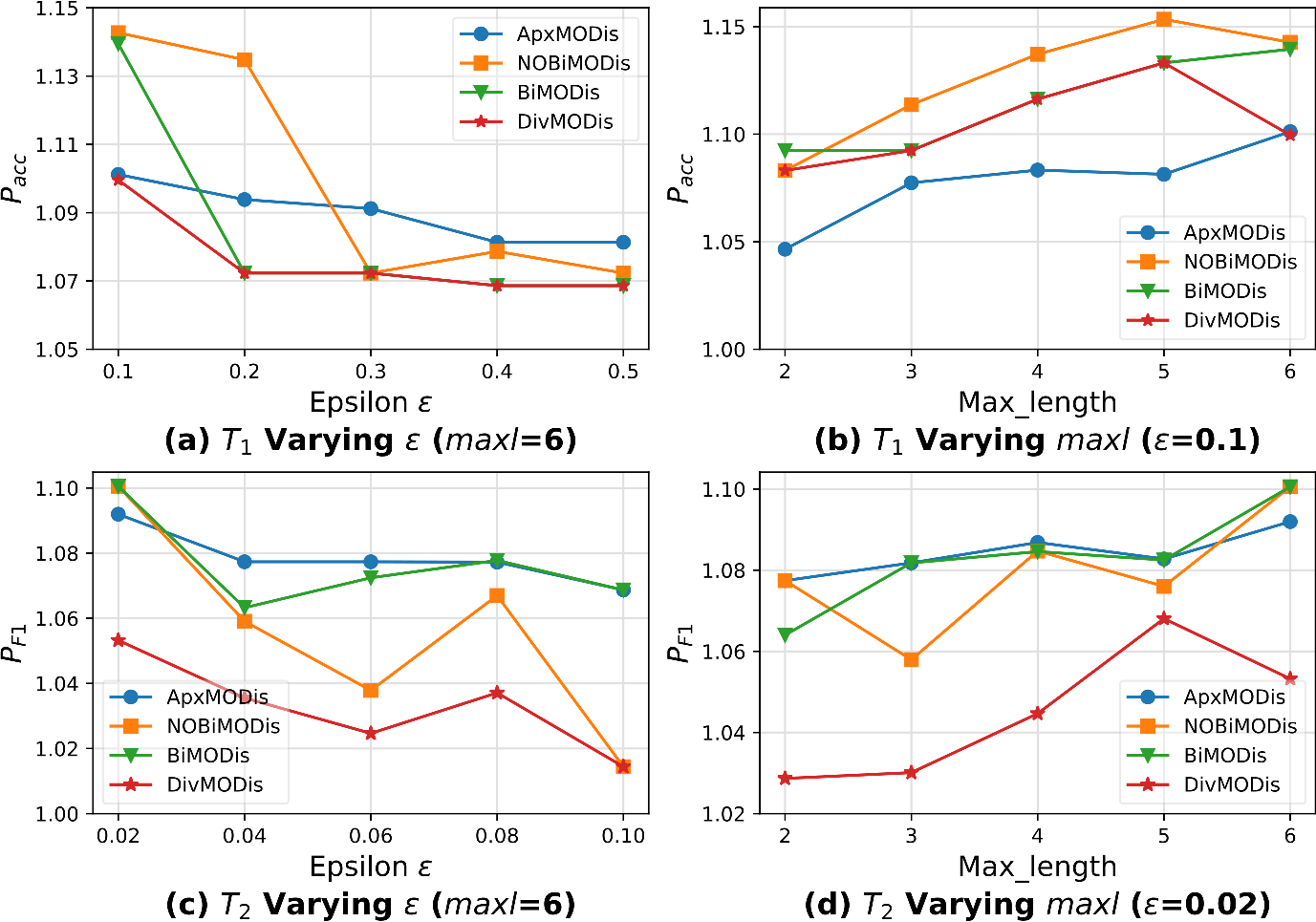}}
\centering
\vspace{-1ex}
\caption{Effectiveness: Impact of Factors}
 \vspace{-4ex}
\label{fig-effective}
\end{figure}


\stitle{Exp-2: Impact factors}. 
We next investigate the \modis methods
under the impact of two factors: $\epsilon$ and the maximum path length (\maxl), as well as the impact of $\alpha$ on \divmodis. 

\eetitle{Varying $\epsilon$}.
Fixing $\maxl$ = 6, we varied $\epsilon$ from $0.5$ to $0.1$ for $T_1$. 
As shown in Fig.~\ref{fig-effective}(a), 
\modis algorithms are able to improve the 
model in $p_{acc}$ better with smaller $\epsilon$, 
as they all ensure to output a $\epsilon$-Skyline set that 
better approximate a Skyline set when $\epsilon$ is set to be 
smaller. In all cases, they achieve a relative improvement 
$\relp(p_{Acc})$ at least 1.07. 
\bimodis and \nomodis perform better 
in recognizing better solutions from both ends in 
reduction and augmentation as smaller $\epsilon$ is enforced. \apxmodis, 
with reduction only, is less sensitive to 
the change of $\epsilon$ due to that larger $\epsilon$ 
may ``trap'' it to  local 
optimal sets from one end. Adding diversification (\divmodis) is able to strike a balance between \apxmodis and \bimodis by enforcing to choose difference datasets out of local 
optimal sets, thus 
improving \apxmodis for smaller $\epsilon$. We choose a smaller range of $\epsilon$ for $T_2$ in Fig.~\ref{fig-effective}(c), as the variance of $p_{F1}$ is small.
As $\epsilon$ varies from $0.1$ to $0.02$, 
\nomodis 
improves F1 score from $0.84$ to $0.91$.

\eetitle{Varying $\maxl$}.  Fixing $\epsilon$ = 0.1, we varied $\maxl$ from $2$ to $6$. Fig.~\ref{fig-effective}(b, d) tells us 
that all \modis algorithms improve the 
task performance
for more rounds of processing.  Specifically, \bimodis and \nomodis benefit most 
as bi-directional search allows both to find 
better solution from wider search space as $\maxl$ becomes larger. \apxmodis is less sensitive, 
as the reduction strategy from dense tables 
incurs smaller loss in accuracy.
\divmodis finds datasets that ensure 
best model accuracy when $\maxl$ = 5, yet may 
``lose chance'' to maintain the accuracy, due to 
that the diversification step may update 
the Skyline set with 
less optimal but more different counterparts in 
future levels (\eg when $\maxl$ = $6$).  



\begin{figure}[tb!]
\centerline{\includegraphics[width =0.5\textwidth]{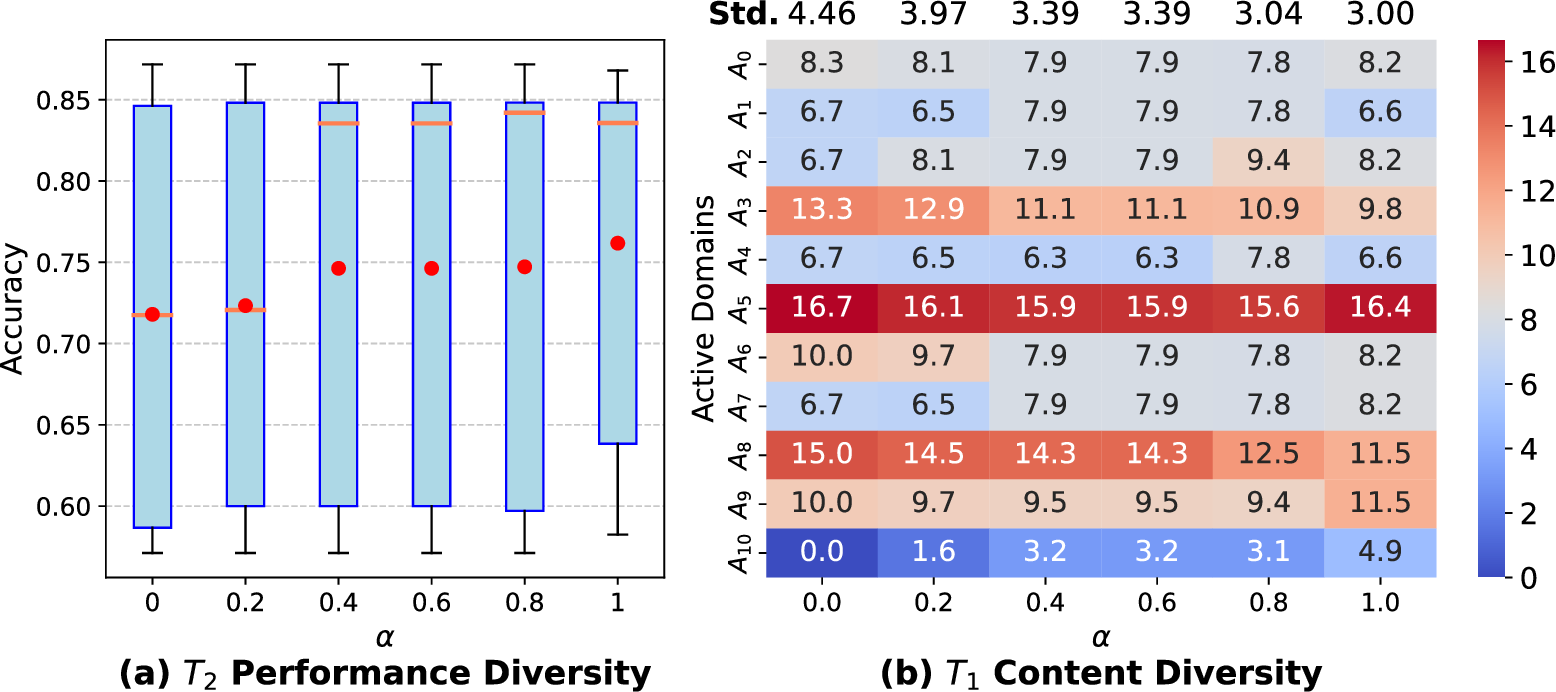}}
\centering
\vspace{-1ex}
\caption{Impact of $\alpha$ for \divmodis}
\vspace{-2ex}
\label{fig:divmodis}
\end{figure}

\eetitle{Varying $\alpha$ in \divmodis}.
We demonstrate the effectiveness of \divmodis by adjusting $\alpha$.
A smaller $\alpha$ prioritizes performance, while a larger $\alpha$ emphasizes content diversity, measured by hamming distance.
Fig.~\ref{fig:divmodis}(a)  illustrates \textit{Performance Diversity}, where smaller 
$\alpha$ results in a wider accuracy range with a balanced and stable distribution. Both the mean and median remain centered. As $\alpha$ increases, the accuracy distribution narrows and shifts toward higher values, reflecting the dominance of high-accuracy datasets in the Skyline set.
Fig.~\ref{fig:divmodis}(b) verifies the impact of \textit{Content Diversity}, visualized as the percentage contribution of each \ad.  Larger 
$\alpha$ leads to more evenly distributed contributions. 
The standard deviation values above the heatmap quantify this trend, showing a consistent decrease as $\alpha$ increases, indicating improved balance.


\eat{
\begin{figure}[tb!]
\centerline{\includegraphics[width =0.3\textwidth]{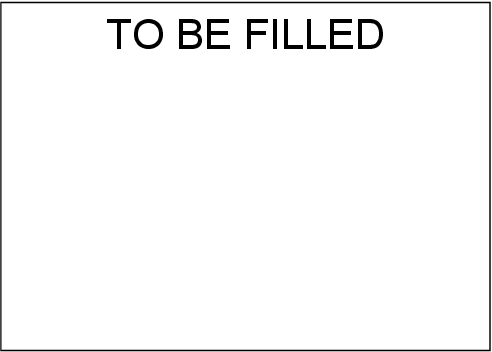}}
\centering
\caption{Impact of factors}
 \vspace{-1ex}
\label{fig:factors}
\end{figure}
}

\eat{
\mengying{As a method for Table Union Search(TUS), Starmie is effective at finding related tables in a data lake, \tbf. However, its algorithm does not consider measures from a downstream data science task. So it is in expect that \tbf. This highlights the importance of goal-driven data discovery.}
}

\begin{figure}[tb!]
\centerline{\includegraphics[width =0.5\textwidth]{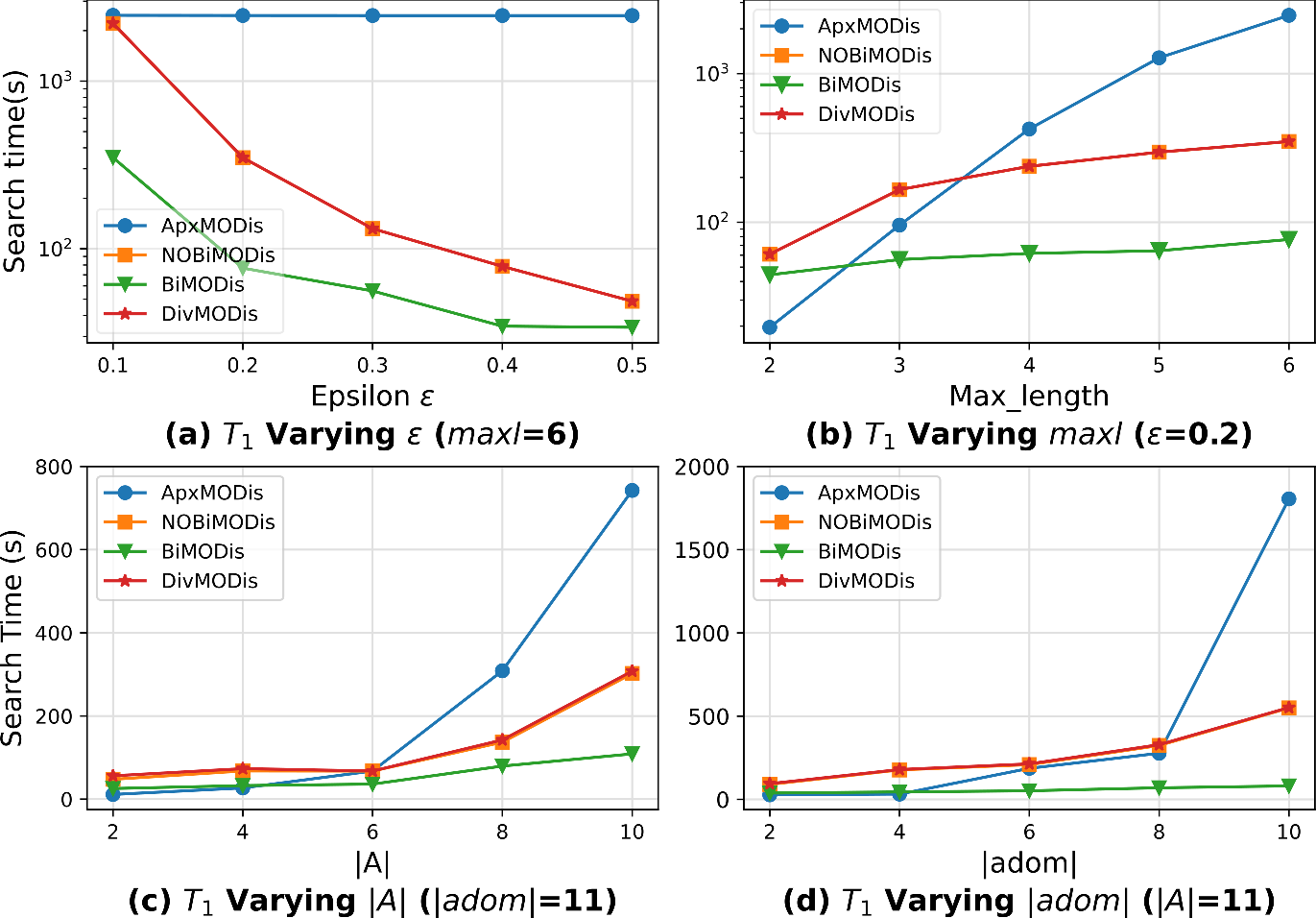}}
\centering
\vspace{-1ex}
\caption{Efficiency and Scalabilitiy}
\vspace{-5ex}
\label{fig-efficiency}
\end{figure}


\stitle{Exp-3: Efficiency and Scalibility}. 
We next report the 
the efficiency of \modis algorithms 
for task $T_1$ and $T_3$ over \kaggle and \hf, respectively, and 
the impact of factors 
$\epsilon$ and $\maxl$. 
We also evaluate their scalability for $T_1$ and $T_5$ in terms of input size.

\eetitle{Efficiency: Varying $\epsilon$}. 
Fixing $\maxl$ = $6$ and varying $\epsilon$ from $0.1$ to $0.5$, 
Fig.~\ref{fig-efficiency} (a)
verifies the following. 
(1) \bimodis, \nomodis and \divmodis take less time as $\epsilon$ increases, as a larger $\epsilon$ provides more chance to prune unnecessary valuations. 
\divmodis has a comparable performance 
with \nomodis, as it mainly benefits from 
the bi-directional strategy, which exploits early pruning and a stream-style placement strategy. 
(2) As shown in Fig.~\ref{fig-efficiency}(a), for  $T_1$, \bimodis, \nomodis, and \divmodis are 2.5, 2, and 2 times faster than \apxmodis on average, respectively. 
\apxmodis takes longer time to explore a larger  universal table with reduct operators. 
It is insensitive to $\epsilon$. We observe that its search from the ``data rich'' end   
may converge faster at high-quality 
$\epsilon$-Skyline sets.

\eat{
This is because in both cases, 
(1) there are more states with non-$\epsilon$-dominance relation 
to existing solution to be resolved; 
and (2) there are more state nodes to be valuated. 
On the other hand, \apxmodis is most sensitive to $\maxl$ due to 
rapid growth of search space, and \bimodis is much less sensitive to $\maxl$ 
as it mitigates the impact better with bi-directional strategy. 
}
\eetitle{Efficiency: Varying $\maxl$}. 
Fixing $\epsilon$ = 0.2 for task $T_1$ and $\epsilon$ = 0.1 for task $T_3$, we varied $\maxl$ from $2$ to $6$, all \modis algorithms take longer as $\maxl$ increases, as shown in Fig.~\ref{fig-efficiency} (b). 
Indeed, larger $\maxl$ results in more states to be valuated, and more non-$\epsilon$-dominance relation to be resolved. \apxmodis is sensitive to $\maxl$  due to the rapid growth of the search space. In contrast, \bimodis mitigates the impact with bi-directional strategy and effective pruning.

\eat{
\eetitle{$T_3$: Varying $\epsilon$ and $\maxl$}. 
We report the efficiency of our algorithms 
for regression task. Our observation is 
consistent with their counterparts for 
$T_1$. This verifies that the efficiency  
of our approach is not very sensitive to
the type of learning tasks or models.
}

\eat{
\begin{figure}[tb!]
\centerline{\includegraphics[width =0.5\textwidth]{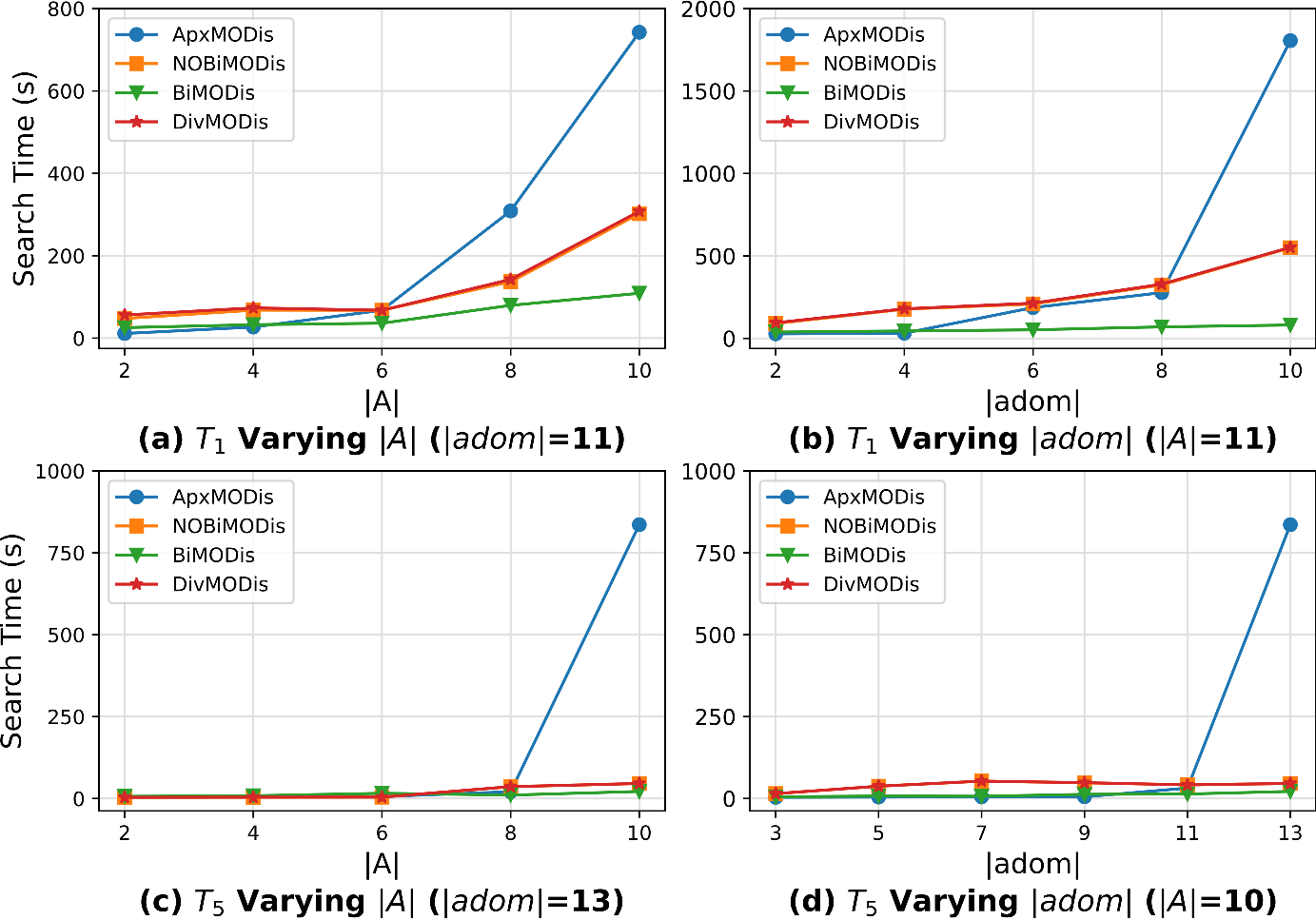}}
\centering
\vspace{-1ex}
\caption{\revise{Scalability with Input Data Size}}
\vspace{-4ex}
\label{fig-scalability}
\end{figure}
}

\eetitle{Scalability}. We varied the number of total 
attributes $|A|$ and size of the largest active domain $|\ad|$. We perform $k$-means clustering over the tuples of the universal table with $k = |\ad|$, and extended operators with range queries to control $|\ad|$. 
Fig.~\ref{fig-efficiency} (c) and (d) show that all \modis algorithms take more time for larger $|A|$ and $|\ad|$. \bimodis scales best due to 
bi-directional strategy. 
\divmodis remains more efficient 
than \apxmodis, indicating affordable 
overhead from diversification.

While our algorithms scale well with $\vert A \vert$ and $\vert \ad \vert$, high-dimensional datasets may present challenges due to the search space growth. Dimensionality reduction such as PCA or feature selection, or correlation-based pruning (to identify and eliminate highly correlated or redundant features), can be tailored to specific tasks to mitigate these challenges.



\eat{
\begin{figure}[tb!]
\centerline{\includegraphics[width =0.5\textwidth]{./fig/movie_radar_chart.png}}
\centering
\caption{Efficiency}
 \vspace{-1ex}
\label{fig:efficiency}
\end{figure}
}

\vspace{1ex}
\stitle{Exp-4: Case study}. We next report two real-world case studies to illustrate the application scenarios of \modis. 

\begin{figure}[tb!]
\centerline{\includegraphics[width=\linewidth]{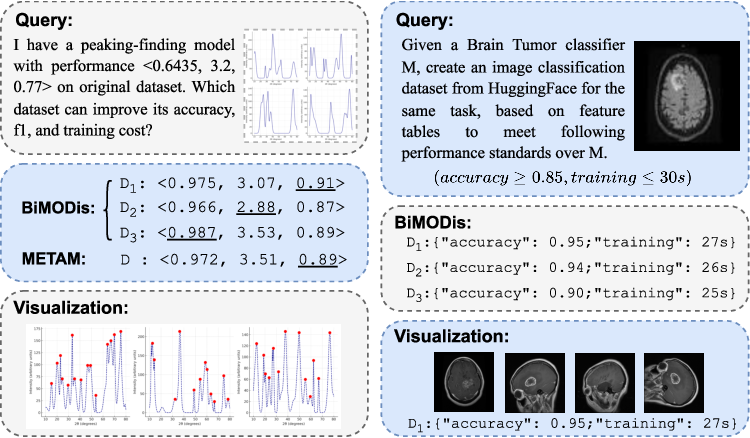}}
\centering
\caption{Case 1 (left): Discover Datasets for Materials Peak Classification Analysis. Case 2 (right): Test Data Generation for Model Performance Benchmarking}
\vspace{-2ex}
\label{fig:cases}
\end{figure}

\eetitle{(1) ``Find data with models''}. A material science team trained a random forest-based classifier to identify peaks in 2D X-ray diffraction data. They seek more datasets to improve the model's {\em accuracy, training cost, and F1 score} for downstream fine-tuning. Original X-ray datasets and models are uploaded to a crowd-sourced X-ray data platform we deployed~\cite{wang2022crux} with 
best performance of $\textless0.6435, 3.2, 0.77\textgreater$.
Within available X-ray datasets, \bimodis created three datasets $\{D_1, D_2, D_3\}$ and achieved the best  performance of 0.987, 2.88, and 0.91, respectively. We set \metam to optimize F1-score, and achieved a performance score of $\textless 0.972, 3.51, 0.89\textgreater$ over its output dataset.
Fig.~\ref{fig:cases} illustrates 
such a case that is manually validated with ground-truth from 
a third-party institution.

\eetitle{(2) Generating test data for model evaluation}. 
We configure \modis algorithms to generate test datasets for model benchmarking, where specific performance criteria can be posed~\cite{ventura2021expand}. Utilizing a trained scientific image classifier from \kaggle, and a pool of image feature datasets $\D$ from \hf with $75$ tables, $768$ columns, and over $1000$ rows. 
We request \bimodis to generate  
datasets over which the classifier demonstrates: ``accuracy $>$ 0.85'' and ``training cost $<$ 30s.'' 
\bimodis successfully generated $3$ datasets to be chosen from within $15$ seconds, with performance $\textless0.95, 0.27\textgreater$, $\textless0.94, 0.26\textgreater$ and $\textless0.90, 0.25\textgreater$, as in Fig.~\ref{fig:cases}.

\section{Conclusion} 
\label{sec:conclude}
We have introduced \modis, 
a framework that generate skyline datasets to improve data science 
models on multiple performance measures. 
We have formalized skyline data generation 
with transducers 
equipped with augment and reduction operators. 
We show the hardness and fixed-parameter 
tractability of the problem. 
We have introduced three algorithms 
that compute approximate Skyline sets 
in terms of $\epsilon$-Skyline set, 
with reduce-from-universal, bi-directional, 
and diversification paradigms. 
Our experiments have verified 
their effectiveness and efficiency. 
A future topic is to 
enhance \modis with query optimization 
techniques to scale it for larger 
input with high-dimensional data.
Another topic is to extend \modis 
for distributed Skyline 
data generation.

\begin{acks} 
This work is supported by NSF under OAC-2104007.
\end{acks}
\bibliographystyle{ACM-Reference-Format}
\bibliography{main}

\newpage
\appendix
\titleformat{\section}[block]
{\normalfont\Large\bfseries}{Appendix \Alph{section}:}{0.5em}{}
\label{sec-appendix}

\vspace{2ex}

\section{Algorithms and Proof}

\subsection{\apxmodis}

\stitle{Proof of Lemma~\ref{lm-approximability}}
{\em For any constant $\epsilon$, \apxmodis correctly computes an $\epsilon$-Skyline set $\Pi$ that approximates a Skyline set defined on the $N$ states it valuated. }

\begin{proof}
We establish the $\epsilon$-approximability of \apxmodis by constructing a reduction from \modis to the multi-objective 
shortest path problem (\mos)~\cite{tsaggouris2009multiobjective}. 

\eetitle{Reduction}.
An instance of \mos consists of an edge-weighted graph $G_w$, where each edge $e_w$ is assigned a $d$-dimensional attribute vector $e_w.c$. The cost of a path $\rho_w$ in $G_w$ is defined as $\rho_w.c$ = $\sum_{e_w\in\rho_w}$ $e_w.c$. 
The dominance relation between two paths $\rho_w$ and $\rho'_w$ is determined by comparing their costs. 
Specifically, $\rho_w$ dominates $\rho'_w$ if $\rho_w$ has equal or lower costs than $\rho'_w$ in all dimensions and is strictly better in at least one dimension. 
The objective is to compute a Skyline set of paths from a start node $u$ to all other nodes in the graph. 

We construct the reduction from our problem to \mos. 
(1) We define $G_w$ as an edge weighted counterpart 
of a running graph $G_\T$. (a) Each vertex in $G_\T$ represents a unique state $s$ during the execution of \apxmodis, with each state corresponding to a specific dataset configuration in the data discovery process. The graph $G_\T$ contains $N$ vertices, corresponding to the $N$ states that \apxmodis has spawned and valuated. 
(b) Each edge $(s, s')$ in $G_\T$ represents a transition from state $s$ to state $s'$, resulting from applying an operation (e.g., reduction or augmentation) that modifies the dataset. The edge is weighted by the difference in performance measures in $\P$ between the two states: $e_w$ = $s.\P - s'.\P$. 
Here, $s.\P$ and $s'.\P$ are the performance vectors of the states $s$ and $s'$, respectively. The edge weight $e_w$ is a $d$-dimensional vector that quantifies how the performance metrics change as a result of the transition. 
A path $\rho \in G_\T$ corresponds to a sequence of transitions between states, starting from the initial state $s_U$. 
Similar to $\rho_w \in G_w$, the cumulative cost of this path $\rho.c$ is defined as the sum of the edge weights along the path, which represents the cumulative change in the performance measures $\P$ as the dataset evolves through different states. 

Given a solution $\Pi_w$ of an instance of \mos, which is an $\epsilon$-Skyline set of paths, we construct a solution for a corresponding 
instance of \modis. 
For each path $\rho_w \in \Pi_w$, we establish a corresponding path $\rho$ in $G_\T$ and identify the final state $s$ that the path reaches.
The final state $s$ corresponds to a specific dataset $D$, which is the result of applying the sequence of operations from $\rho$. 
We then include $D$ in the set $\D_F$. 
This forms a set of datasets as the solution to \modis. 

We next prove 
that 
$\Pi_w$ is an $\epsilon$-Skyline set 
of paths $\Pi_w$ in $G_w$. 
if and only 
if $\D_F$ is an $\epsilon$-Skyline set 
of $\D_S$. 

\etitle{If condition}. 
Let $\D_F$ be an $\epsilon$-Skyline set of $\D_S$.  
By the definition given in sec~\ref{sec-problem}, this means that for every dataset $D' \in (\D_S\setminus\D_F)$, there exists at least one dataest $D \in \D_F$ such that $D$ $\epsilon$-dominates $\D'$. 
Specifically, this means that $D$ has costs that are at most $(1+\epsilon)$ times the costs of $D'$ in all performance measures in $\P$, and $D$ has a strictly lower cost in at least one measure.
From the reduction, each path $\rho_w \in \Pi_w$ corresponds to a sequence of transitions in $G_\T$ leading to a final state $s$, which represents a dataset $D \in \D_F$. Similarly, $\rho'_w \notin \Pi_w$ corresponds to a dataset $D' \in (\D_S\setminus\D_F)$. 
Since $D$ $\epsilon$-dominates $D'$, the corresponding path 
$\rho_w$ $\epsilon$-dominates $\rho'_w$. 
This dominance is preserved because the performance measures $\P$, directly corresponding to the edge weights in $G_w$. 
Therefore, $\Pi_w$ is an $\epsilon$-Skyline set 
of paths $\Pi_w$ in $G_w$ if $\D_F$ is an $\epsilon$-Skyline set 
of $\D_S$.

\etitle{Only If condition}. Conversely, we assume the following:
(1) $\Pi_w$ is an $\epsilon$-Skyline set of paths $\Pi_w$ in $G_w$, but 
(2) the induced $\D_F$ is not an $\epsilon$-Skyline set of $\D_S$. 
Assumption (2) implies one of the following two cases:
\bi
\item \textbf{Case 1}:
There exists a dataset $D' \in (\D_S\setminus\D_F)$ that is not $\epsilon$-dominated by any dataset in $\D_F$. This means there is a corresponding path $\rho'_w$ in $G_w$ that is not $\epsilon$-dominated by any path in $\Pi_w$. This contradicts assumption (1)  because $\rho'_w$ should be $\epsilon$-dominated by at least one path in $\Pi_w$.
\item \textbf{Case 2}: There exists a dataset $D \in \D_F$ that is $\epsilon$-dominated by a dataset $D'' \in (\D_S\setminus\D_F)$. 
This means there exists a path $\rho''_w$ corresponding to $D''$ in $G_w$ that $\epsilon$-dominates the path $\rho_w$ corresponding to $D$ in $\Pi_w$. However, this would imply that $\Pi_w$ does not fully capture the $\epsilon$-Skyline set because $\rho_w$ should not be in $\Pi_w$ if it is $\epsilon$-dominated by $\rho''_w$. Thus, it contradicts assumption (1).
\ei
Both cases lead to a contradiction with the assumption (1) that $\Pi_w$ is an $\epsilon$-Skyline set of paths $\Pi_w$ in $G_w$. Therefore, the initial assumption (2) must be false, meaning $\Pi_w$ is an $\epsilon$-Skyline set 
of paths $\Pi_w$ in $G_w$, only 
if $\D_F$ is an $\epsilon$-Skyline set 
of $\D_S$. 

By proving both directions, we establish the equivalence that $\Pi_w$ is an $\epsilon$-Skyline set 
of paths $\Pi_w$ in $G_w$, if and only 
if $\D_F$ is an $\epsilon$-Skyline set 
of $\D_S$.

\eetitle{Correctness}. 
We then show that 
algorithm~\apxmodis 
is an optimized process of 
the algorithm in~\cite{tsaggouris2009multiobjective}, 
which correctly computes 
$\Pi_w$ for $G_w$.
Specifically, this means that in $G_w$, for any path $\rho'_w \notin \Pi_w$ with a corresponding state $s'$, there exists a path $\rho_w \in \Pi_w$ with a corresponding state $s$, such that for every performance measure $p_i$ (where $1\leq i \leq d$, and $d=|\P|$), the condition $s.\P(p_i) \leq (1+\epsilon)s'.\P(p_i)$ holds.

We prove the correctness of this result by induction.

\sstab
(1) \textbf{Base case}.
After the first iteration in the main procedure of \apxmodis, and due to the ``merge'' steps in the \upi procedure, the position $pos(s')$ in $\Pi^1_w$ will be occupied by a path $\rho_w$, for which: (i) $pos(s) = pos(s')$; and (ii) $s.\P(p_d) \leq s'.\P(p_d)$. From (i) and based on the Equation~(\ref{eq:pos}), for $1\leq i \leq d-1$, we have $\left\lfloor\log _{1+\epsilon} \frac{s.\P(p_i)}{p_{l_i}}\right\rfloor = \left\lfloor\log _{1+\epsilon} \frac{s'.\P(p_i)}{p_{l_i}}\right\rfloor$. This implies $\log _{1+\epsilon} \frac{s.\P(p_i)}{p_{l_i}} - 1 \leq \log _{1+\epsilon} \frac{s'.\P(p_i)}{p_{l_i}}$, so that $s.\P(p_i) \leq (1+\epsilon)s'.\P(p_i)$ for $1\leq i < d$. Combined with (ii), we conclude that $s.\P(p_i) \leq (1+\epsilon)s'.\P(p_i)$ holds for $1\leq i \leq d$.

\sstab
(2) \textbf{Induction}. 
Assume that after $i-1$ iterations, \apxmodis correctly computes the $\epsilon$-Skyline set $\Pi^{i-1}_w$ for all paths from the source node $s_U$ that contain up to $i-1$ edges. This means that for every path $\rho'_w \notin \Pi^{i-1}_w$ with at most $i-1$ edges, there exists a path $\rho_w \in \Pi^{i-1}_w$ such that the corresponding states $s$ and $s'$ satisfy: 
$$s.\P(p_i) \leq (1+\epsilon)s'.\P(p_i), \forall 1\leq i \leq d$$

We next prove that after $i$ iterations,  \apxmodis correctly computes the $\epsilon$-Skyline set $\Pi^i_w$ for all paths from the source node $s_U$ that contain up to $i$ edges.

By induction, every path in $\Pi^{i-1}_w$ $\epsilon$-dominates any other paths of up to $i-1$ edges not included in $\Pi^{i-1}_w$, so we only need to ensure the correctness of the $i$th iteration. 
In this iteration, paths are expanded to include $i$ edges. As seen in the base case, after the ``merge'' step in procedure \upi, \apxmodis ensures that for any state $s'$ corresponding to a path not included in $\Pi^i_w$, there exists at least one state $s$ with corresponding path in $\Pi^i_w$, such that:
$$s.\P(p_i) \leq (1+\epsilon)s'.\P(p_i), \forall 1\leq i \leq d$$
Thus, after $i$ iterations, $\Pi^i_w$ covers all paths with up to $i$ edges that should be included in the $\epsilon$-Skyline set.

Putting these together, we show that \apxmodis correctly computes the $\epsilon$-Skyline set $\Pi^i_w$ for all $i > 0$. This verifies the correctness of \apxmodis.
\end{proof}

\stitle{Proof of Lemma~\ref{cor-fptas}}. 
{\em Given $\T$ with configuration $C$, 
if $|\D_S|$ has a size in $O(f(|D_U|))$, 
where $f$ is a polynomial, 
then \apxmodis is an FPTAS for \modis. }

\begin{proof} We consider the reduction 
of an instance of \modis to its counterpart 
of \mos as detailed in the proof of Lemma~\ref{lm-approximability}.
\mos is known to be solvable by an FPTAS. That is, 
there is an algorithm that can compute an $\epsilon$-Skyline set in polynomial time relative to the size of the input graph 
and $\frac{1}{\epsilon}$~\cite{tsaggouris2009multiobjective}. 

We configure \apxmodis to run in  
a $(|\D_S|, \epsilon)$-approximation, 
which is a simplified implementation of an  
FPTAS in~\cite{tsaggouris2009multiobjective} with multiple 
rounds of ``replacement'' 
strategy following path dominance. 
In the proof of Lemma~\ref{lm-approximability}, we have already shown that \apxmodis correctly computes the $\epsilon$-Skyline set for $G_w$,  which is equivalent to the $\epsilon$-Skyline set for $\D_S$ in \modis.
Meanwhile, as $|\D_S|$ is bounded by a polynomial of  
the input size $|D_U|$, 
the time complexity of \apxmodis is $O\left(f(|D_U|)\cdot \left(\left(\frac{\log(p_m)}{\epsilon}\right)^{|\P|-1}+I\right)\right)$, where 
$f$ is a polynomial. This ensures that \apxmodis approximates the Skyline set for all datasets within PTIME.
\end{proof}

\stitle{Space cost}. 
We also report the space cost. 
(1) It takes a vector of length in $O(|P|-1)$ to encode the 
position $pos(\rho)$. The replacement strategy 
in \apxmodis keeps one copy of position per path at runtime and 
``recycles'' the space once it is verified to be dominated. 
According to Equation~\ref{eq:pos}, there 
are at most $\prod_{i=1}^{|\P|-1}\left(\left\lfloor\log _{1+\epsilon}\frac{p^{max}_i}{p^{min}_i}\right\rfloor+1\right)$ paths to be remained 
in a $(|\P|-1)$-dimensional array 
at runtime, 
hence the total space cost is in 
$O\left(\prod_{i=1}^{|\P|-1}\left(\left\lfloor\log _{1+\epsilon}\frac{p^{max}_i}{p^{min}_i}\right\rfloor+1\right)\right)$.

\eat{
For each valuedated path $\rho$, we use $(|\P|-1)$ measures in $|\P|$ (except the deterministic measure $p*$) to calculate its ``position'' $pos(\rho)$ in a ($|\P|-1$)-ary space, which we use as the Skyline set.
Here, the last measure in $\P$ is set as $p*$ by default.
When multiple paths are allocated to the same position in the Skyline set, only the one with the lowest $p^*$ is retained.
This ensures that each position in the Skyline set holds at most one path,
so the {\em Space Complexity} equals the size of the array for the Skyline set. Assuming all performance metrics are costs, according to Equation~\ref{eq:pos}, the Space Complexity is $O\left(\prod_{i=1}^{|\P|-1}\left(\left\lfloor\log _{1+\epsilon}\frac{p^{max}_i}{p^{min}_i}\right\rfloor+1\right)\right)$.
}

\subsection{\bimodis}

\stitle{Correlation based Pruning}. 
We present the details of Correlation-based Pruning. 
We first introduce a monotonicity property as the 
condition for the applicability of the pruning. 

\eetitle{Monotonicity property}. 
Given the {\em current} 
historical performances over valuated
states, we say a state $s$ (resp. $s'$) with  
a performance measure $p$ at a path $\rho$ 
has a {\em monotonicity property}, if for 
any state $s''$ reachable from $s$ (resp. can reach $s'$) via $\rho$, 
$s.\hat{p_u}\textless \frac{s''.\hat{p_l}}{1+\epsilon}$ (resp. $s'.\hat{p_u}\textless \frac{s''.\hat{p_l}}{1+\epsilon}$). 

\eat{
The pruning strategy is not applicable 
if the above condition no longer holds 
due to updates of new records. The overhead of 
checking whether the property holds for 
measurement pairs, upon the arrival 
of new tests are small. 
}

\eetitle{Pruning rule}. 
We next specify Correlation-based pruning with 
a {\em pruning rule} as follows. 
First, recall that \bimodis dynamically maintains, 
for each performance $p\in \P$ and each 
state $s$, an estimated range 
$[\hat{p_l}, \hat{p_u}]\subseteq [p_l, p_u]$. 
The bounds $\hat{p_l}$ (resp. $\hat{p_u}$ 
are updated with runtime performance estimation 
of $s$ upon the changes of correlated 
performance measures. 

Specifically, for any  
state $s'$ on a path $\rho$  
obtained by augmented features 
of its ``ancestor'' state $s$ on $\rho$, 
where $s$ has a performance $p$ ``learning cost'' 
$s.p$ with lower bound $s.\hat{p_l}$ = $0.4$, 
and an ``accuracy'' with estimated 
upperbound $s.\hat{p'_u}$ = $0.8$, 
then (1) $s'$ has an estimated running cost 
initialized as $s.\hat{p_l}$ = $0.4$, indicating 
a learning cost no smaller than the counterpart $s$ 
with smaller dataset; 
and (2) $s'$ has an estimated accuracy 
with an upperbound $s.\hat{p'_u}$ = $0.8$, 
as $p$ and $p'$ are statistically 
negatively correlated with a rule specified 
as: for every current valuated $s$,  
$p$ as ``learning cost'', and $p'$ as ``accuracy'', 
if $p$ is larger, then $p'$ is smaller. 
The algorithm \bimodis dynamically 
maintains a bounds list for 
all created states $s$ in the 
bidirectional search. 

Given two states $s$ and $s'$, 
where $s' \succapprox_{\epsilon} s$, 
a state $s''$ on a path $\rho$ from $s$ or to $s'$ 
{\em can be pruned according to Correlation-Based Pruning} if 
for every $p\in \P$, $s''$ has $p$ 
at $\rho$ with a monotonicity property 
\wrt $s$ (resp. $s'$). 

Note that the above rule is checkable in PTIME 
in terms of input size $|\D_S|$. When $|\D_S|$ 
is large, one can generate a path with 
all states unevaluated, 
and check at runtime if the condition holds between 
two evaluated states and any unevaluated 
state in betwen in PTIME, to 
prune the unevaluated states. 

We are now ready to show Lemma~\ref{lm-prune}. 

\stitle{Proof of Lemma~\ref{lm-prune}}
{\em Let $s \in Q_f$ and $s' \in Q_b$. 
If $s' \succapprox_{\epsilon} s$, 
then any state node $s''$ 
on a path from $s$ or to $s'$, 
that can be pruned according to Correlation-Based Pruning, 
$D_{s''}$ is not in $\epsilon$-Skyline sets 
of the datasets from valuated states.}

\eat{
As the measures are normalized as ``costs'' (to be minimized), $\hat{p_l}$ (resp. $\hat{p_u}$) 
refer to an upper bound (resp. lower bound) estimation, 
both within the user specified range $[p_l, p_u]$. 
This property ensures that the estimated ranges for unvaluated measures accurately reflect each state's best and worst possible outcomes. 
}


We next perform a case study of 
$s$ and $s'$ as follows, subject to 
the monotonicity property.  

\stitle{Case 1: Both $s'.\P(p)$ and $s.\P(p)$ are valuated.}
If $s' \succapprox_{\epsilon} s$, then by definition, $s'.\P(p)\leq (1+\epsilon) s.\P(p)$ for all $p \in \P$. This 
readily leads to $\epsilon$-dominance, \ie $s'\succeq_\epsilon s$. As $s''$ has every performance measures 
$p\in \P$ with a monotonicity property \wrt $s$, 
$s\succeq_\epsilon s''$. Hence $s''$ can be safely pruned 
without valuation. 

\sstab
\textbf{Case 2: Neither $s'.\P(p)$ nor $s.\P(p)$ is valuated.}
By definition, as $s' \succapprox_{\epsilon} s$, 
then for every $p\in \P$, $s'.\hat{p_u}\leq (1+\epsilon) s.\hat{p_l}$. 
Given that $s''$ has every performance measures 
$p\in \P$ with a monotonicity property \wrt $s$, 
then by definition, for each $p\in \P$, we have 
$s'.p\leq s'.\hat{p_u}\leq (1+\epsilon) s.\hat{p_l}\leq (1+\epsilon)s.\hat{p_u}\textless (1+\epsilon)\frac{s''.\hat{p_l}}{1+\epsilon} \leq s''.p$, for 
every $p\in \P$. By definition of state dominance, $s' \succ s''$, for unevaluated $s''$.  
Following a similar proof, 
one can infer that $s \succ s''$ 
for a state $s$ in the forward front 
of \bimodis. 
Hence $s''$ can be safely pruned.

\sstab
\textbf{Case 3: One of $s'.\P(p)$ or $s.\P(p)$ is valuated.}
Given that $s' \succapprox_{\epsilon} s$, we have  
\bi
\item (a) $s'.\P(p) \leq (1+\epsilon) s.\hat{p_l}$, if only $s'.\P(p)$ is valuated; or 
\item (b) $s'.\hat{p_u} \leq (1+\epsilon) s.\P(p)$, if only $s.\P(p)$ is valuated.
\ei
Consider case 3(a). 
As $s$ can reach $s''$ via a path $\rho$, and 
$s''$ satisfiies the pruning condition, 
we can infer that 
$s'.\P(p) \leq (1+\epsilon) s.\hat{p_l} \leq 
(1+\epsilon) s.\hat{p_u} \textless (1+\epsilon) 
\frac{s''.\hat{p_l}}{1+\epsilon}\leq s''.p$, 
hence $s'\succ s''$. 

Similarly for case 3(b), we can infer that   
$s'.\hat{p_u} \leq (1+\epsilon) s.p \leq 
(1+\epsilon) s.\hat{p_u} \textless (1+\epsilon) 
\frac{s''.\hat{p_l}}{1+\epsilon}\leq s''.p$. 
hence $s'\succ s''$.  
For both cases, $s''$ can be pruned 
without evaluation. 

Lemma~\ref{lm-prune} hence follows. 


We present the details of the 
algorithm \bimodis in Fig.~\ref{alg:com-bimodis}. 

\begin{figure}
\centering
\begin{algorithm}[H]
\caption{\bimodis
}
\label{alg:bimodis:complete}
\begin{algorithmic}[1]
\algtext*{EndFor}
\algtext*{EndIf}
\algtext*{EndWhile}
\algtext*{EndFunction}
\algtext*{EndProcedure}

\State \textbf{Input:} 
    Configuration $C$ = $(s_U, \O, M, T, \E)$, 
    Records $Rec$,
    a constant $\epsilon>0$;
\State \textbf{Output:} 
     $\epsilon$-Skyline set $\mathcal{D}_F$.
     \vspace{1ex}

\State \textbf{Set} $\D_F := \varnothing$, 
    \swb $:= \varnothing$,
    PrunS $:= \varnothing$;
    $s_b = \text{BackSt}(s_E, s_U)$; \label{a2:inis}
\State \textbf{queue} $Q_f := \{(s_U, 0)\}$,  
    \textbf{queue} $Q_b := \{(s_b, 0)\}$;
\State $\D_F^0[pos(s_U)] = \text{CorrFP}(s_U, Rec, \E)$;
\State $\D_F^0[pos(s_b)] = \text{CorrFP}(s_b, Rec, \E)$; \label{a2:inie}

\While{$Q_f \neq \varnothing$, $Q_b \neq \varnothing$ \textbf{and} $Q_f \cap Q_b = \varnothing$}
\State $(s', d)= Q_f$.dequeue(), ${\D_F}^{d+1} = {\D_F}^d$; \Comment{Forward Serach} \label{a2:fss}
\For{all $s \in$ \opg($s'$, `F')}
    \State $\P_s = \text{CorrFP}(s, Rec, \E)$; \textbf{set} $\rho_s$  \textbf{with} $\P_s$; \label{a2:ss}
    \If{$pos(s) \in$ PrunS} \textbf{continue;} \EndIf \label{a2:pos_prune}
        \State pruned = False
        \For{bound \textbf{in} SandwBs}
        \State pruned = SandwPrun($\rho_s$, bound, SandwBs)
    \If{pruned} \textbf{break;} \EndIf
    \EndFor
    \State pruned = \upi($\D_F^{d+1}$, PrunS, pos(s), $\epsilon$) \label{a2:se}
    \If{not pruned} $Q_f$.enqueue((s, d+1)) \EndIf \label{a2:enqueue}
\EndFor \label{a2:fse}
\State $(s', d)= Q_b$.dequeue(), ${\D_F}^{d+1} = {\D_F}^d$; \Comment{Backward Serach} \label{a2:bss}
\For{all $s \in$ \opg($s'$, `B')}
\State same with line~\ref{a2:ss} to \ref{a2:se} in Forward Search
\eat{
    \State $\P_s = \text{CorrFP}(s, Rec, \E)$; \textbf{set} $\rho_s$ \textbf{with} $\P_s$;
    \If{$\rho_s \in$ PrunS} \textbf{continue;} \EndIf
        \State pruned = False
        \For{bound \textbf{in} \swb}
        \State pruned = \swp($\rho_s$, bound, SandwBs)
    \EndFor
    \If{pruned} \textbf{break;} \EndIf
    \State pruned = UPareto($\D_F^{d+1}$, PrunS, pos[$\rho_s$], $\epsilon$)}
    \If{not pruned} $Q_b$.enqueue((s, d+1)) \EndIf
\EndFor \label{a2:bse}
\EndWhile

\vspace{1ex}
\Procedure{CorrFP}{$s$, $Rec$, $\E$} \label{a2:corrs}
\State Build $G_c$ for measures recorded in $Rec$;
\State StrongRs = GetSR($G_c$)
    \If{s \textbf{in} $Rec$.keys()} \Comment{Case 1: By $Rec$}
        \State $\P_s=Rec[s]$;
        \If{$|valid(\P_s)| \geq 0.8|\P_s|$} 
            \Return $\P_s$;
        \EndIf
    \EndIf

    \For{all missing $p_i^s$ in $\P_s$}  \Comment{Case 2: By $G_{c}$}
        \If{$(p_i, p_j) \in$ StrongRs \textbf{and} $p_j^s \in Rec[s]$} 
        \State \textbf{find} closed $p_j^l$ and $p_j^u$ with $p_j^s$ \textbf{in} $Rec$;
        \State $p_i^s = (p_i^l + p_i^u)/2$
        \EndIf
    \EndFor

    \If{$|valid(\P_s)| < 0.8|\P_s|$}  \Comment{Case 3: By $\E$}
    \State \textbf{fill} missing $p_s \in \P_s$ by invoking $\E$; \textbf{update} $Rec[s] = \P_s$
    \EndIf
\State \Return $\P_s$; 
\EndProcedure \label{a2:corre}
\end{algorithmic}
\end{algorithm}
\vspace{-3ex}
\caption{Complete Version of \bimodis}
\vspace{-3ex}
\label{alg:com-bimodis}
\end{figure}

\subsection{\divmodis}

\stitle{Proof of Lemma~\ref{lemma:div}}
{\em Given $N$ and $\epsilon$, 
\divmodis achieves a $\frac{1}{4}$ approximation for 
diversified \modis, \ie 
(1) it correctly computes a $\epsilon$-Skyline set $D^P_F$
over $N$ valuated datasets, and 
(2) $\kw{div}(D^P_F)\geq \frac{1}{4}\kw{\kw{div}(\D^*_F)}$.}

We here present a detailed analysis
for the lemma~\ref{lemma:div}.

\stitle{Monotone submodularity}.  
We first show that 
the diversification function $\kw{div}(\cdot)$
is a monotone submodular function.
Given a set of datasets $\D_F$, 
we show that for any 
set of datasets 
$Y \subseteq X \subseteq \D_F$,

\tbi
\item $\kw{div}(Y) \leq \kw{div}(X)$; and  
\item $\forall x \in \D_F \setminus X$,
$\kw{div}(X \cup \{x\})-\kw{div}(X) \leq \kw{div}(Y \cup \{x\})-\kw{div}(Y)$.
\ei

\sstab
(1) To see $\kw{div}(Y) \leq \kw{div}(X)$, we have 

\begin{small}
\begin{equation*}
\kw{div}(X) - \kw{div}(Y)
= \sum_{i=1}^{k-1}
\sum_{j=i+1}^{k} \kw{dis}(D_i^X, D_j^X) - \sum_{i=1}^{k-1} \sum_{j=i+1}^{k} \kw{dis}(D_i^Y, D_j^Y) 
\end{equation*}
\end{small}
Given $Y \subseteq X$, 
we have 
\\
\begin{small}
\begin{equation*}
\kw{div}(X) -\kw{div}(Y) = \sum_{D\in X\setminus Y, D\in Y}\kw{dis}(D,D')\geq 0;  
\end{equation*}
\end{small}

\eat{
Given that for any pair of datasets,
\[\kw{dis}(D_i, D_j) = 
\alpha\frac{1-\kw{cos}(s_i.L, s_j.L)}{2}  + 
(1-\alpha)\frac{\kw{euc}(t_i.\P, t_j.\P)}{\kw{euc_{m}}}\]
Since score function $\kw{dis(\cdot)}$ returns a
non-negative value, it ensures $\sum\kw{dis}(D_l,D_m) \geq 0$. In this case, the above analysis proves that $\kw{div}(\cdot)$ is monotone.
}
(2) We next show the submodularity of the function $\kw{div}(\cdot)$. 
To simplify the presentation, 
we introduce a notation {\em marginal gain}. 
For any $x \in \D_F \setminus X$ and $Y \subseteq X$, the marginal gain
of diversification score for  $X \cup \{x\}$ and  $Y \cup \{x\}$, denoted as $\margin(X,x)$ and $\margin(Y,x)$, 
are defined as: 


\eat{
By definition, given any pair of datasets,
the diversification score
is computed by a linear combination 
of accumulated euclidian distance from the historical
performance and pairwise cosine similarities. 
Hence, the marginal gain can be calculated as
the gain of euclidian distance from the historical
performance (denoted as $\Delta\kw{relevance}$) plus the gain of the diversity (denoted as $\Delta\kw{diversity}$) of datasets,respectively.
In this case, we can have
}

\begin{small}
\begin{equation*}
\begin{aligned}
&\margin(X,x) = \\
&\sum_{i=1}^{k-1} \sum_{j=i+1}^{k} \alpha\frac{1-\kw{cos}(s_i.L, s_j.L)}{2} - \sum_{i=1}^{k-1} \sum_{j=i+1}^{k} \alpha\frac{1-\kw{cos}(s_l.L, s_m.L)}{2}\\
&+ \\
&\sum_{i=1}^{k-1} \sum_{j=i+1}^{k}(1-\alpha)\frac{\kw{euc}(t_i.\P, t_j.\P)}{\kw{euc_{m}}} - \sum_{i=1}^{k-1} \sum_{j=i+1}^{k}(1-\alpha)\frac{\kw{euc}(t_l.\P, t_m.\P)}{\kw{euc_{m}}}\\
\end{aligned}
\end{equation*}
\end{small}
where $D_i, D_j \in {X \cup \{x\}}$ and $D_l, D_m \in X$.

\begin{small}
\begin{equation*}
\begin{aligned}
&\margin(Y,x) = \\
&\sum_{i=1}^{k-1} \sum_{j=i+1}^{k} \alpha\frac{1-\kw{cos}(s_i.L, s_j.L)}{2} - \sum_{i=1}^{k-1} \sum_{j=i+1}^{k} \alpha\frac{1-\kw{cos}(s_l.L, s_m.L)}{2}\\
&+ \\
&\sum_{i=1}^{k-1} \sum_{j=i+1}^{k}(1-\alpha)\frac{\kw{euc}(t_i.\P, t_j.\P)}{\kw{euc_{m}}} - \sum_{i=1}^{k-1} \sum_{j=i+1}^{k}(1-\alpha)\frac{\kw{euc}(t_l.\P, t_m.\P)}{\kw{euc_{m}}}\\
\end{aligned}
\end{equation*}
\end{small}
where $D_i, D_j \in Y \cup x$ and $D_l, D_m \in Y$.

\eat{
\begin{mdframed}
\textbf{Condition:}
We assume that
the size of the $\epsilon$-Skyline sets are at least $k$, $|\kw{div}(D^P_F)| >= k$. 
For any pair of $\epsilon$-Skyline sets $X, Y$ such that $Y \subseteq X$ and $|Y| \geq k$, the  function $\kw{div}(\cdot)$
computes the diversification scores over
the same subset of datasets with size $k$.
\end{mdframed}
}

In our problem,we only consider $\epsilon$-Skyline sets with size 
at most $k$. With this condition, we observe that
given the dataset $x$, $\margin(Y,x)$ and 
$\margin(X,x)$ measure the marginal gain of 
diversification scores by replacing
a dataset $x' \in Y$ with $x$ and
$x' \in X$ with $x$. $\margin(Y,x)$ and 
$\margin(X,x)$ measure the margin gain by replacing $x$ with a same dataset $x'$ 
over a same $\epsilon$-Skyline set with size $k$.
In this case, we can have
\begin{small}
\begin{equation*}
\begin{aligned}
\margin(X,x)  = \margin(Y,x)
\end{aligned}
\end{equation*}
\end{small}
due to \divmodis replace the
same dataset $x'$ that are in $X$ and $Y$. Thus, we can see that
the marginal gain of $X$ is no larger than marginal gain of $Y$
by including $x$.
This analysis completes the proof of diversification function $\kw{div}(\cdot)$ is a monotone submodular function.


\eat{
\warn{
To compute $\kw{\kw{div}(\D^*_F)}$,
$\divmodis$ iterativly adds the dataset
$d^*$ that maximizes the diversification score: 
WRONG PROOF. Show the function is submodular - 
it has nothing to do with how an algorithm optimize it!}
\begin{equation*}
d^* = \argmax_{d \in D \setminus D_F} \kw{div}(D_F \cup d^*) - \kw{div}(D_F)
\end{equation*} 
since $Y \subseteq X$, $Y$ selects 
the $d$ to the $D_F$ at the earlier iteration
than $X$. Therefore, for any $ x \in \D_F
\setminus X$,
$\kw{div}(X \cup x)-\kw{div}(X) \leq
\kw{div}(Y \cup x)-\kw{div}(Y)$.
}

\stitle{Approximability}. 
We next prove that \divmodis ensures a $\frac{1}{4}$-approximation of diversified size-$k$ Skyline set.
We verify this by proving an invariant that 
the approximation holds  
for any size-$k$ $\epsilon$-Skyline set $D^P_F$ 
generated at every level $i$.
By integrating a greedy selection and replacement policy, \divmodis keeps a 
-set with the most diverse and representative datasets to mitigate the biases in the Skyline set at each level.
Consider a set of datasets $D^{i-1}_F$
at level $i$, and a new batch of
datasets $D^i_F$ arrives. \divmodis
aims to maintain the set of datasets 
$D^i_F$ such that, at level $i$, $|D^i_F| \leq k$, and $\kw{div}(D^i_F)$ is maximized. At any level $i$,
\divmodis approximates the global 
optimal solution upon the newly generated datasets.
Consider the global optimal solution 
at level $i$, over $D_F^i$ as $D_F^{*P}$,
we can show that \divmodis maintains
$D^P_F$ at any level $i$ by solving 
a streaming submodular maximization
problem~\cite{chakrabarti2015submodular}.

\eetitle{Reduction}.
We show there exists an approximation at any level by a 
reduction from the diversification phase of \modis problem 
 to {\em stream submodular maximization problem}~\cite{badanidiyuru2014streaming,chakrabarti2015submodular}.
 Given a streaming of elements $E = \{e_0, \ldots e_m\}$, an integer $k$, and a submodular 
score function $f$, it
computes a set of elements $S$ with size $k$ 
with maximized $f(S)$.
Given the $\epsilon$-Skyline set $\D_F^P$, and integer $k$, the diversification of \modis problem
aims to compute an $\epsilon$-Skyline set ${\D_F}^{P}$ such that (1) $|{\D_F}^{P}| \leq k$
and $\kw{div}({\D_F}^{P})$ is maximized.
Given an instance of diversification of \modis
problem at any level $i$, we construct 
an instance of stream submodular maximization problem by setting (1) $f = \kw{div}$; 
(2) $E =  D_F^i$; (3) integer $k$ is equal
to the value of k in the instance of diversification of \modis.

\eetitle{Correctness}.\divmodis approximates
$D_F^{*P}$ with a ratio $\frac{1}{4}$ follows a greedy selection and replacement policy that integrates the "replace" strategy given the  ${\D_F}^{i}$ at level $i$.
\divmodis always terminates
when no datasets are generated at level $i$ by procedure \upi (See lines 1-3 in Fig.~\ref{alg:divmodis}).

\eetitle{Approximation}.
$\D^{P}_F$ approximates $\D^{*P}_F$ 
with a ratio $\frac{1}{4}$ 
when terminates at level $i$. 
\divmodis exploits the greedy selection 
as in~\cite{chakrabarti2015submodular} but specifies diversification function $\kw{div}(\D^{P}_F)$ to maintain the $\epsilon$-Skyline set of datasets and replaces newly arrived datasets whenever possible.  
It returns the $\epsilon$-Skyline set of datasets those corresponding 
elements in $E$ are selected by the instance stream submodular maximization problem. This ensures a $\frac{1}{4}$ 
approximability by this consistent construction from the solution 
for stream submodular maximization.

The above analysis completes the proof of Lemma~\ref{lemma:div}.

\begin{figure}[tb!]
\centerline{\includegraphics[width =0.5\textwidth]{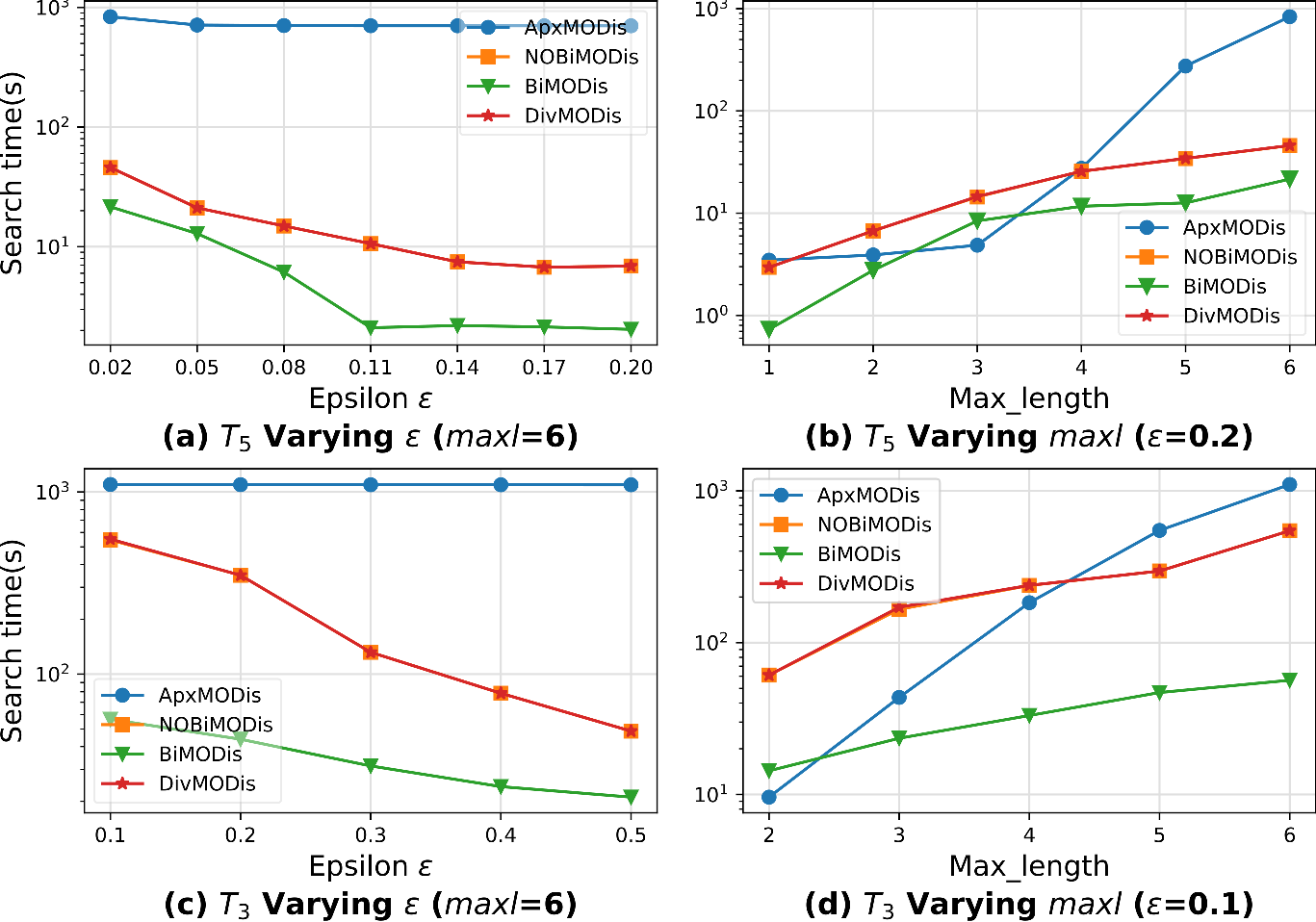}}
\centering
\vspace{-1ex}
\caption{Efficiency Analysis on $T_5$ and $T_3$}
\label{fig:effict53}
\end{figure}

\begin{figure}[tb!]
\centerline{\includegraphics[width =0.5\textwidth]{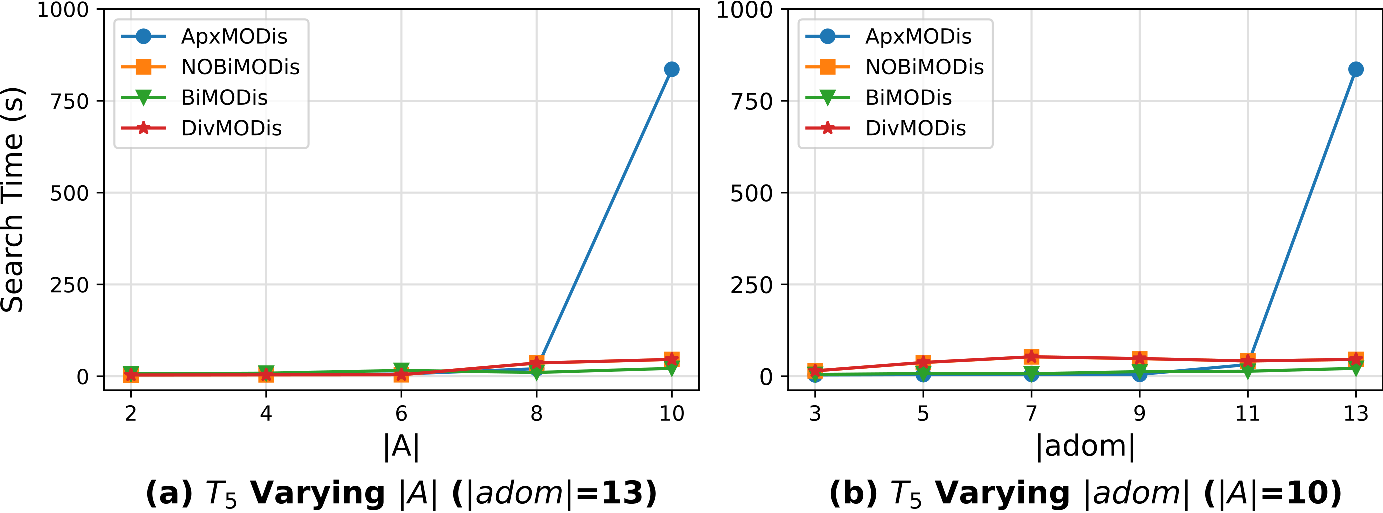}}
\centering
\caption{Scalability on $T_5$}
\label{fig:scala5}
\end{figure}

\begin{table*}[tb!]
\customsize
\centering
\renewcommand{\arraystretch}{1.1}
\begin{small}
\begin{tabular}{|c|c|c|c|c|c|c||c|c|c|c|}
\hline
$T_1$: Movie & Original & \metam & \metammo & \starmie & \sklearn & \ho & \apxmodis & \nomodis & \bimodis & \divmodis \\ \hline
$p_{Acc}$ & 0.8560 & 0.8743 & 0.8676 & 0.8606 & 0.8285 & 0.8545 & 0.9291 & \textbf{0.9874} & \ul{\textit{0.9755}} & 0.9427 \\ \hline
$p_{Train}$ & 1.4775 & 1.6276 & 1.1785 & 1.2643 & \textbf{0.6028} & 0.9692 & 0.9947 & 0.8766 & \ul{\textit{0.8027}} & 0.8803  \\ \hline
$p_{Fsc}$ & 0.0824 & 0.0497 & 0.0801 & 0.1286 & 0.7392 & 0.3110 & 0.6011 & \ul{\textit{0.7202}} & \textbf{0.9240} & 0.8010\\ \hline
$p_{MI}$ & 0.0538 & 0.0344 & 0.0522 & 0.1072 & 0.3921 & 0.1759  & \textbf{0.4178} & 0.3377 & 0.3839 & \ul{\textit{0.4165}}\\ \hline
Output Size  & (3264, 10) & (3264, 11) & (3264, 11) & (3264, 23) & (3264, 3) & (3264, 8) & (2958, 9) & (1980, 12) & (1835, 11) & (2176, 10) \\ \hline

\hline
\hline

$T_3$: Avocado & Original & \metam & \metammo & \starmie & \sklearn & \ho & \apxmodis & \nomodis & \bimodis & \divmodis \\ \hline
MSE & 0.0428 & 0.0392 & 0.0312 & 0.036152 & 0.050903 & 0.0442 & 0.029769 & \textbf{0.022821} & \ul{0.027511} & \ul{0.027511} \\
\hline
MAE & 0.1561 & 0.1497 & 0.1452 & 0.145259 & 0.173676 & 0.1592 & 0.127916 & \textbf{0.115326} & \ul{0.123200} & \ul{0.123200} \\
\hline
Training Time & 0.0280 & 0.0178 & 0.0350 & 0.043600 & 0.008618 & 0.0156 & 0.006516 & \textbf{0.003293} & \ul{0.004366} & \ul{0.004366} \\
\hline
Output Size & (9999, 11) & (9999, 12) & (9999, 12) & (9999, 12) & (9999, 3) & (9999, 5) & (1589, 10) & (817, 5) & (1310, 9) & (1310, 9) \\
\hline

\end{tabular}
\end{small}
\caption{Comparison of Data Discovery Algorithms in Multi-Objective Setting ($T_1$, $T_3$)}
\label{tab:comparison2}
\end{table*}

\eat{
\stitle{Proof of Proposition~\ref{prop-simulate}} 
{\em A data discovery system 
$\T$ can be configured to express (1) data augmentation, 
and (2) feature selection. }
\begin{proof}
\end{proof}

\stitle{Proof of Lemma~\ref{prop-non-blocking}}
{\em The operators of type $\oplus$ and 
$\ominus$ are non-blocking. 
}
\begin{proof}
\end{proof}

\stitle{Proof of Proposition~\ref{prop-equivalent}}
{\em Given a set of operators $\O$ with non-blocking 
operators $\oplus$ and $\ominus$,  
Type 3, Type 4, and Type 5 systems 
have the same expressiveness. 
\ie $L(\T_3)$ = $L(\T_4)$ = 
$L(\T_5)$.} 
\begin{proof}
\end{proof}

\stitle{Proof of Proposition~\ref{thm-cr}}
{\em Given a configuration $C$ with non-blocking operators 
of types $\{\oplus, \ominus\}$, 
and a data discovery 
system $\T$, the running of 
$\T$ is terminating and confluent. }
\begin{proof}
\end{proof}
}

\section{Additional Experiments}
\label{sec:add:exp}

We have performed more complementary 
experimental studies.

\stitle{Efficiency}. 
\textbf{Fig.~\ref{fig:effict53}
(a, b)} evaluates the 
efficiency of \modis algorithms for task $T5$ on generating graph 
data for the link regression task. The observation is consistent with our findings for their counterparts over tabular data. In particular, \bimodis is quite feasible for generating graph data for GNN-based link regression task, with around $20$ seconds in all settings, and consistently 
outperforms other \modis algorithms.
\textbf{Fig.~\ref{fig:effict53}
(c, d)} presents the efficiency results for task $T_3$, which involves avocado price prediction. Similar to other tasks, \bimodis demonstrates superior efficiency, maintaining significantly lower search times compared to other methods. The results reinforce the scalability and practicality of \bimodis across diverse datasets and tasks, including both graph-based and tabular data.

\stitle{Scalability}.
\textbf{Fig.~\ref{fig:scala5}} presents the scalability test results for $T_5$. 
With a universal graph size of $(7925, 34)$, we performed $k$-means clustering on edges, setting $5$ as the minimum number of clusters, $30$ as the maximum, and identifying $13$ as the optimal number of clusters based on performance. For node features, we leveraged the graph's structure to reduce the input feature space from $34$ to $10$ by aggregating attributes from similar types of relations, such as combining multiple training records of an ML model, while preserving all augmented information.
Across all settings, methods applied bi-directional search (\bimodis, \nomodis, and \divmodis) consistently achieve superior efficiency, handling both increasing attributes and active domain sizes effectively. In contrast, \apxmodis exhibits slower performance as $\vert A \vert$ and
$\vert \ad \vert$ grows, highlighting the scalability of the bi-directional search strategy in managing large and complex graph datasets.

\stitle{Effectiveness}. The effectiveness results for $T_1$ and $T_3$ are reported in \textbf{Table~\ref{tab:comparison2}}, where we select the best results from the Skyline set based on the first metric for each task. 
These results align with those observed for other tasks, consistently showing that \modis methods outperform baseline approaches in most cases. Notably, \nomodis and \bimodis secure the first and second positions across the majority of metrics.

\begin{figure*}[h]
\centerline{\includegraphics[width =\textwidth]{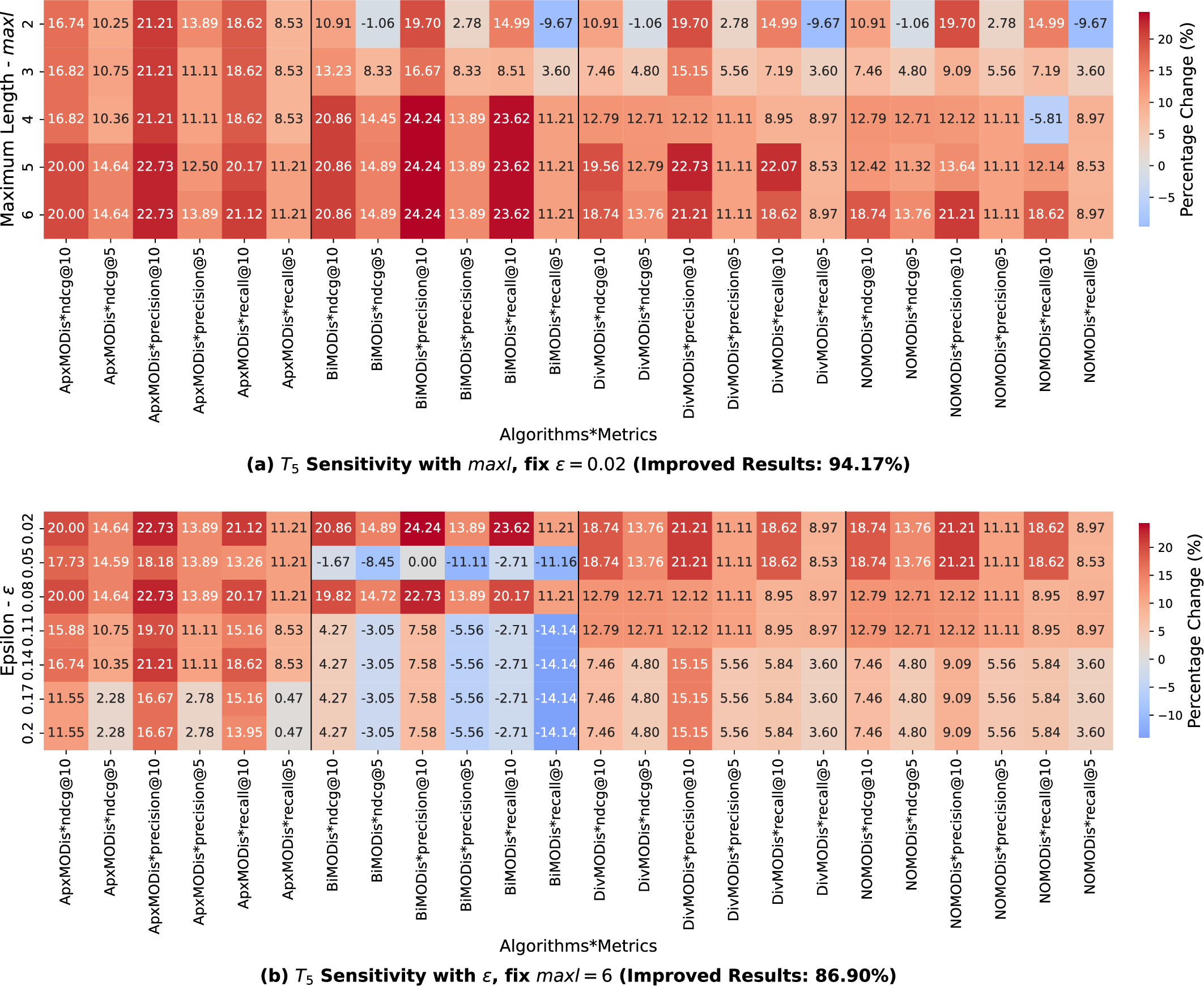}}
\centering
\caption{Sensitivity Analysis for Parameters on $T_5$}
\label{fig:modsnet-sense}
\end{figure*}

\stitle{Sensitivety analysis}.
\textbf{Fig.~\ref{fig:modsnet-sense}}
reports the impact of critical factors including the maximum 
length of paths, and $\epsilon$ (as in $\epsilon$-Skyline set) to 
the accuracy measures. The larger the ``Percentage Change'' is, the better the generated Skyline set can improve the prformance of the input model. We found that all the \modis algorithms benefit from larger maximum length and 
smaller $\epsilon$ in terms of percentage of accuracy improvement. This is consistent with our observation over the tests that report absolute 
accuracy measures. Moreover, \modis algorithms are relatively more sensitive to the maximum length, compared with the changes to $\epsilon$.

\end{document}